\documentclass[twocolumn]{aastex63}
\pdfoutput=1 
\usepackage{amsmath,amstext}
\usepackage[T1]{fontenc}
\usepackage{apjfonts} 
\usepackage[figure,figure*]{hypcap}
\usepackage{longtable}
\usepackage{threeparttable}

\usepackage{amssymb}
\usepackage{amsmath}
\usepackage{fontawesome}
\usepackage{gensymb}
\usepackage{mathrsfs}
\usepackage{upgreek}
\usepackage{fontenc}
\usepackage{hyperref}
\usepackage{float}
\usepackage{tabularx}
\usepackage{booktabs}

\usepackage{comment}

\renewcommand{\d}{\ensuremath{\rm d}}

\graphicspath{{./}{figures/}}

\begin{document}


\title{JWST Reveals Compact Nuclear Starbursts Masquerading as AGNs in Metal-Poor Dwarfs:
Where Are the Accreting Intermediate-Mass Black Holes?}

\author[0000-0003-3937-562X]{William Matzko}
\affiliation{George Mason University, Department of Physics and Astronomy, MS3F3, 4400 University Drive, Fairfax, VA 22030, USA}

\author[0000-0003-2277-2354]{Shobita Satyapal}
\affiliation{George Mason University, Department of Physics and Astronomy, MS3F3, 4400 University Drive, Fairfax, VA 22030, USA}

\author[0000-0002-0913-3729]{Jeffrey D. McKaig}
\affiliation{X-ray Astrophysics Laboratory, NASA Goddard Space Flight Center, Code 662, Greenbelt, MD 20771, USA}
\affiliation{Oak Ridge Associated Universities, NASA NPP Program, Oak Ridge, TN 37831, USA}

\author[0000-0003-3152-4328]{Sara Doan}
\affiliation{George Mason University, Department of Physics and Astronomy, MS3F3, 4400 University Drive, Fairfax, VA 22030, USA}

\author{Michael McDonald} 
\affiliation{Department of Physics and Astronomy, University of California, Riverside, 900 University Avenue, Riverside, CA 92521, USA}

\author[0000-0001-7578-2412]{Archana Aravindan}
\affiliation{The University of Texas at Austin, 2515 Speedway, Austin, TX 78712, USA}

\author[0000-0003-4693-6157]{Gabriela Canalizo}
\affiliation{Department of Physics and Astronomy, University of California, Riverside, 900 University Avenue, Riverside, CA 92521, USA}

\author[0000-0003-1051-6564]{Jenna M. Cann}
\affiliation{X-ray Astrophysics Laboratory, NASA Goddard Space Flight Center, Code 662, Greenbelt, MD 20771, USA}
\affiliation{Center for Space Sciences and Technology, University of Maryland, Baltimore County, 1000 Hilltop Circle, Baltimore, MD 21250}

\author{Nicholas P. Abel} 
\affiliation{College of Applied Science, University of Cincinnati, Cincinnati, OH, 45206, USA}

\author[0000-0003-2722-8841]{Omkar Bait}
\affiliation{National Radio Astronomy Observatory, 520 Edgemont Road, Charlottesville, VA 22903, USA}
\affiliation{The NSF-Simons AI Institute for Cosmic Origins, 201 E. 24th Street, POB 4.102, Austin, Texas 78712-1229, USA}

\author[0000-0002-2183-1087]{Laura Blecha}
\affiliation{University of Florida, Department of Physics, P.O. Box 118440, Gainesville, FL 32611-8440}

\author[0000-0002-5666-7782]{Torsten B\"oker}
\affiliation{European Space Agency, c/o STSCI, 3700 San Martin Drive, Baltimore, MD 21218, USA}

\author[0000-0002-4375-254X]{Thomas Bohn}
\affiliation{ Hiroshima Astrophysical Science Center, Hiroshima University, 1-3-1 Kagamiyama, Higashi-Hiroshima, Hiroshima 739-8526, Japan}

\author[0000-0001-6697-7808]{Jacqueline Fischer}
\affiliation{George Mason University, Department of Physics and Astronomy, MS3F3, 4400 University Drive, Fairfax, VA 22030, USA}

\author[0000-0002-5907-3330]{Stephanie LaMassa}
\affiliation{Space Telescope Science Institute, 3700 San Martin Drive, Baltimore, MD 21218, USA}

\author[0000-0003-3229-2899]{Suzanne C. Madden}
\affiliation{Département d’Astrophysique, Université Paris-Saclay, Gif-sur-Yvette, France}

\author[0000-0001-8440-3613]{Mallory Molina}
\affiliation{Department of Physics \& Astronomy, Vanderbilt University, Nashville, TN 37235, USA}

\author[0000-0003-2283-2185]{Barry Rothberg}
\affiliation{U.S. Naval Observatory, 3450 Massachusetts Avenue NW, Washington, DC 20392, USA}
\affiliation{George Mason University, Department of Physics and Astronomy, MS3F3, 4400 University Drive, Fairfax, VA 22030, USA}

\author[0000-0001-7144-7182]{D. Schaerer}
\affiliation{Observatoire de Genève, Université de Genève, Chemin Pegasi 51, 1290 Versoix, Switzerland}
\affiliation{CNRS, IRAP, 14 Avenue E. Belin, 31400 Toulouse, France}

\author[0000-0003-0248-5470]{Anil Seth}
\affiliation{Department of Physics and Astronomy, University of Utah, 115 South 1400 East, Salt Lake City, UT 84112, USA}

\author[0000-0003-3432-2094]{Remington O. Sexton}
\affiliation{U.S. Naval Observatory, 3450 Massachusetts Avenue NW, Washington, DC 20392, USA}

\correspondingauthor{Shobita Satyapal}
\email{ssatyapa@gmu.edu}

\begin{abstract}
We present JWST/NIRSpec spectroscopy of the low-mass, metal-poor galaxy SDSS~J160135.95+311353.7 (J1601), selected for its extreme mid-infrared colors and compact nuclear emission, placing it within widely used WISE color diagnostics for active galactic nuclei (AGNs). Despite this selection, we find no evidence for coronal lines, X-ray emission, or variability typically associated with accretion activity.

We compare J1601 to SDSS~J120122.30+021108.3 (J1201), a similar but lower-mass, more metal-poor system studied previously \citep{Doan2025}. Both galaxies host compact nuclear starbursts but differ in their stellar populations and dust properties: J1601 shows CO bandhead absorption indicative of red supergiants, weak nuclear Wolf--Rayet features, and a circumnuclear PAH ring, consistent with a more developed recent starburst, while J1201 is more dust-enshrouded and chemically primitive. Despite these differences, neither system shows evidence for AGN activity, indicating that the absence of accretion is not simply due to evolutionary timing.

Photoionization models show that the weakness of high-ionization emission cannot be explained by low metallicity alone, implying a genuine deficit of hard ionizing photons. Crucially, the red mid-infrared colors in both systems originate from compact, unresolved nuclear emission confined to the nuclear star cluster.

These results demonstrate that compact nuclear starbursts can mimic AGN-like mid-infrared colors without accretion, and that commonly used AGN diagnostics may not uniquely identify accreting black holes in metal-poor dwarf galaxies. Our findings suggest that such systems may not provide the conditions required for efficient black hole growth and/or may lie near or below the regime where black hole seeds can form.

\end{abstract}

\section{Introduction}

JWST has revolutionized our view of the early universe, uncovering a surprisingly abundant population of compact broad-line sources extending to redshifts as high as $z \approx 9$--11 \citep[e.g.,][]{2024Natur.627...59M, 2024NatAs...8..126B, 2023ApJ...959...39H, 2025A&A...693A..50N}. These discoveries imply that massive black holes were already in place when the universe was only a few hundred million years old, challenging current theories of black hole formation, growth, and co-evolution with galaxies. Yet despite several years of intensive follow-up, the physical nature of these sources remains uncertain. Many appear extremely compact, often lack clearly detected host galaxies, and show unusual `V-shaped'' optical continua, giving rise to the class of sources dubbed `little red dots'' (LRDs; \citealt{2024ApJ...963..129M}). Most show no X-ray or radio counterparts \citep[e.g.,][]{2024ApJ...969L..18A,2025MNRAS.538.1921M,2024arXiv241204224M}, high-ionization emission lines \citep[e.g.,][]{2025arXiv250818358W}, variability \citep[e.g.,][]{2024arXiv240704777K}, or signatures of a hot dusty torus \citep[e.g.,][]{2025ApJ...991L..10S}. Some even exhibit Balmer breaks atypical of local active galactic nuclei (AGNs) \citep[e.g.,][]{2025arXiv250113082J}. As a result, the astronomical community remains divided as to whether all of these compact sources are powered by accreting black holes, extreme stellar populations, or more exotic physical processes. Recent models propose that at least some LRDs may represent early stages of black hole growth in which an accreting black hole is embedded in a dense gas envelope or quasi-star-like cocoon, producing unusual continua, weak high-energy signatures, and non-standard line diagnostics \citep[e.g.,][]{2026ApJ...996...48B,2025arXiv250316596N,2025A&A...701A.168D,2026arXiv260606575G}.

Despite this growing range of interpretations, the apparent abundance of compact red and/or broad-line sources at high redshift raises a fundamental question: why are analogous systems not observed with comparable frequency in the local universe? Rare low-redshift objects with compact morphologies and unusual optical continua, broad Balmer emission, or high-ionization emission lines have been identified in metal-poor dwarf galaxies \citep[e.g.,][]{2007ApJ...671.1297I,2008ApJ...687..133I,2016A&A...596A..64S,2020ApJ...895..147C,2021MNRAS.504..543B,2026arXiv260521574L,2026MNRAS.545f2235J,2026arXiv260514233P}, but such systems do not share all the properties of high-redshift broad-line objects or LRDs and remain extremely rare in the local universe. If the compact sources uncovered by JWST represent a common early phase of black hole growth, their local counterparts or descendants should be sought in nearby low-mass, metal-poor galaxies, where chemically primitive gas, compact star formation, and dense nuclear environments may reproduce some of the physical conditions present in the early universe. The path forward may therefore not lie solely in pushing to ever higher redshift, but also in using spatially resolved observations of nearby metal-poor dwarfs to test whether compact nuclear emission in such systems is powered by accreting black holes or by extreme star formation.

Nearby metal-poor dwarf galaxies provide a critical testing ground for this picture. If massive black holes were already common in compact, chemically primitive galaxies at early cosmic times, then their lower-mass counterparts may be expected to reside in low-mass galaxies in the local universe. This makes such systems natural places to search for intermediate-mass black holes (IMBHs), with characteristic masses of $\sim10^2$--$10^5,M_\odot$, which are central to models of supermassive black hole seed formation and early growth \citep[see review by][]{Greene2020}. Yet observational evidence for accreting black holes in the lowest-mass, most metal-poor galaxies remains sparse, leaving open the question of whether IMBHs are intrinsically rare in this regime or simply difficult to detect. Because kinematic detections are not currently feasible at these masses beyond the Local Group, accretion signatures remain the only practical pathway for identifying IMBHs.

Uncovering accretion activity in dwarf galaxies, however, is intrinsically challenging. Expected accretion luminosities are low, massive metal-poor stars can dominate the ionizing radiation field, and both gas-phase metallicities and black hole masses are small---conditions that severely compromise the effectiveness of traditional AGN diagnostics. Optical spectroscopic surveys \citep{Groves2006,Cann2019}, X-ray searches \citep{2015ApJ...798...38S}, radio observations \citep{1991ApJ...378...65C,2022ApJ...933..160S}, and broad-band mid-infrared selections \citep{satyapal2018} are incomplete in the low-mass regime and tend to favor more massive black holes in higher-mass, metal-rich host galaxies, where dilution by star formation and obscuration by dust are less severe. The vast majority of currently identified AGNs therefore reside in galaxies that do not resemble the chemically primitive systems thought to dominate the early universe \citep{2019NatAs...3....6M}.

The challenge is further exacerbated at low metallicity. Optical narrow-line diagnostics fail to reliably identify AGNs in chemically primitive environments \citep[e.g.,][]{Groves2006,Cann2019}, producing line ratios indistinguishable from those of star-forming galaxies. This limitation applies not only locally, but also to JWST's high-redshift AGN candidates, where rest-frame optical recombination lines dominate the spectra \citep[e.g.,][]{2024A&A...691A.145M,2025A&A...700A..12M}. Moreover, for black hole masses below $\sim10^{5},M_\odot$, the expected widths of broad emission lines are comparable to those of H~\textsc{ii} regions, rendering broad-line searches ineffective. These effects leave open the possibility that the absence of detected AGNs in metal-poor dwarf galaxies may not reflect a true absence of black holes, but rather the failure of conventional diagnostics in precisely the regime most relevant for early black hole growth.

Motivated by these challenges, we initiated a JWST program to search for accreting IMBHs in nearby metal-poor dwarf galaxies that lack optical spectroscopic and X-ray evidence for AGN activity. Because broad-line searches and X-ray surveys become increasingly ambiguous in the IMBH regime, and optical narrow-line diagnostics can fail at low metallicity, we adopted mid-infrared color selection as a complementary route into a parameter space that remains poorly explored by current AGN searches in dwarf galaxies. The all-sky coverage of the Wide-field Infrared Survey Explorer (WISE; \citealt{2010AJ....140.1868W}) enables a wide-area search for low-mass, metal-poor galaxies with unusually hot dust emission. Such red mid-infrared colors may signal dust heated by accretion, but they could also arise from compact, embedded star formation under physical conditions not fully captured by standard templates. A key goal of our program is therefore to test whether extreme mid-infrared colors in the lowest-mass, lowest-metallicity galaxies uniquely trace accretion, or whether compact nuclear starbursts can occupy the same diagnostic space.

In the first paper of this series \citep{Doan2025}, we presented JWST/NIRSpec integral field spectroscopy and XMM-Newton observations of SDSS~J120122.30+021108.3 (hereafter J1201), one of the lowest-mass and lowest-metallicity galaxies in the local universe known to display AGN-like mid-infrared colors. Despite its unresolved nuclear continuum source, extreme compactness, and red mid-infrared colors, J1201 showed no evidence for coronal line emission, X-ray activity, or variability. Instead, the data pointed toward an unusually compact, dusty nuclear starburst as the dominant power source.

In this paper, we extend our investigation to SDSS~J160135.95+311353.7 (hereafter J1601), a more massive and slightly less metal-poor dwarf galaxy that lies in a region of mid-infrared color space traditionally associated with AGNs and exhibits similarly compact nuclear emission. Like J1201, J1601 is a stringent test case for mid-infrared-selected IMBH searches: it has low stellar mass, low metallicity, compact morphology, and extreme nuclear conditions. We combine JWST/NIRSpec spectroscopy, which resolves the near-infrared nuclear continuum and provides sensitive constraints on infrared coronal lines, with Keck/KCWI optical integral-field spectroscopy, which probes the larger-scale ionized gas, metallicity, extinction, and optical high-ionization diagnostics, including He,\textsc{ii}. As we show in this work, J1601 lacks coronal line emission and other signatures of accreting black holes, and instead appears dominated by a compact nuclear starburst with hot dust.

The comparison between J1201 and J1601 provides a direct test of whether extreme mid-infrared colors in compact, metal-poor dwarfs uniquely identify accreting IMBHs. The key questions addressed in this paper are: (1) whether J1601 hosts a coronal line emitting region, or shows any other signatures expected from an accreting black hole; (2) whether the red mid-infrared colors in these systems originates from extended star formation or compact nuclear sources; and (3) what these systems imply for the environments in which IMBHs can form, grow, and become observable. This work constrains both the reliability of AGN diagnostics in chemically primitive galaxies and the physical nature of compact nuclear starbursts in the low-mass regime.

This paper is organized as follows. Section~2 describes the target selection and places J1601 and J1201 in the context of dwarf AGN candidates, metal-poor high-ionization emitters, and WISE color diagnostics. Sections~3 and 4 summarize the JWST/NIRSpec and Keck/KCWI observations and data reduction. Section~5 presents the main results, including the morphology of the ionized gas and continuum emission, the nuclear emission-line spectrum, metallicity and extinction constraints, stellar-population diagnostics, and multiwavelength limits on AGN activity. Section~6 discusses the implications for compact nuclear starbursts, the detectability of accreting IMBHs in metal-poor dwarfs, and the reliability of mid-infrared and high-ionization AGN diagnostics. Section~7 summarizes the main conclusions.

For J1601, we assume a flat $\Lambda$CDM cosmology with $H_0=70$~km~s$^{-1}$~Mpc$^{-1}$, $\Omega_\mathrm{M}=0.3$, and $\Omega_\Lambda=0.7$. The resulting luminosity distance is $D_L=135$~Mpc, consistent with that obtained using the Cosmicflows-3 local velocity flow model \citep{2019MNRAS.488.5438G,2020AJ....159...67K}.

\section{Target Selection}
The observations presented in this work were obtained as part of a Cycle~1 JWST General Observer program (GO~1983; PI: Satyapal) designed as a pilot study to probe the lowest-mass and most metal-poor regime in the search for accreting intermediate-mass black holes (IMBHs). Rather than conducting a broad survey, the strategy of this program was to make a decisive advance in the IMBH mass frontier with a modest investment of observing time by targeting a small number of nearby dwarf galaxies selected to maximize sensitivity to accretion signatures inaccessible to previous facilities.

\begin{figure*}
\centering
\includegraphics[width=\columnwidth]{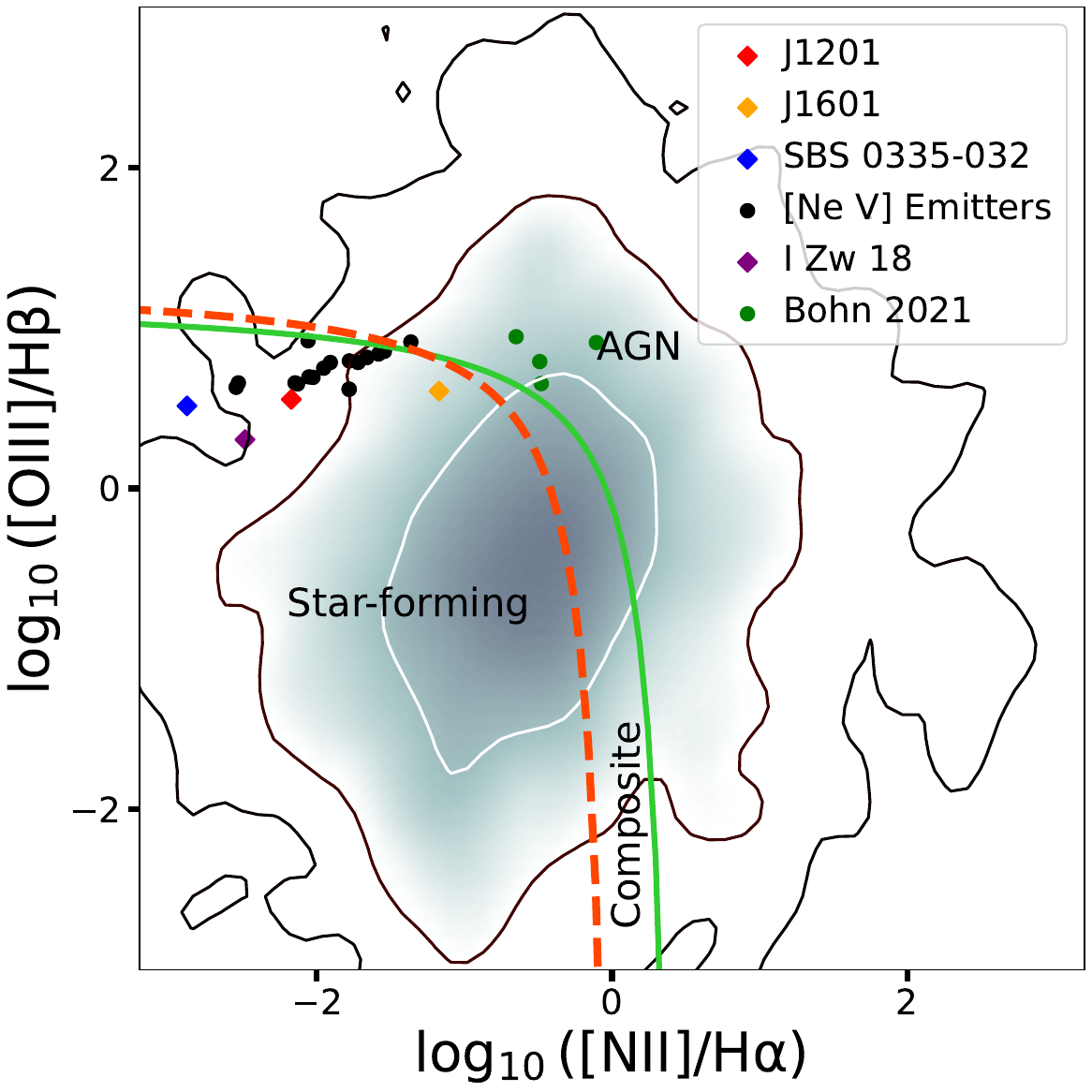}
\includegraphics[width=\columnwidth]{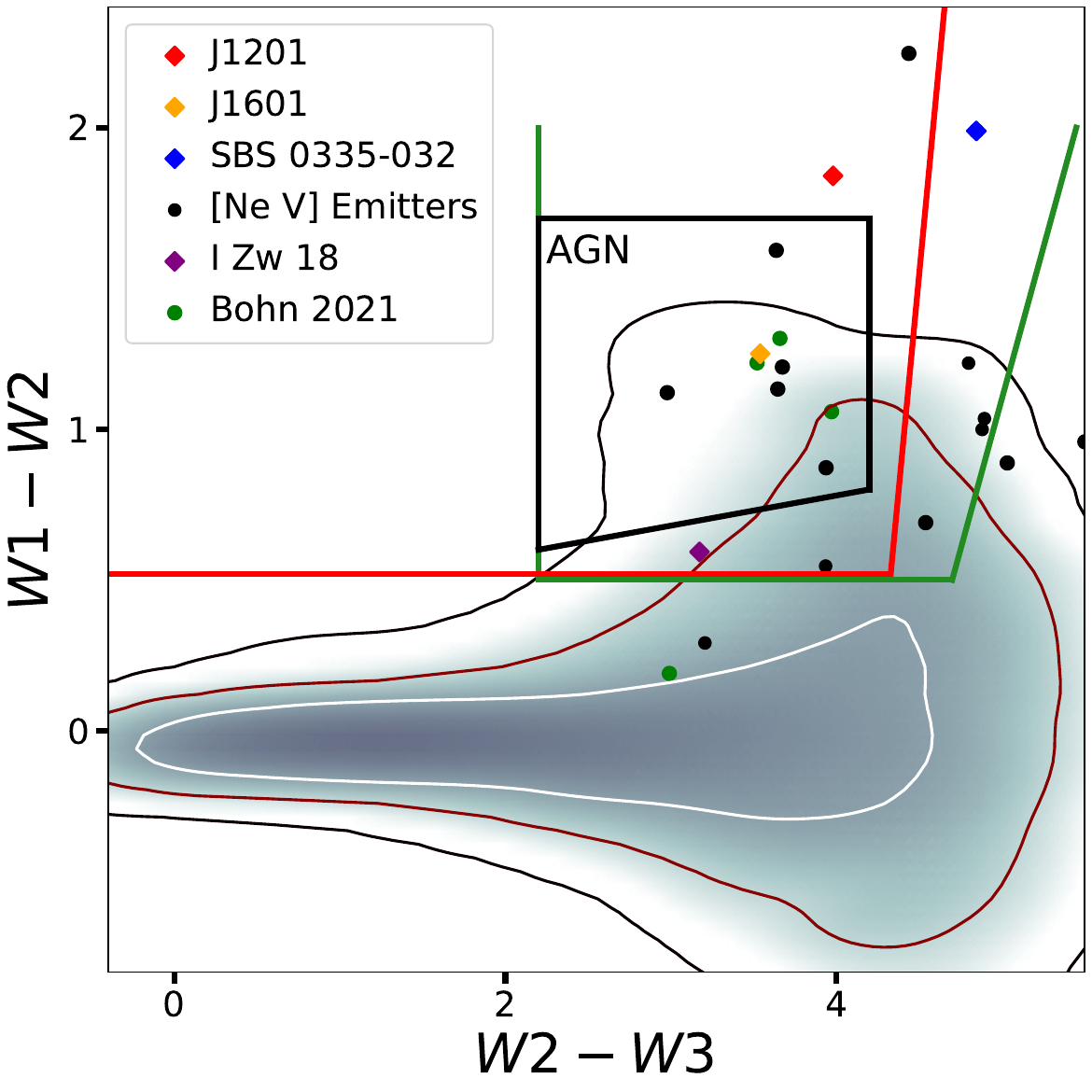}

\caption{\textit{Left}: A BPT ratio plot showing the narrow line ratios of J1601 and J1201 from SDSS indicated by the orange and red diamonds, respectively. We also display the narrow line ratios for the sample of BCDs showing [\ion{Ne}{5}] emission from \citep{2012MNRAS.427.1229I,2021MNRAS.508.2556I,2024ApJ...966..170H}, which includes SBS 0335-032, another metal--poor dwarf with extreme MIR colors and MIR variability suggestive of accretion activity \citep{2023arXiv230403726H}. The star-forming, ``composite,'' and AGN regions are marked with text and separated by the curves defined in \citet{2003MNRAS.346.1055K} (solid green line) and \citet{2001ApJ...556..121K} (dashed red line).  The grey shading displays the BPT values for the entire sample of dwarf galaxies from \citet{2013ApJ...775..116R}. We also show the BPT ratios of the dwarf galaxies with AGNs displaying outflows from \citep{Bohn2021}.
\textit{Right}: A WISE color-color plot with the value for J1601 displayed by the orange diamond. We also show the colors of the [\ion{Ne}{5}] emitting dwarfs and the dwarfs from \citep{Bohn2021} as in the BPT plot. The AGN demarcation box from \citet{jarrett2011} is shown by the black line, the demarcation box from \citet{satyapal2018} is shown by the red line, and the demarcation box from \citet{2018MNRAS.478.3056B} is shown in green.}
\label{fig:bptwise}
\end{figure*}

The program targeted two compact, nearby dwarf galaxies with extreme mid-infrared colors suggestive of AGN activity, but with no optical spectroscopic evidence for accretion. Both galaxies are significantly less massive than the vast majority of AGN candidates identified in dwarf galaxy surveys based on optical emission-line diagnostics \citep[e.g.,][]{2013ApJ...775..116R, 2020ARA&A..58..257G}, as well as more recent samples identified with DESI \citep[e.g.,][]{2024ApJ...961..173Z, Pucha2025}. Their proximity enables spatial scales and sensitivities that are unattainable for comparable systems at higher redshift, while their compact morphologies and low metallicities make them compelling local analogs of the galaxies uncovered by JWST at early cosmic times.

Mid-infrared color selection was adopted as a complementary route for identifying candidate accreting IMBHs. AGNs can heat dust to high temperatures, producing red infrared colors that are difficult to reproduce with stellar populations alone \citep[e.g.,][]{2007ApJ...660..167D,2012ApJ...753...30S,2013MNRAS.434..941M}. The two targets were selected from the subset of low-mass ($M_* \lesssim 5\times10^8,M_\odot$), low-metallicity ($Z \lesssim 0.1,Z_\odot$) dwarf galaxies that satisfy the three-band mid-infrared color criteria of \citet{satyapal2018}. Under the assumptions of those photoionization and stellar-population models, such colors were not expected from even extreme metal-poor stellar populations. The JWST observations therefore provide a direct test of whether the hot dust emission in these systems is powered by accretion or by compact nuclear star formation.

\par

The target selection was intentionally designed to be complementary to previous IMBH searches. We deliberately excluded galaxies exhibiting optical broad emission lines or narrow-line ratios indicative of AGN activity. Broad-line searches become increasingly ambiguous for black holes with masses $\lesssim10^{5},M_\odot$, where expected line widths are comparable to those of \ion{H}{2} regions, while theoretical work predicts that the harder radiation fields produced by accreting IMBHs can alter nebular ionization in ways that render standard optical diagnostics ineffective \citep{Cann2019,2025ApJ...994..146C}. We also did not pre-select targets with detected optical [\ion{Ne}{5}] emission. Although optical [\ion{Ne}{5}] is highly suggestive of a hard ionizing source, its origin in metal-poor systems remains uncertain, with proposed contributions from extreme stellar populations, X-ray binaries, and shocks \citep[e.g.,][]{2005ApJS..161..240T,2012MNRAS.421.1043S,2021MNRAS.508.2556I}. Moreover, optically identified [\ion{Ne}{5}] emitters are rare, and selecting only such objects would limit the search to a small and already unusual subset of metal-poor dwarfs. Our goal was instead to test whether mid-infrared-selected galaxies without conventional optical AGN signatures could reveal a larger population of accreting IMBH candidates through infrared coronal lines, which are intinsically more luminous, less affected by extinction , and can provide more sensitive probes of the hard ionizing radiation associated with accreting IMBHs \citep{Cann2018} than optical lines in dusty or compact nuclear environments.

The two targets were selected to span complementary regions of parameter space within this pilot program. J1601 was selected because previous near-infrared spectroscopy suggested a tentative detection of [\ion{Si}{6}] $\lambda1.96,\mu$m \citep{Cann2021}, making it an ideal test case for confirming or refuting a candidate coronal-line detection with JWST's improved sensitivity and spatial resolution.
The second target, J1201, was selected to push the stellar-mass frontier to the lowest values ever probed in previous surveys (Paper I). With a stellar mass of $\log(M_*/M_\odot)=6.09$, J1201 is more than three orders of magnitude less massive than the Large Magellanic Cloud and occupies a region of parameter space unexplored by previous AGN surveys.

Figure~\ref{fig:bptwise} shows the locations of J1201 and J1601 on the BPT diagram and in WISE color--color space. Both galaxies lie in the star-forming region of the BPT diagram, confirming that they would not be identified as AGNs by standard optical narrow-line diagnostics. In contrast, both have unusually red mid-infrared colors for low-mass, metal-poor galaxies. J1601 falls within the widely used WISE AGN selection regions, while J1201 lies near the Jarrett et al. wedge and satisfies the Satyapal et al. three-band MIR selection. Their location in this diagnostic space makes them useful test cases for determining whether hot dust in metal-poor dwarfs is powered by accretion or by compact nuclear star formation. We also compare to other JWST dwarf-galaxy AGN candidates with coronal-line emission where relevant.



The \textit{WISE} color--color diagram further highlights that many metal-poor dwarf galaxies displayed in Figure~\ref{fig:bptwise}, including several [\ion{Ne}{5}] emitters, are characterized by red mid-infrared colors suggestive of hot dust. However, even among this population, the locations of J1201 and J1601 are on the extreme end compared to the overall population. Their combination of compact morphology, low metallicity, and AGN-like mid-infrared colors makes them exceptional targets for testing whether hot nuclear dust in metal-poor systems is powered by accretion onto an intermediate-mass black hole or by compact nuclear starbursts.

\begin{figure}
    \centering
    \includegraphics[width=\columnwidth]{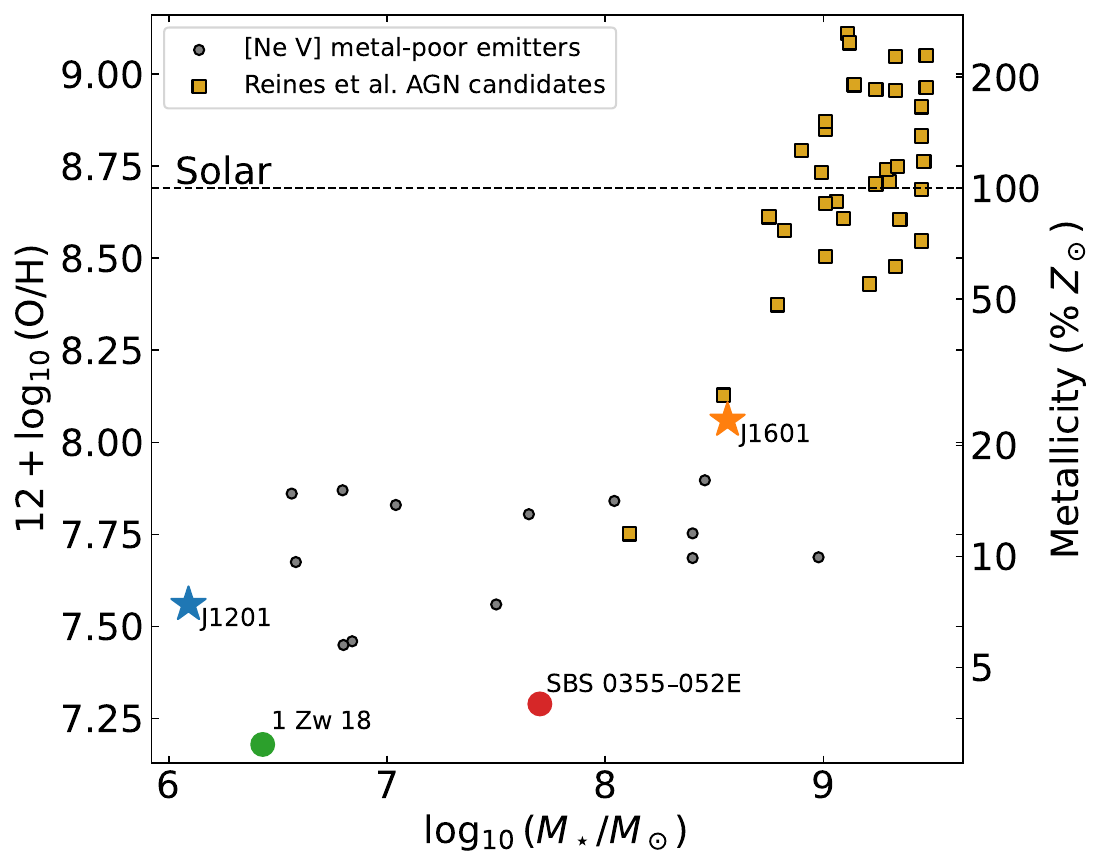}
    \caption{
Stellar mass and metallicity of J1201 and J1601 in relationship with previously identified dwarf AGN candidates and metal-poor systems. Optically selected dwarf AGNs from \citet{2013ApJ...775..116R} are shown for comparison, along with metal-poor [\ion{Ne}{5}] emitters from \citet{2001ApJ...560..630L,2012MNRAS.427.1229I, 2019ApJ...885...96J, 2021MNRAS.508.2556I, 2024ApJ...966..170H}. Benchmark extremely metal-poor galaxies SBS~0335$-$052E and I~Zw~18 are labeled explicitly. The solid horizontal line marks solar metallicity, with the right-hand axis indicating metallicity in units of percent solar.
J1201 and J1601 occupy the low-mass, low-metallicity regime at or below the bulk of known dwarf AGN candidates, highlighting their unique leverage for probing black hole activity in chemically primitive environments.
}

    \label{fig:mass_metallicity}
\end{figure}

In Figure~\ref{fig:mass_metallicity}, we show the stellar mass and metallicity of the targets, together with dwarf AGNs from the literature, [\ion{Ne}{5}] emitters, and benchmark extremely metal-poor systems such as SBS~0335$-$052E and I~Zw~18. Both targets reside at lower stellar masses and metallicities than the bulk of known dwarf AGN candidates, highlighting their unique leverage for probing accretion physics in chemically primitive environments. We note that we intentionally do not include coronal-line–selected dwarf galaxies from the CLASS survey \citep{2022ApJ...936..140R} or the [\ion{Fe}{10}] emitter sample of \citet{2021ApJ...922..155M} in Figure~\ref{fig:mass_metallicity}. While the CLASS survey identifies 24 dwarf galaxies with stellar masses at or below that of J1601, the majority of these sources are classified based on [\ion{Fe}{10}] emission, and one corresponds to a [\ion{Ne}{5}] emitter already identified in previous studies. Recent work has shown that [\ion{Fe}{10}] $\lambda6374$ can be misidentified as \ion{Si}{2} $\lambda6371$ in low-resolution or low signal-to-noise spectra \citep{2023RNAAS...7...99H}, introducing additional uncertainty in the interpretation of [\ion{Fe}{10}]–selected samples in metal-poor systems. Similarly, the [\ion{Fe}{10}] emitter sample of \citet{2021ApJ...922..155M} primarily probes more massive galaxies, with only 12 out of 81 objects having stellar masses below that of J1601. Given both the potential ambiguity in [\ion{Fe}{10}] identifications and the focus of this work on the lowest-mass, most metal-poor regime, we restrict the comparison in Figure~\ref{fig:mass_metallicity} to dwarf AGN candidates selected via optical diagnostics, metal-poor [\ion{Ne}{5}] emitters identified in the literature. A summary of their global properties is provided in Table~\ref{tab:globalprops}. 

\begin{table*}[t]
\centering
\caption{Global Properties of J1201 and J1601}
\label{tab:globalprops}
\begin{tabular}{lcc}
\hline\hline

\multicolumn{3}{c}{Basic Properties} \\
\hline
Property & J1201 & J1601 \\
\hline
Right Ascension & 12:01:22.326 & 16:01:35.950 \\
Declination & +02:11:08.67 & +31:13:53.69 \\
Redshift $z$ & 0.0035 & 0.0309 \\
Physical Scale (pc arcsec$^{-1}$) & 72.2 & 604 \\
Upper Limit on Nuclear Size (pc) & $\lesssim 7$ & $\lesssim 60$ \\

\hline
\multicolumn{3}{c}{Host Galaxy Properties} \\
\hline
$r$-band Half-light Radius $R_e$ (arcsec) & 6.66 & 2.20 \\
$r$-band Half-light Radius $R_e$ (kpc) & 0.48 & 1.33 \\
$\log(M_\star/M_\odot)$ & 6.09 & 8.56 \\
$\log(\mathrm{SFR}/M_\odot\,\mathrm{yr^{-1}})$ & $-1.99$ & $-0.44$ \\
$\log(\mathrm{sSFR}/\mathrm{yr^{-1}})$ & $-8.15$ & $-9.04$ \\
$12 + \log(\mathrm{O/H})$ & 7.67 & 8.08 \\
$Z/Z_\odot$ (\%) & 9.6 & 24.5 \\
Relative to LMC Mass & $\sim10^{-3}$ & $\sim10^{-1}$ \\

\hline
\multicolumn{3}{c}{Mid-Infrared Properties (WISE)} \\
\hline
$W1 - W2$ (mag) & 1.835 & 1.243 \\
$W2 - W3$ (mag) & 3.984 & 3.543 \\
Jarrett et al. WISE AGN Wedge\tablenotemark{a} & No / Marginal & Yes \\
Satyapal et al. MIR AGN Region\tablenotemark{b} & Yes & Yes \\

\hline
\multicolumn{3}{c}{AGN Diagnostics} \\
\hline
BPT Classification & Star-forming & Star-forming \\
Optical [\ion{Ne}{5}] & No & No \\
Broad Emission Lines & No & No \\

\hline
\multicolumn{3}{c}{Context Within Comparison Samples} \\
\hline
Position Relative to Dwarf AGN Samples & Among the lowest-mass known & At the low-mass end \\
Position Relative to Metal-poor AGN Candidates & Lowest-mass, lowest-metallicity regime & Low-mass, low-metallicity regime \\

\hline
\end{tabular}

\tablecomments{
Distances assume a standard $\Lambda$CDM cosmology with $H_0 = 70~\mathrm{km~s^{-1}~Mpc^{-1}}$. 
Metallicity relative to solar is computed using $(\mathrm{O/H})/(\mathrm{O/H})_\odot = 10^{X - 8.69}$, where 
$X = 12 + \log(\mathrm{O/H})$. At $\sim2\,\mu$m, the JWST/NIRSpec IFU point-spread function (PSF) has a full width at half maximum of $\approx0\farcs07$. The data are sampled with $0\farcs1$ spaxels, so the effective spatial resolution of the final data cubes is of order $0\farcs1$. The quoted upper limits on the nuclear sizes assume that the sources are unresolved at this scale, corresponding to $\lesssim7$~pc and $\lesssim60$~pc for J1201 and J1601, respectively. J1201 has a stellar mass more than three orders of magnitude below that of the Large Magellanic Cloud (LMC), while J1601 is approximately an order of magnitude less massive. \tablenotemark{a}\ \citet{jarrett2011}. \tablenotemark{b}\ \citet{satyapal2018}.
}
\end{table*}

\section{JWST NIRSpec Observations and Data Reduction}

J1601 was observed with the NIRSpec integral field unit (IFU) \citep{2022A&A...661A..82B, 2022A&A...661A..80J} on 1 June 2023 as part of our Cycle~1 program. Observations were obtained using the medium-resolution gratings G235M/F170LP and G395M/F290LP in the NSIRS2RAPID readout mode with 60 groups and 2 integrations at four dither positions, resulting in a total on-source exposure time of $\sim7100$~s per grating/filter combination. This setup provides continuous wavelength coverage from $\sim1.6$ to $5.2~\mu$m with a resolving power of $R \sim 1000$. The $3\arcsec \times 3\arcsec$ field of view corresponds to $\sim2 \times 2$~kpc at the distance of J1601.

To maximize observing efficiency, a single LeakCal exposure was obtained to correct for light leakage from permanently open microshutters.

The raw data were downloaded from the Barbara A. Mikulski Archive for Space Telescopes (MAST) and processed using the JWST Science Calibration Pipeline (version 1.14.0; \citealt{jwst_pipeline}) with Calibration Reference Data System (CRDS) context \texttt{jwst\_1100.pmap}. Detector-level corrections were applied using the \texttt{Detector1} stage, followed by calibration and cube construction using the \texttt{calwebb\_spec2} and \texttt{calwebb\_spec3} pipelines. The final data cubes were generated using the drizzle algorithm with a spaxel scale of $0\farcs1$.

The resulting data cubes exhibit low-level sinusoidal modulations in individual spaxels, commonly referred to as NIRSpec IFU ``wiggles,'' caused by undersampling of the point-spread function. These features are negligible for spectra extracted over apertures larger than $\sim0\farcs2$, but can affect analyses at the spaxel level. At the time the Cycle~1 data were reduced and analyzed, the newer public implementations for mitigating these artifacts were not yet available. We therefore applied a custom spaxel-level correction procedure before constructing line maps. A full description of the correction method is provided in Appendix~\ref{sec:wiggles}. We have since verified that applying the newer wiggle-correction implementations produces consistent results and does not change the aperture-integrated spectra, line identifications, or scientific conclusions presented in this work.

\subsection{Spectral Extraction,  Line Identification, and Fitting}
\label{subsec:line_identification_fitting}

\begin{figure}
\centering
\includegraphics[width = \columnwidth]{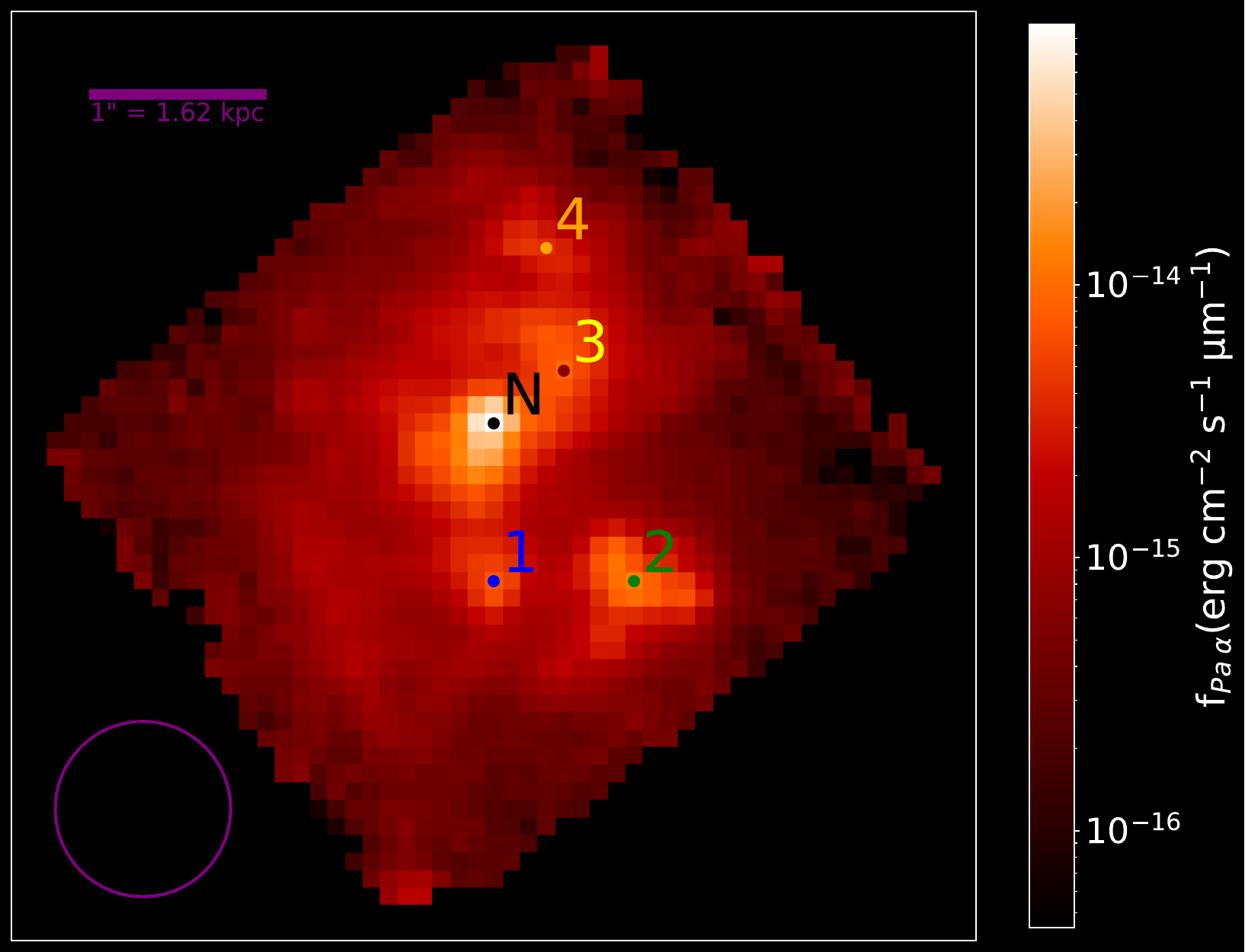}
\caption{Pa$\alpha$ emission map of J1601 from JWST/NIRSpec. A bright central (nuclear) source is observed, along with two knots of emission south of the nucleus and a faint extension toward the northwest. The dots indicate the centers of the spectral extraction apertures, each with a radius of $0\farcs5$ (shown in the lower left). North is up and east is to the left.}
\label{fig:J1601_aperture_image}
\end{figure}

To probe spatial variations in the emission properties, we extracted spectra from five circular apertures of radius $0\farcs5$ centered on representative morphological regions within the field of view (Figure~\ref{fig:J1601_aperture_image}). These apertures were chosen to balance spatial resolution and signal-to-noise, and to sample the nuclear source, the southern emission knots, and lower-surface-brightness extended emission.

To ensure robust line identification, we visually inspected the spectra from each aperture and from the individual dither exposures. A feature was retained only if it was detected in the combined aperture spectrum at the expected wavelength and was independently recovered in at least two of the four dither exposures. We adopted this requirement as a reproducibility criterion rather than requiring detection in all four dithers, because the individual dither spectra have lower signal-to-noise than the combined cube and sample different detector pixels and IFU slices; weak but real emission lines can therefore fall below the detection threshold in one or more individual dithers. Features seen in only a single dither, or not recovered in the combined aperture spectrum, were classified as spurious and excluded from the analysis. The line list used for quantitative measurements and AGN constraints is therefore based on features detected in the aperture spectra, reproducible in independent dithers, and consistent with known atomic or molecular transitions.
\par
Because a compact nuclear coronal line could be diluted in the larger regional apertures, we performed a separate targeted search for coronal-line emission at the expected wavelengths in the nuclear spaxels, in small apertures centered on the continuum peak, and in the corresponding line maps. This search was not limited to features identified in the $0\farcs5$ aperture spectra. Candidate coronal-line features were required to appear at the expected wavelength and be reproducible in independent dither subsets or in the combined cube at a significance consistent with the local noise. No such features were detected. The coronal-line upper limits reported below are therefore based on targeted nuclear extractions rather than on the broader aperture line-identification procedure alone.
\par

Spectral fitting was performed using the open-source Python code \textsc{BADASS} \citep{sexton_2020}, adapted for the JWST wavelength range. Each spectrum was modeled with Gaussian line profiles and a third-order Legendre polynomial continuum. Initial parameter estimates were obtained via likelihood maximization, followed by Markov Chain Monte Carlo (MCMC) sampling using the affine-invariant \textsc{emcee} algorithm \citep{2013PASP..125..306F} to derive uncertainties.

Emission lines were initially identified manually in the aperture spectra and matched to known atomic and molecular transitions within $\pm100~\mathrm{km,s^{-1}}$. Flux maps generated from these fits were used to verify that the identified lines produced coherent spatial structures and to distinguish between atomic and molecular emission components. Comparisons with photoionization models based on low-metallicity stellar populations further guided line identification.

For spaxel-level analysis, we restricted the fitting to lines already identified in the aperture spectra, except for the targeted coronal-line searches described above. Candidate features identified only in individual spaxels were inspected for spatial coherence, but were not used for the primary nuclear line list or physical interpretation unless independently supported by the aperture spectra or by the targeted nuclear coronal-line analysis.

\section{Keck KCWI Observations and Data Reduction}
\label{sec:kcwi}

\begin{figure*}
\centering
\includegraphics[width=\textwidth]{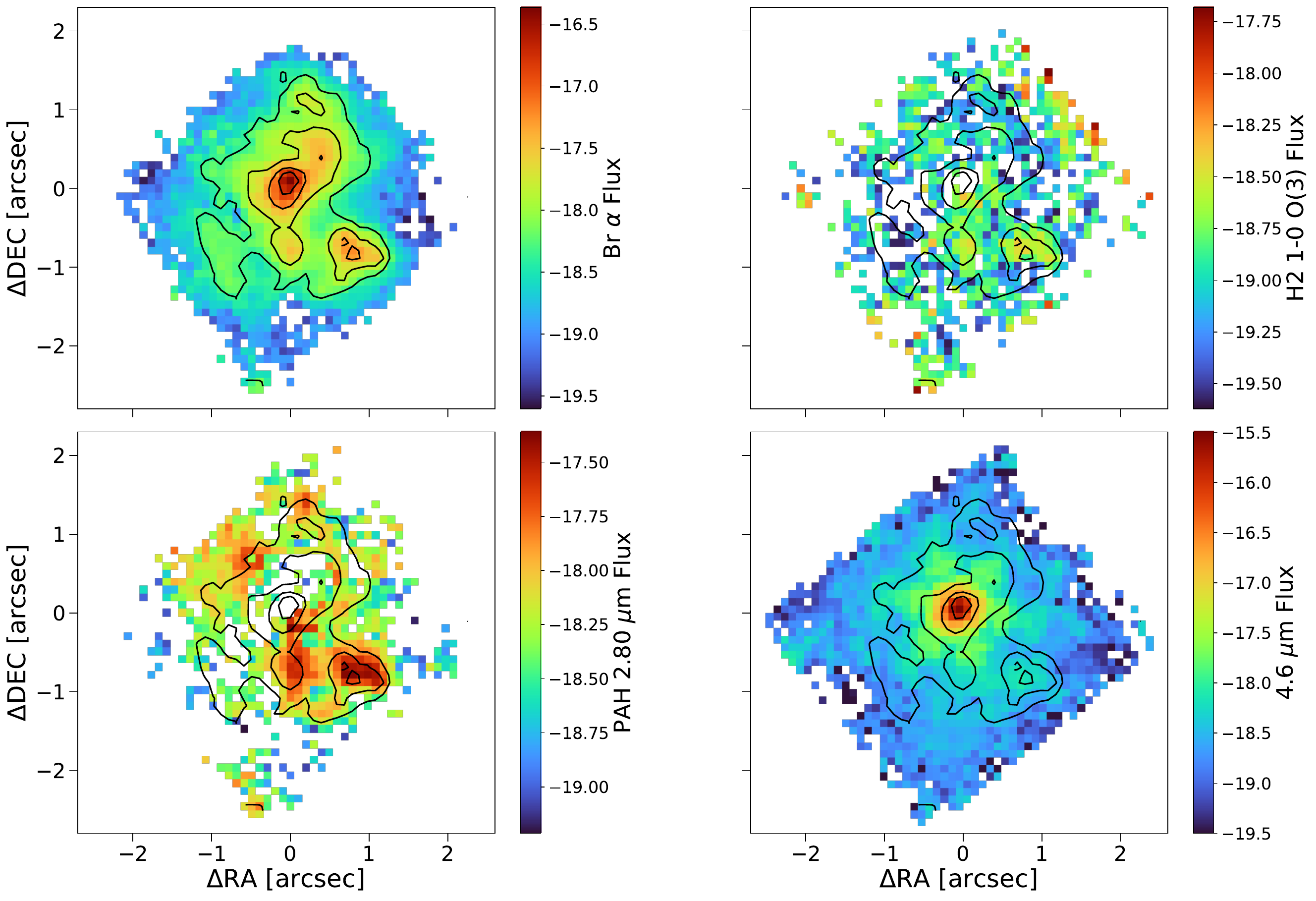}
\caption{Spatial maps of J1601 showing representative emission from ionized gas, molecular gas, PAH emission, and the line-free 4.6~$\mu$m continuum. The panels show Br$\alpha$, H$_2$ 1--0 O(3), PAH emission, and the 4.6~$\mu$m continuum. The contours in each panel trace the Pa$\alpha$ emission shown in Figure~\ref{fig:J1601_aperture_image}. Only spaxels with S/N $\geq 1$ are displayed. The hydrogen recombination emission is dominated by a bright nuclear source, with additional knots to the south and a faint stream extending from the nucleus. In contrast, the H$_2$ emission is weak or absent in the central resolution element and is strongest in the southern knots, demonstrating a spatial separation between the ionized and molecular gas. The PAH emission is detected primarily on circumnuclear scales and is relatively weak at the position of the compact nuclear continuum source. The 4.6~$\mu$m continuum is dominated by a compact unresolved nuclear source, and no clear counterpart to the southern knots, indicating that the hot dust responsible for the red mid-infrared colors is highly centrally concentrated. Fluxes are in log units of $\mathrm{~erg~cm^{-2}~s^{-1}.}$}
\label{fig:J1601_4flux}
\end{figure*}

We observed J1601 using the Keck Cosmic Web Imager (KCWI; \citealt{2018ApJ...864...93M}) on 2024 August 5 as part of Keck Program 2024B--U230 (PI: Canalizo). KCWI provides optical integral-field spectroscopy over a field of view large enough to probe the ionized gas on galaxy-wide scales. In this work, the KCWI data are used to measure the larger-scale optical emission-line properties, gas-phase metallicity, extinction, optical high-ionization lines including \ion{He}{2} $\lambda4686$, and upper limits on optical coronal-line emission.

\begin{figure*}
\centering
\includegraphics[width=\textwidth]{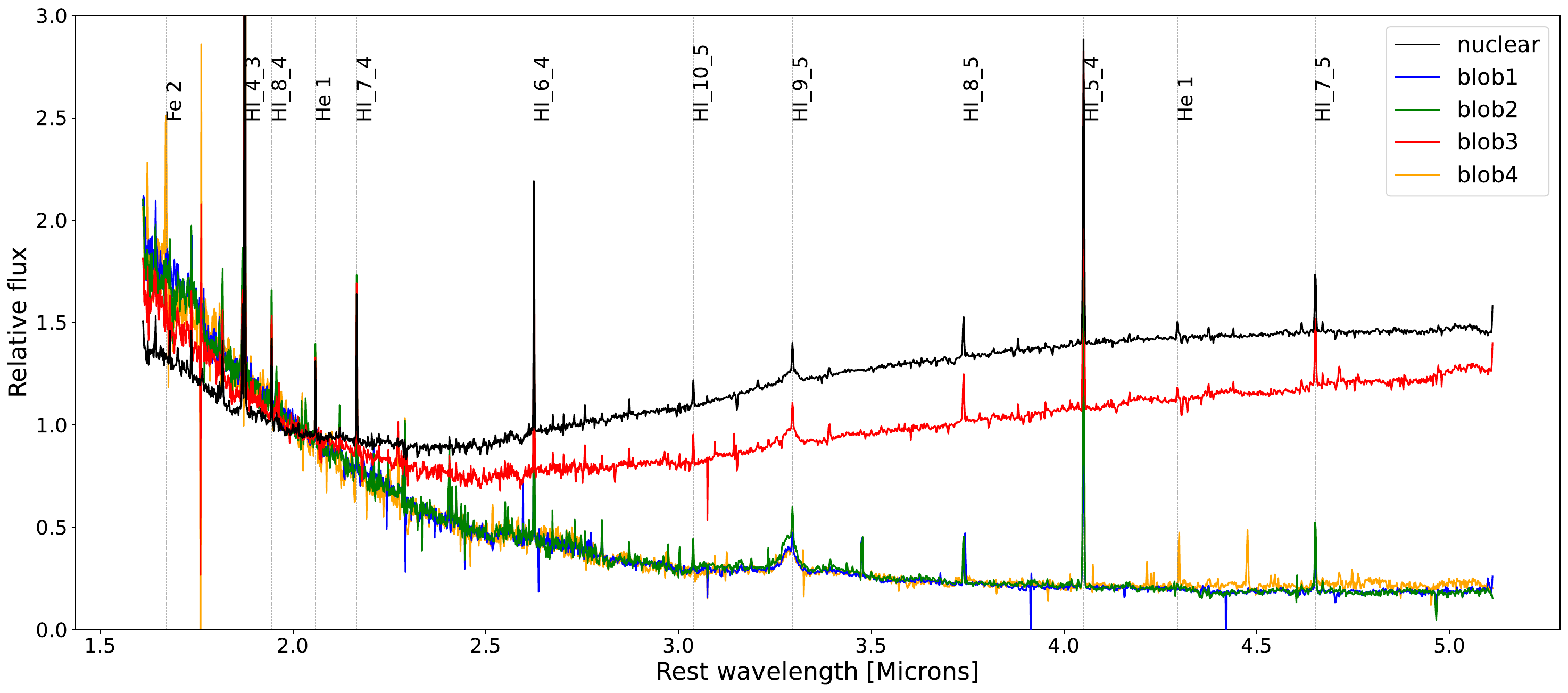}
\caption{Extracted JWST/NIRSpec spectra for the apertures shown in Figure~\ref{fig:J1601_aperture_image}, normalized at 2~$\mu$m. The nuclear spectrum (black) exhibits a steeply rising near-infrared continuum, while the off-nuclear apertures show significantly flatter spectral slopes. The red aperture (``blob 3'') lies close enough to the nucleus that it partially captures the nuclear continuum and should not be interpreted as an independent hot-dust source.}
\label{fig:J1601_aperture_spectra}
\end{figure*}

The observations were obtained using the small slicer and the BL and RL gratings for the blue and red arms, respectively. The BL grating provides fixed wavelength coverage of $3500$--$5600$~\AA, while the red-arm grating can be tuned to observe a $\sim3500$~\AA-wide window between $5400$ and $10,800$~\AA. We selected a central wavelength of $7150$~\AA\ for the red arm to achieve nearly continuous wavelength coverage between the blue and red channels. This configuration resulted in an effective wavelength range of approximately $3500$--$8800$~\AA, a field of view of $8\farcs4 \times 20\farcs4$, and spectral resolving powers of $R \approx 3600$ and $R \approx 2000$ for the blue and red arms, respectively.

The data were acquired under clear sky conditions with an estimated seeing of $\sim1\farcs0$. The total integration times were $2000$~s for the blue arm and $1800$~s for the red arm.

The KCWI data were reduced using the Python-based KCWI Data Reduction Pipeline (DRP; version~1.1.0)\footnote{\url{https://github.com/Keck-DataReductionPipelines/KCWI_DRP}}. The pipeline performs the standard reduction steps described in the DRP documentation, including flat-fielding, sky subtraction, wavelength calibration, flux calibration, cosmic-ray removal, and final assembly of the three-dimensional data cubes. We used custom Python routines to align the blue- and red-arm data cubes to a common spatial grid prior to stitching them together.

Several coronal lines of interest, most notably [\ion{Ne}{5}], fall near the edge of the usable KCWI wavelength range at the redshift of J1601, where the instrumental sensitivity declines sharply. In particular, [\ion{Ne}{5}] $\lambda3426$ falls close to the blue edge of the KCWI coverage, while [\ion{Ne}{5}] $\lambda3346$ lies outside or at the extreme edge of the usable range. Including these edge regions in the inverse-sensitivity-function fitting introduced significant uncertainties. To mitigate this issue, we performed a specialized reduction restricted to the spectral region surrounding [\ion{Ne}{5}] $\lambda3426$. This auxiliary cube was used only for the targeted search for optical [\ion{Ne}{5}] emission and for estimating the corresponding upper limits; all other analyses were carried out using the data cubes produced with the standard reduction procedure.

Because KCWI spaxels are rectangular, we resampled the data cubes to square $0.15\arcsec \times 0.15\arcsec$ spaxels using the \texttt{IFSR\_KCWIRESAMPLE} routine from the \texttt{IFSRED} library \citep{2014ascl.soft09004R}.

Representative KCWI spectra and the optical emission-line diagnostics derived from these cubes are presented in Section~\ref{sec:results}. The KCWI data are used below to measure the optical line ratios, metallicity, extinction, , \ion{He}{2} $\lambda4686$ emission, Wolf--Rayet features, and optical coronal-line upper limits.

\section{Results}
\label{sec:results}

\subsection{Morphology of the Continuum, Ionized, and Molecular Gas in NIRSpec}
\label{subsec:morphology}

The JWST/NIRSpec observations of J1601 reveal a morphologically complex system on parsec scales, with $\sim40$ emission lines detected between $\sim1.7$--$5.2~\mu$m. Figure~\ref{fig:J1601_aperture_image} shows the Pa$\alpha$ emission map derived from the IFU data cube. The galaxy is dominated by a bright central nuclear source, accompanied by two compact knots of emission located to the south, as well as a faint stream extending from the nucleus. The overplotted aperture centroids indicate the locations used for spectral extraction, each with a radius of $0\farcs5$, chosen to balance spatial resolution and signal-to-noise.


Figure~\ref{fig:J1601_4flux} shows representative spatial maps of the main emission components discussed below, together with the line-free continuum at 4.6~$\mu$m. The hydrogen recombination lines share a common morphology, characterized by a bright, compact nuclear source and extended emission in the southern knots and faint stream. This morphology is seen in both Pa$\alpha$ and Br$\alpha$, indicating that the ionized gas is strongly concentrated toward the nucleus but also extends over surrounding star-forming regions.

\startlongtable
\begin{deluxetable*}{ccccc}
\tabletypesize{\footnotesize}
\tablecaption{J1601 Emission Line Fit Parameters for the Nuclear Aperture. The quoted parameter errors represent the 68-th percentile of the parameter distribution generation by \textsc{emcee}.}

\tablehead{\colhead{Line} & \colhead{$\lambda_{\rm rest}$} & \colhead{Flux} & \colhead{FWHM} & \colhead{$V_{\rm off}$} \\ \colhead{} & \colhead{($\mu$m)} & \colhead{(10$^{-17}$ erg cm$^{-2}$ s$^{-1}$)} & \colhead{(km s$^{-1}$)} & \colhead{(km s$^{-1}$)}}

\decimals
\startdata
\midrule 
Hydrogen Recombination Lines & & & & \\
\midrule
Br $\eta$ 11-4 & 1.68065 & ${2.73}^{0.67}_{0.73}$ & ${425.11}^{37.16}_{26.09}$ & ${-50.82}^{11.61}_{18.11}$\\
Br $\zeta$ 10-4 & 1.73621 & ${4.39}^{0.73}_{0.64}$ & ${456.65}^{21.60}_{22.48}$ & ${-12.24}^{9.11}_{8.29}$\\
Br $\epsilon$ 9-4 & 1.81741 & ${5.03}^{0.59}_{0.46}$ & ${414.63}^{17.54}_{21.53}$ & ${-19.49}^{5.88}_{5.88}$\\
Pa $\alpha$ 4-3 & 1.8751 & ${149.90}^{1.14}_{1.00}$ & ${374.31}^{2.25}_{2.22}$ & ${-16.38}^{1.20}_{1.04}$\\
Br $\delta$ 8-4 & 1.94456 & ${7.85}^{0.47}_{0.47}$ & ${382.96}^{13.69}_{19.38}$ & ${-19.86}^{7.89}_{8.90}$\\
Br $\gamma$ 7-4 & 2.16553 & ${12.86}^{0.35}_{0.31}$ & ${356.91}^{10.15}_{9.70}$ & ${-21.90}^{3.92}_{4.08}$\\
Pf $\iota$ 14-5 & 2.61194 & ${0.93}^{0.22}_{0.19}$ & ${306.08}^{65.67}_{64.19}$ & ${-0.63}^{30.32}_{28.90}$\\
Br $\beta$ 6-4 & 2.62515 & ${23.40}^{0.30}_{0.31}$ & ${314.57}^{3.56}_{3.79}$ & ${-19.27}^{1.65}_{1.98}$\\
Pf $\theta$ 13-5 & 2.6744 & ${0.93}^{0.28}_{0.21}$ & ${281.41}^{52.33}_{43.95}$ & ${-35.29}^{27.73}_{24.29}$\\
Pf $\eta$ 12-5 & 2.75752 & ${1.44}^{0.31}_{0.26}$ & ${289.35}^{54.71}_{67.63}$ & ${-11.41}^{27.25}_{23.89}$\\
Pf $\zeta$ 11-5 & 2.87221 & ${1.70}^{0.39}_{0.35}$ & ${265.72}^{45.32}_{71.56}$ & ${-15.48}^{24.51}_{24.79}$\\
Pf $\epsilon$ 10-5 & 3.03837 & ${3.35}^{0.69}_{0.58}$ & ${315.72}^{34.89}_{39.88}$ & ${-5.02}^{26.50}_{24.48}$\\
Pf $\delta$ 9-5 & 3.29609 & ${3.92}^{0.31}_{0.28}$ & ${386.32}^{24.15}_{28.45}$ & ${-24.81}^{10.77}_{10.91}$\\
Pf $\gamma$ 8-5 & 3.73954 & ${6.38}^{0.29}_{0.29}$ & ${385.95}^{19.67}_{20.54}$ & ${-26.44}^{8.11}_{8.09}$\\
Br $\alpha$ 5-4 & 4.05115 & ${42.63}^{0.42}_{0.38}$ & ${307.21}^{4.05}_{4.19}$ & ${-18.25}^{1.48}_{1.36}$\\
Pf $\beta$ 7-5 & 4.65251 & ${9.32}^{0.30}_{0.26}$ & ${305.66}^{7.72}_{8.15}$ & ${-21.67}^{4.18}_{4.12}$\\
Hu $\epsilon$ 11-6 & 4.67124 & ${1.21}^{0.29}_{0.26}$ & ${258.77}^{47.95}_{71.33}$ & ${-14.71}^{27.18}_{22.73}$\\
\midrule
$H_{2}$ lines & & & & \\
\midrule
1-0 S(1) & 2.12125 & ${0.49}^{0.22}_{0.20}$ & ${204.78}^{37.65}_{68.65}$ & ${6.76}^{9.46}_{7.92}$\\
1-0 Q(1) & 2.40594 & ${0.57}^{0.27}_{0.23}$ & ${289.03}^{85.09}_{152.90}$ & ${6.06}^{42.64}_{52.58}$\\
1-0 O(3) & 2.80175 & ${0.58}^{0.26}_{0.22}$ & ${233.31}^{44.80}_{97.18}$ & ${-35.80}^{26.19}_{37.24}$\\
\midrule
Other lines & & & & \\
\midrule
$$[Fe II]$$ & 1.6435 & ${3.85}^{1.07}_{0.78}$ & ${619.90}^{24.87}_{25.82}$ & ${-50.61}^{16.16}_{17.27}$\\
$$[S V]$$ & 1.70025 & ${3.06}^{0.87}_{0.60}$ & ${713.75}^{1.84}_{2.78}$ & ${58.29}^{13.24}_{11.25}$\\
$$[Fe II]$$ & 1.80939 & ${0.95}^{0.45}_{0.32}$ & ${306.47}^{42.60}_{26.75}$ & ${2.53}^{19.44}_{14.82}$\\
He I & 1.86854 & ${10.71}^{0.63}_{0.62}$ & ${479.22}^{21.21}_{21.47}$ & ${25.69}^{9.57}_{8.84}$\\
He I & 2.05813 & ${5.95}^{0.46}_{0.38}$ & ${338.01}^{18.59}_{15.20}$ & ${-11.19}^{7.46}_{9.05}$\\
He I & 3.2064 & ${1.56}^{0.26}_{0.25}$ & ${484.57}^{90.45}_{97.82}$ & ${-58.90}^{33.79}_{31.24}$\\
He I & 4.29459 & ${1.64}^{0.18}_{0.17}$ & ${258.30}^{13.45}_{7.29}$ & ${-52.12}^{1.63}_{1.85}$\\
$$[K III]$$ & 4.61683 & ${1.07}^{0.20}_{0.17}$ & ${234.30}^{37.45}_{22.26}$ & ${-5.47}^{2.27}_{1.86}$\\
\midrule
Unknown lines & & & & \\
\midrule
? & 1.8041 & ${0.49}^{0.54}_{0.35}$ & ${137.01}^{0.87}_{0.88}$ & $---$\\
? & 2.7292 & ${0.34}^{0.26}_{0.33}$ & ${143.04}^{91.36}_{6.91}$ & $---$\\
\enddata
\label{table:nuc_line_flux}
\end{deluxetable*}

In contrast, the molecular hydrogen emission exhibits a markedly different spatial distribution. The H$_2$ emission is weak or absent in the central resolution element, while remaining prominent in the southern knots where it overlaps with the extended ionized gas. This spatial segregation between ionized and molecular gas indicates that the nuclear and off-nuclear regions differ in their excitation conditions and molecular-gas content.

The PAH emission also differs from the hydrogen recombination-line morphology. It is detected primarily on circumnuclear scales and is relatively weak at the position of the compact nuclear continuum source. This spatial distribution suggests that the PAH-emitting material is not co-spatial with the hottest nuclear dust component traced by the 4.6~$\mu$m continuum.

The 4.6~$\mu$m continuum is dominated by a compact nuclear source, with no significant counterpart in the southern knots. The source is unresolved in the NIRSpec data. Using the radial profile of the continuum integrated over the W2-band wavelength range, we place a conservative upper limit of $\lesssim160$~pc of this source at the distance of J1601. This indicates that the red mid-infrared colors originate from a highly concentrated nuclear component rather than from galaxy-wide star formation.

The key result from these maps is that the hot-dust continuum is much more centrally concentrated than the recombination-line, molecular-gas, and PAH emission. We next examine the spectra extracted from these regions to determine whether the compact nuclear source shows spectroscopic evidence for accretion.

\subsection{JWST/NIRSpec Aperture Spectra and Nuclear Emission Line Detections}
\label{subsec:spectral_features}

Figure~\ref{fig:J1601_aperture_spectra} shows the extracted one-dimensional JWST/NIRSpec spectra corresponding to the apertures in Figure~\ref{fig:J1601_aperture_image}. The nuclear spectrum exhibits a red, steeply rising near-infrared continuum, while the off-nuclear apertures have substantially flatter spectral slopes. The red aperture (``blob 3'') lies close enough to the nucleus that its spectrum is partially contaminated by the nuclear continuum; the rising continuum in this aperture should therefore not be interpreted as evidence for an independent hot-dust source in blob 3.



\begin{figure*}
\centering
\includegraphics[width=\textwidth]{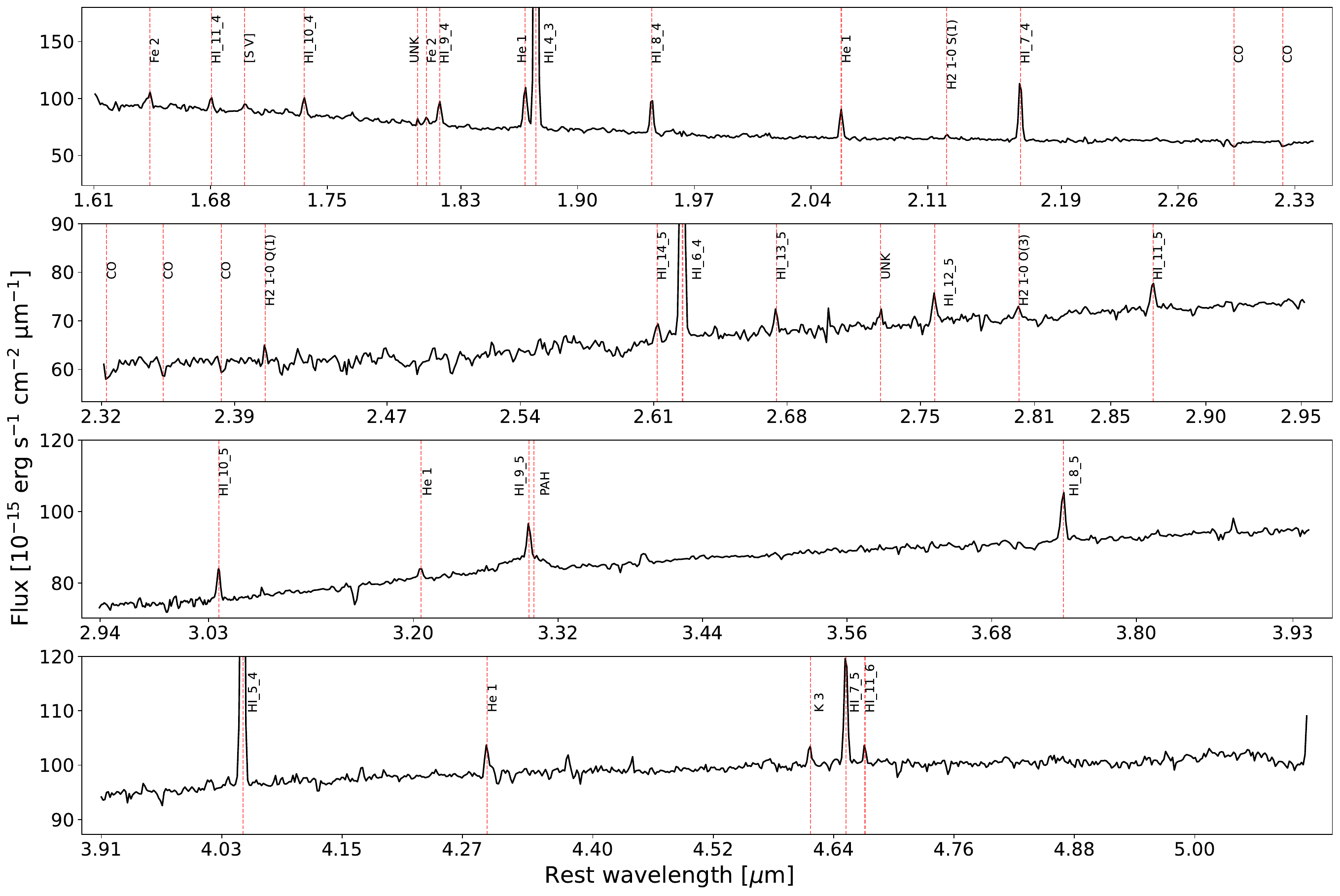}
\caption{Full nuclear JWST/NIRSpec spectrum with identified emission lines labeled. The spectrum is dominated by hydrogen recombination lines, molecular hydrogen transitions, and He,\textsc{i} emission. Unlabeled features were not considered robust line detections because they are not recovered consistently in the independent dither exposures. No infrared coronal lines or broad near-infrared emission-line components are observed.}
\label{fig:J1601_nuclear_full}
\end{figure*}

The nuclear emission-line spectrum is dominated by hydrogen recombination lines, ro-vibrational transitions of molecular hydrogen, and \ion{He}{1} emission. The measured line fluxes and kinematic properties for the nuclear extraction are listed in Table~\ref{table:nuc_line_flux}. Figure~\ref{fig:J1601_nuclear_full} shows the full nuclear spectrum with the robustly identified lines labeled. We find no evidence for broad near-infrared emission-line components in the nuclear aperture.

The spectrum also shows CO bandhead absorption and PAH emission, which are discussed in Section~\ref{subsec:stellar_population}. No nebular \ion{He}{2} emission is detected in the NIRSpec nuclear spectrum.

No infrared coronal lines are detected in the JWST/NIRSpec data. In particular, we do not detect [\ion{Si}{6}], [\ion{Ca}{8}], [\ion{S}{6}], or [\ion{Si}{7}] at the expected wavelengths. The corresponding upper limits are listed in Table~\ref{table:jwst_cl_uls}. These non-detections provide direct constraints on high-ionization infrared emission from the compact nuclear source; the implications of the infrared coronal-line upper limits limits are discussed in Section~\ref{subsec:agn_constraints}.

\startlongtable
\begin{deluxetable}{ccc}
\tabletypesize{\footnotesize}
\tablecaption{J1601 infrared coronal line upper limits from JWST/NIRSpec using the nuclear aperture shown in Figure \ref{fig:J1601_aperture_image}.}
\tablehead{\colhead{Line} & \colhead{$\lambda_{\rm rest}$} & \colhead{Upper Limit} \\ \colhead{} & \colhead{($\mu$m)} & \colhead{(10$^{-19}$ erg cm$^{-2}$ s$^{-1}$)}}
\decimals
\startdata
\midrule
{[S VI]} & 1.887 & < 129.85\\
{[Si XI]} & 1.934 & < 126.21\\
{[Si VI]} & 1.962 & < 141.43\\
{[Al IX]} & 2.044 & < 66.85\\
{[Ca VIII]} & 2.3211 & < 50.50\\
{[Si VII]} & 2.4807 & < 94.51\\
{[Al V]} & 2.904 & < 76.62\\
{[Ca IV]} & 3.206 & < 65.11\\
{[Al VI]} & 3.659 & < 77.83\\
{[Si IX]} & 3.928 & < 90.24\\
{[Mg IV]} & 4.487 & < 36.57\\
{[Ar VI]} & 4.528 & < 45.99\\
\enddata
\label{table:jwst_cl_uls}
\end{deluxetable}

\subsection{KCWI Optical Morphology, Spectrum, and Emission-Line Diagnostics}
\label{subsec:kcwi_diagnostics}

\begin{figure}
\centering
\includegraphics[width=\columnwidth]{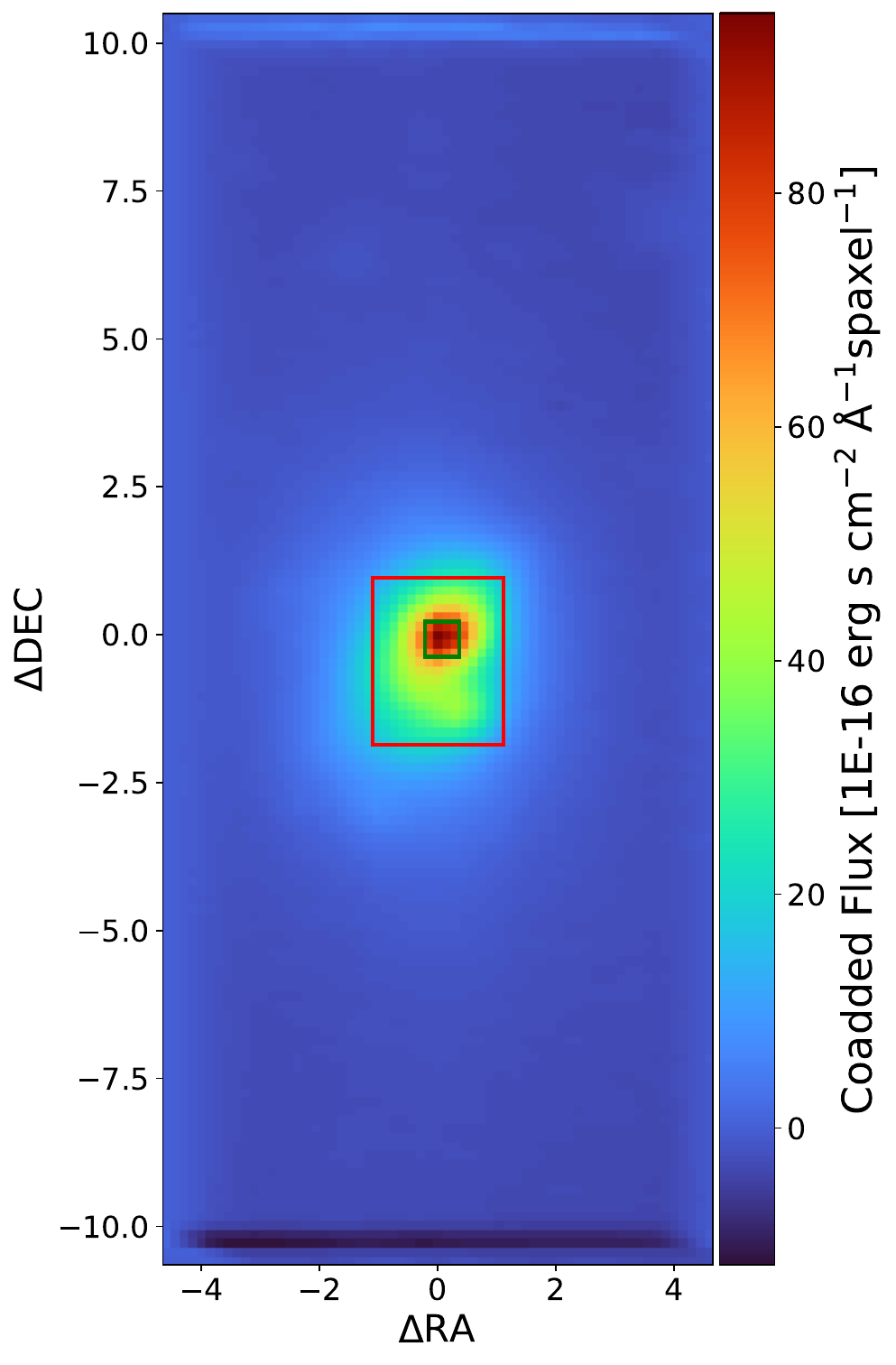}
\caption{Large-scale KCWI view of J1601 showing the region used for the matched-field maps (red box). This region covers approximately the same projected field as the JWST/NIRSpec IFU, allowing direct comparison between the optical KCWI maps and the higher spatial resolution NIRSpec maps. The green box shows the nuclear extraction aperture used to search for optical coronal lines. The image displays the total integrated optical flux in the wavelengths covered by the KCWI data. As can be seen, there is some broader scale low level emission, but the brightest region containing the young stellar clusters are fully contained within the NIRSpec field of view.}
\label{fig:KCWI_extraction_apertures}
\end{figure}

The Keck/KCWI data allows us to explore the optical emission of J1601 and compare it to the higher spatial resolution near-infrared emission probed by JWST/NIRSpec. The KCWI field of view is substantially larger than the NIRSpec IFU field, allowing us to examine the larger-scale optical emission in relation to the compact nuclear region captured by JWST. Figure~\ref{fig:KCWI_extraction_apertures} shows the larger-scale KCWI image with the region used for the matched-field maps overplotted, as well as the nuclear extraction aperture used to search for optical coronal lines. The matched-field region was chosen to cover approximately the same projected area sampled by the JWST/NIRSpec IFU, enabling a direct comparison between the optical and near-infrared morphologies.

\begin{figure*}
\centering
\includegraphics[width=\textwidth]{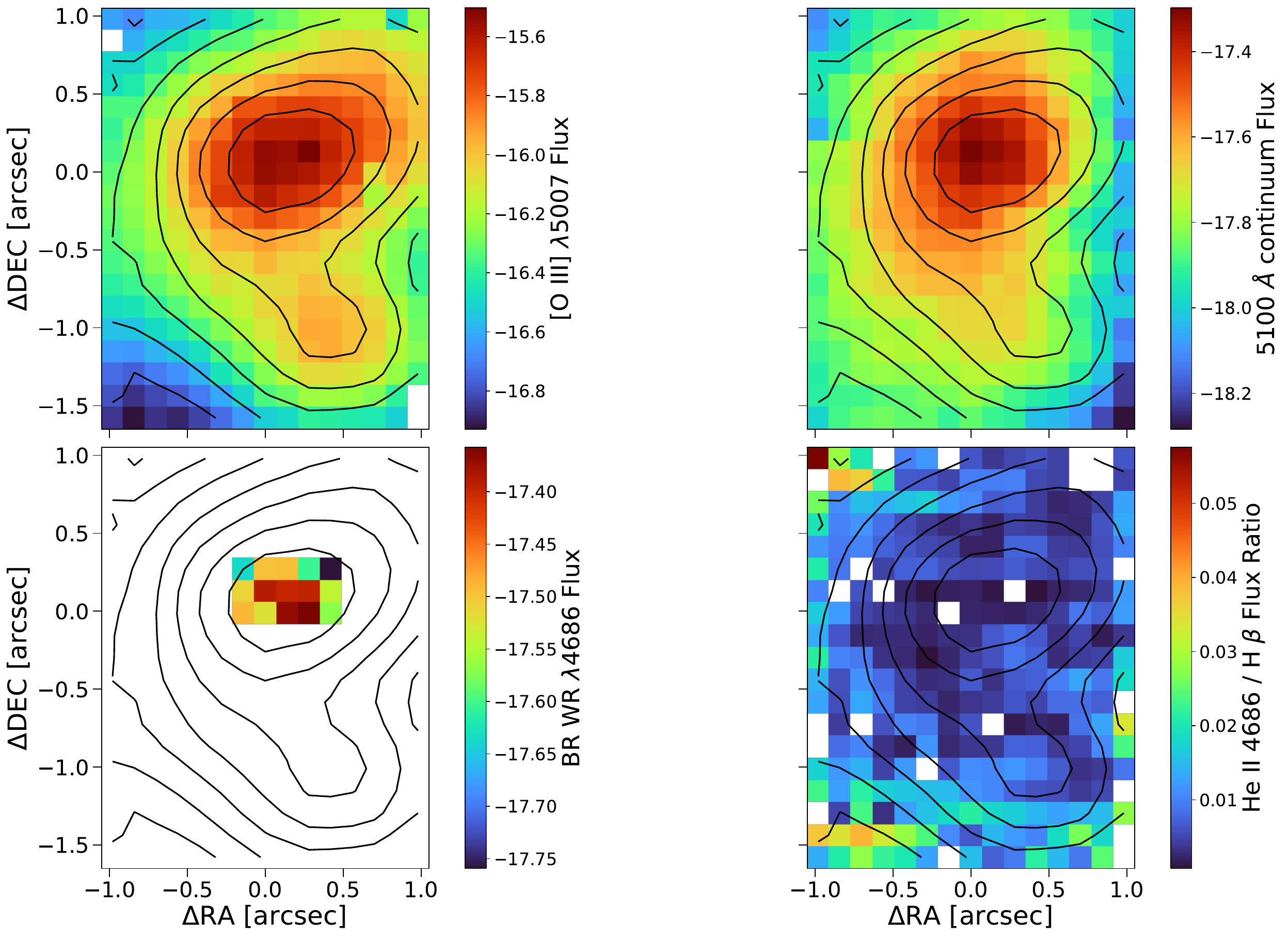}
\caption{KCWI maps of J1601 extracted over a field of view comparable to the JWST/NIRSpec IFU. The panels show the [\ion{O}{3}] $\lambda5007$ emission, the rest-frame 5100~\AA\ continuum, the optical Wolf--Rayet feature, and the narrow \ion{He}{2} $\lambda4686$/H$\beta$ ratio. The [\ion{O}{3}] emission is extended and clumpy, broadly following the large-scale ionized-gas morphology seen at higher spatial resolution in the JWST/NIRSpec recombination-line maps. The 5100~\AA\ continuum is more extended than the compact 4.6~$\mu$m continuum, while the narrow \ion{He}{2}/H$\beta$ ratio remains weak across the field. Only spaxels with S/N $\geq 1$ are displayed. Fluxes are in log units of $\mathrm{erg~s^{-1}~cm^{-2}}$. The line ratio is not in log space.}
\label{fig:J1601_kcwi_maps}
\end{figure*}

Figure~\ref{fig:J1601_kcwi_maps} shows KCWI maps extracted over this matched field of view. The panels show the [\ion{O}{3}] $\lambda5007$ emission, the rest-frame 5100~\AA\ continuum, the optical Wolf--Rayet feature, and the narrow \ion{He}{2} $\lambda4686$/H$\beta$ ratio. The [\ion{O}{3}] emission is extended and clumpy, with a bright central region and lower-surface-brightness emission across the field. Its morphology broadly follows the same large-scale ionized-gas distribution seen at higher spatial resolution in the JWST hydrogen recombination-line maps, including the bright nuclear region and extended emission associated with the surrounding star-forming structure.

The 5100~\AA\ continuum is centrally concentrated but more spatially extended than the compact 4.6~$\mu$m continuum seen in the NIRSpec data. This indicates that the optical stellar continuum and the hot-dust continuum do not trace identical components. The optical Wolf--Rayet feature is detected near the central star-forming region and is spatially associated with the region of bright optical line emission, and is discussed further in Section~\ref{subsec:stellar_population}. The narrow \ion{He}{2} $\lambda4686$/H$\beta$ ratio remains weak across the field, with no localized enhancement at the nucleus.

\begin{figure*}
\centering
\includegraphics[width=\textwidth]{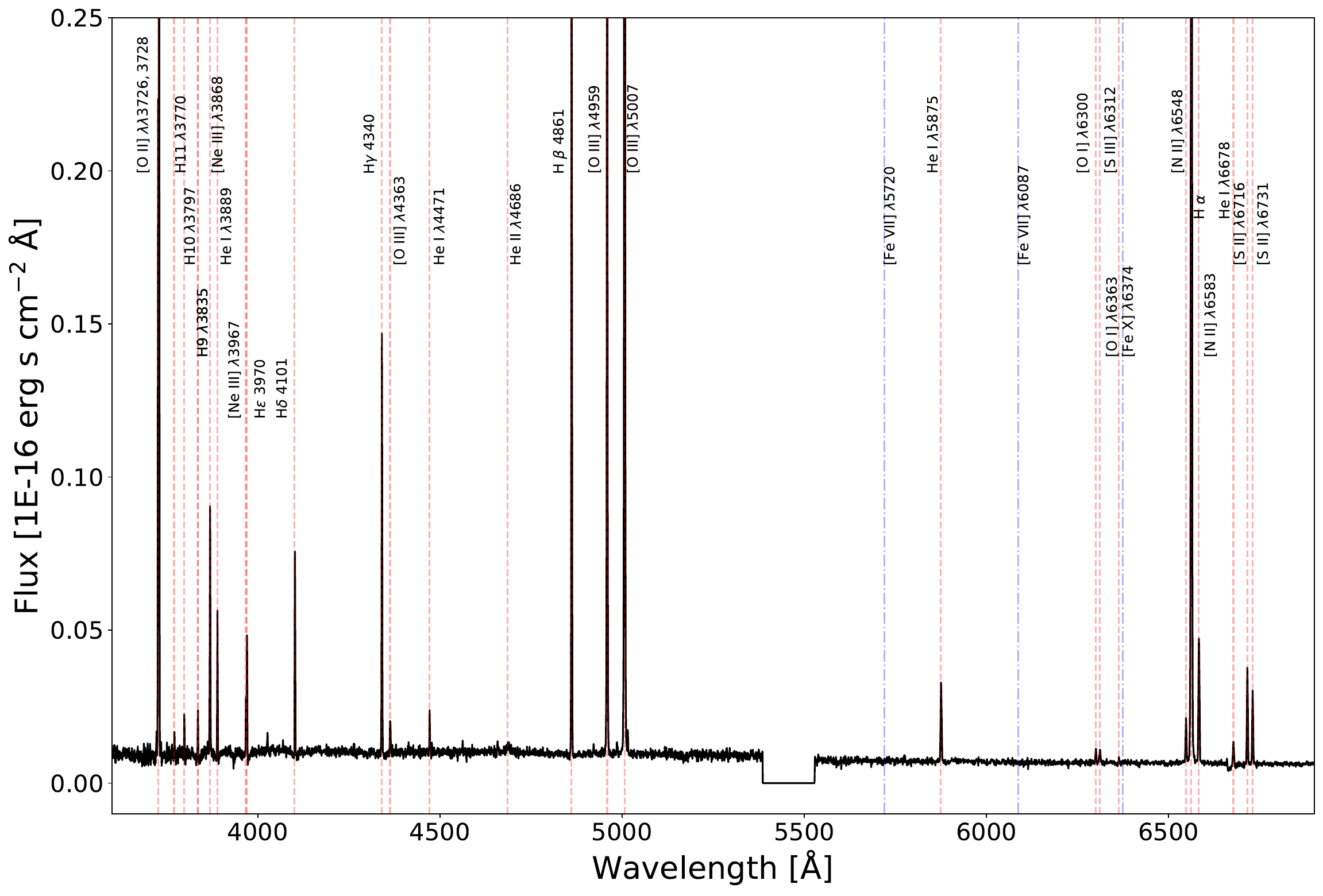}
\caption{Nuclear KCWI spectrum of J1601 extracted from the nuclear aperture shown in Figure~\ref{fig:KCWI_extraction_apertures}. The spectrum is dominated by optical nebular emission lines characteristic of star-forming gas. Weak narrow \ion{He}{2} $\lambda4686$ emission is detected. The expected location of the brighter optically coronal lines in this wavelength range are displayed by the dotted lines. No coronal lines are detected in the entire optical spectrum, including the [\ion{Ne}{5}] $\lambda3426$ line, which falls outside of the wavelength range shown here.}
\label{fig:J1601_kcwi_spectrum}
\end{figure*}

Figure~\ref{fig:J1601_kcwi_spectrum} shows a representative KCWI spectrum extracted from the nuclear aperture shown in Figure~\ref{fig:KCWI_extraction_apertures}. The optical spectrum is dominated by strong nebular emission lines, including [\ion{O}{2}], H$\beta$, [\ion{O}{3}], H$\alpha$, [\ion{N}{2}], and [\ion{S}{2}], characteristic of photoionized star-forming gas. Weak narrow \ion{He}{2} $\lambda4686$ emission is present, and the broader Wolf--Rayet feature is seen and discussed further in Section~\ref{subsec:stellar_population}. No optical coronal lines are detected in the nuclear spectrum. In particular, we find no evidence for [\ion{Ne}{5}] $\lambda3426$ emission at the expected observed wavelength.

\begin{figure}
\centering
\includegraphics[width=\columnwidth]{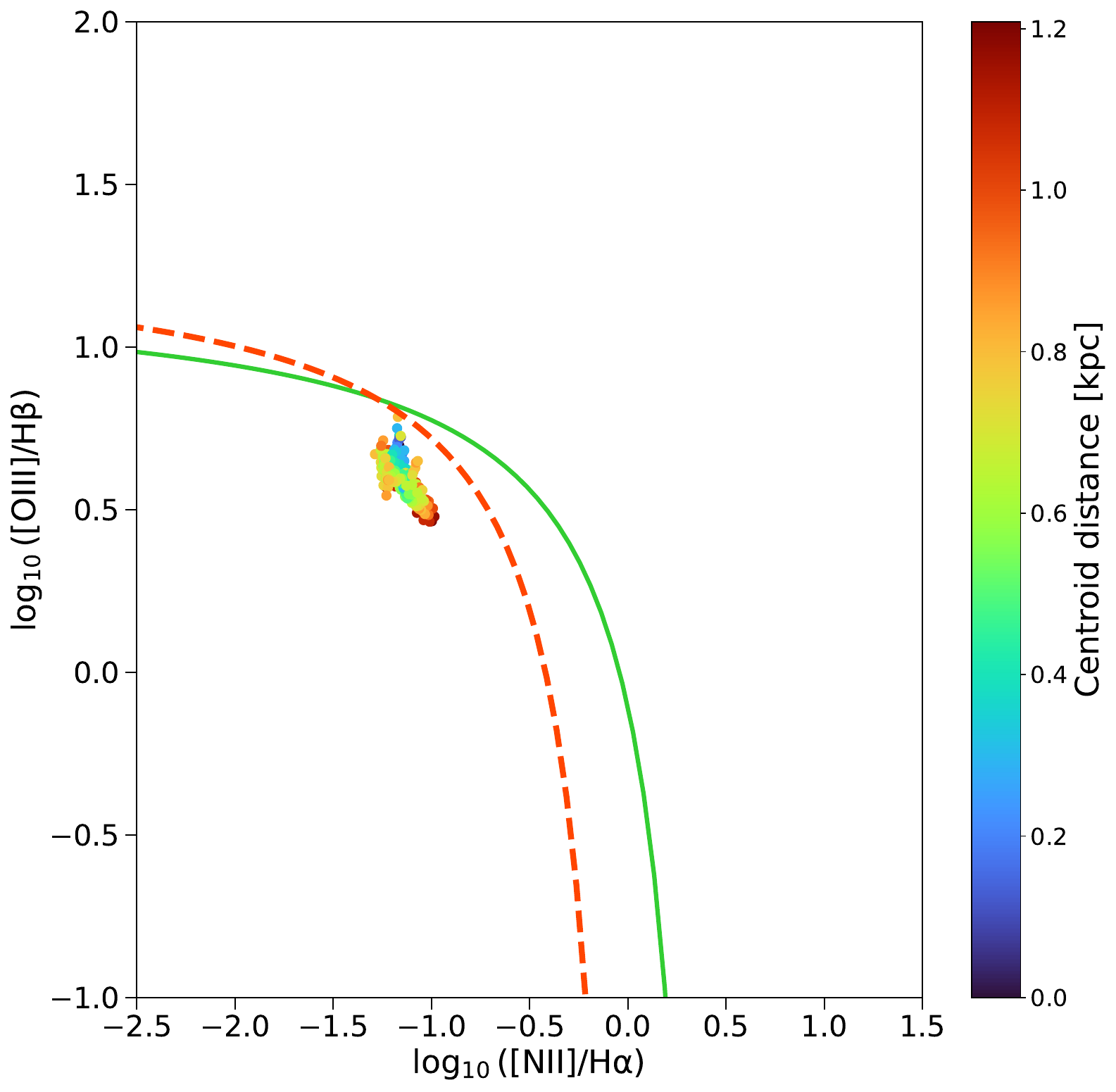}
\caption{Spatially resolved BPT diagram for J1601 using KCWI spaxels with sufficient signal-to-noise. All measured spaxels fall in the star-forming region, with no spaxels occupying the composite or AGN region.}
\label{fig:J1601_kcwi_bpt}
\end{figure}

We used the spatially resolved KCWI line measurements to test whether any regions of the galaxy exhibit optical line ratios characteristic of AGN photoionization. Figure~\ref{fig:J1601_kcwi_bpt} shows the standard BPT diagram based on [\ion{O}{3}] $\lambda5007$/H$\beta$ and [\ion{N}{2}] $\lambda6584$/H$\alpha$. All spaxels with S/N $\geq 1$ fall within the star-forming region of the diagram. The nuclear spaxels do not occupy the composite or AGN region, indicating that the optical narrow-line ratios are dominated by stellar photoionization throughout the field of view.

We also examined optical high-ionization diagnostics that are more sensitive to hard radiation fields in low-metallicity systems. Following the diagnostic proposed by \citet{2012MNRAS.421.1043S}, we compared narrow \ion{He}{2} $\lambda4686$/H$\beta$ with [\ion{N}{2}] $\lambda6584$/H$\alpha$ on a spaxel-by-spaxel basis. The median values are $\log\left(\mathrm{He,II},\lambda4686/\mathrm{H}\beta\right)=-2.14$ and $\log\left([\mathrm{N,II}],\lambda6584/\mathrm{H}\alpha\right)=-1.16$, with little variation across the KCWI field. These ratios place J1601 within the star-forming galaxy locus in this diagnostic space.

As an additional test, we considered the diagnostic used by \citet{Mazzolari2025}, which compares [\ion{O}{3}] $\lambda4363$/H$\gamma$ with [\ion{O}{3}] $\lambda5007$/[\ion{O}{2}] $\lambda3727$ to identify an ``AGN-only'' region. The median values for J1601 are $\log\left([\mathrm{O,III}],\lambda4363/\mathrm{H}\gamma\right)=-1.0$ and $\log\left([\mathrm{O,III}],\lambda5007/[\mathrm{O,II}],\lambda3727\right)=0.31$. The [\ion{O}{3}] $\lambda4363$/H$\gamma$ map is relatively flat, while [\ion{O}{3}] $\lambda5007$/[\ion{O}{2}] $\lambda3727$ shows more spatial structure. The measured ratios remain outside the AGN-only region and are consistent with star-forming galaxies. 

We searched the KCWI data for optical coronal-line emission, including [\ion{Ne}{5}] $\lambda3426$. No optical coronal lines are detected. The resulting upper limits are listed in Table~\ref{table:kcwi_cl_uls}. The flux limits were estimated by integrating a Gaussian at the expected observed wavelength, with the FWHM fixed to the instrumental resolution and the amplitude set by the local continuum rms.

\startlongtable
\begin{deluxetable}{ccc}
\tabletypesize{\footnotesize}
\tablecaption{J1601 optical coronal line upper limits from Keck/KCWI using the nuclear aperture shown in Figure \ref{fig:KCWI_extraction_apertures}.}
\tablehead{\colhead{Line} & \colhead{$\lambda_{\rm rest}$} & \colhead{Upper Limit} \\ \colhead{} & \colhead{(\AA)} & \colhead{(10$^{-19}$ erg cm$^{-2}$ s$^{-1}$)}}
\decimals
\startdata
\midrule
{[Ne V]} & 3346.0 & < 28.23\\
{[Ne V]} & 3426.0 & < 22.18\\
{[Fe V]} & 3839.0 & < 13.11\\
{[Fe V]} & 3891.0 & < 13.87\\
{[Fe V]} & 3911.0 & < 8.14\\
{[Fe V]} & 4071.0 & < 9.91\\
{[Fe V]} & 4180.0 & < 7.02\\
{[Fe V]} & 4227.0 & < 9.17\\
{[Ar XIV]} & 4412.0 & < 9.59\\
{[Fe VII]} & 4893.0 & < 8.73\\
{[FeVI]} & 5146.0 & < 8.54\\
{[FeVII]} & 5159.0 & < 10.75\\
{[Fe VI]} & 5176.0 & < 12.84\\
{[Fe XIV]} & 5303.0 & < 8.87\\
{[Fe VI]} & 5335.0 & < 9.40\\
{[Fe VII]} & 5720.0 & < 9.98\\
{[Fe VII]} & 6087.0 & < 11.77\\
{[Fe X]} & 6374.0 & < 16.45\\
\enddata
\label{table:kcwi_cl_uls}
\end{deluxetable}

\subsection{Physical Conditions of the Nuclear Region}
\label{physical_conditions}
\subsubsection{Gas-Phase Metallicity}
\label{metallicity}

\begin{figure}
\centering
\includegraphics[width=\columnwidth]{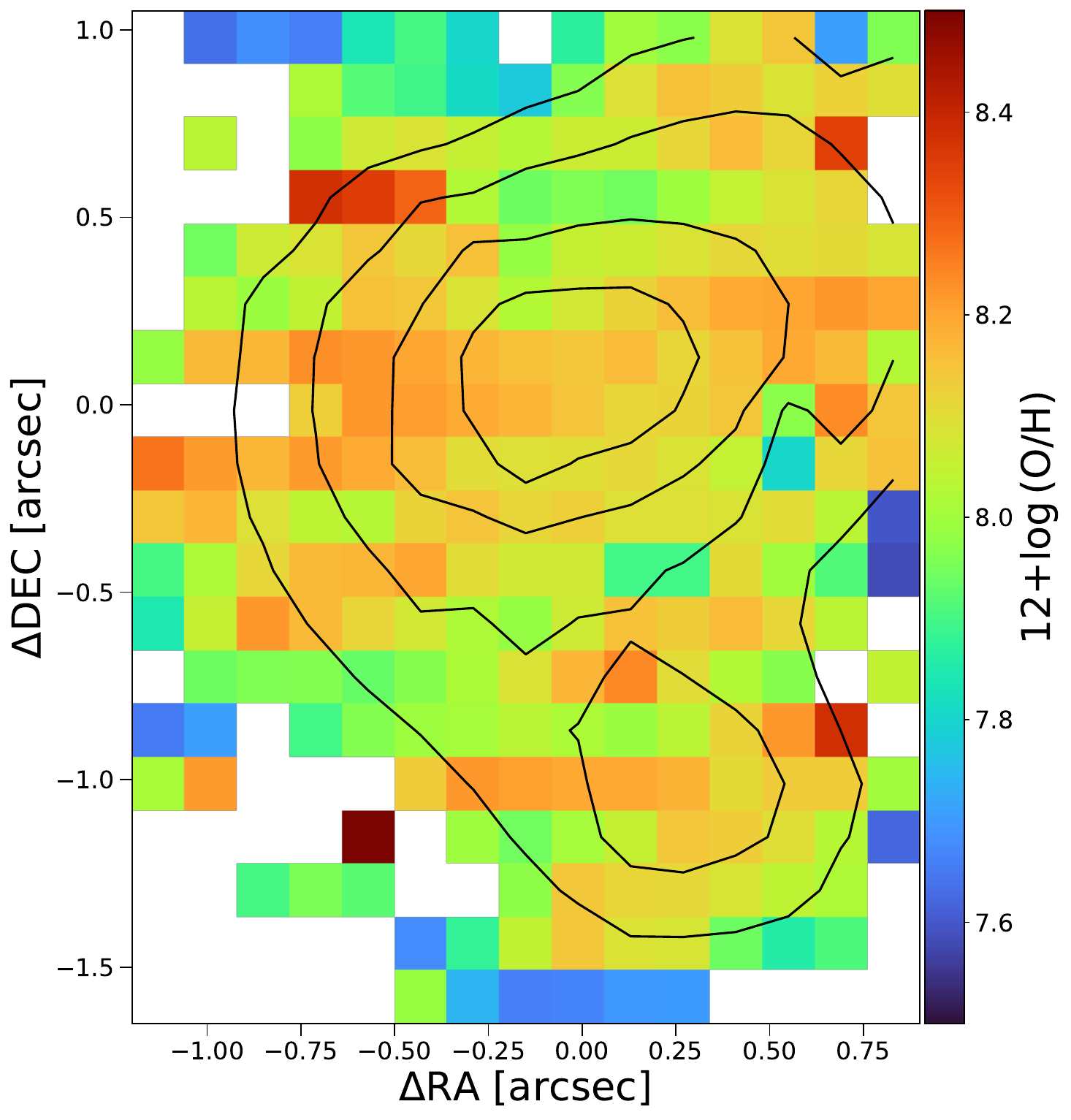}
\caption{Gas-phase metallicity map of J1601 derived from the Keck/KCWI IFU data using the direct method. Contours indicate the [\ion{O}{3}] $\lambda5007$ flux. The metallicity distribution is approximately uniform across the field of view, with no significant radial gradient or localized nuclear enhancement.}
\label{fig:J1601_metallicity}
\end{figure}

Figure~\ref{fig:J1601_metallicity} shows the gas-phase metallicity map of J1601 derived from the Keck/KCWI IFU data. The metallicity was computed using the direct method following \citet{Jiang2019}, with the electron temperature and density determined using the iterative procedure of \citet{Izotov2006} and the prescriptions of \citet{Sanders2016}. The electron temperature is constrained by the auroral-to-nebular [\ion{O}{3}] line ratio, including [\ion{O}{3}] $\lambda4363$ and [\ion{O}{3}] $\lambda5007$, and is of order $T_e\sim10^4$~K in the central region. The [\ion{S}{2}] $\lambda6717/\lambda6731$ doublet gives a central electron density of order $n_e\sim100~{\rm cm^{-3}}$, consistent with the density adopted for the Case~B recombination-line extinction estimates below. The [\ion{O}{3}] $\lambda5007$ flux contours are overlaid for reference.

The metallicity distribution is approximately uniform across the KCWI field of view, with no significant radial gradient and no localized enhancement at the nucleus. The characteristic metallicity is $12+\log({\rm O/H})\simeq8.06$--$8.08$, consistent with a low-metallicity dwarf galaxy. The spatial uniformity of the abundance map indicates that the ionized gas is well mixed on the scales probed by KCWI.

\subsubsection{Dust Extinction and Obscuration}
\label{subsubsec:extinction}

We estimate the extinction toward the ionized gas in J1601 using both optical and near-infrared hydrogen recombination-line ratios. For the optical extinction, we use the Balmer decrement measured from the KCWI data. For the near-infrared extinction, we use the Pa$\alpha$ ($1.875~\mu$m) and Br$\alpha$ ($4.051~\mu$m) lines measured from the JWST/NIRSpec data, following \citet{Doan2025}. We assume Case~B recombination, using the electron temperature and density derived from the KCWI optical emission lines above, with values consistent with $T_e \sim 10^4$~K and $n_e \sim 100~\mathrm{cm^{-3}}$. Intrinsic line ratios are adopted from \citet{Hummer1987}. For the near-infrared extinction calculation, we assume a foreground-screen geometry and adopt a near-infrared extinction law of the form $\tau_{\lambda} \propto \lambda^{-1.85}$ \citep{Fitzpatrick2009}, combined with the average Small Magellanic Cloud extinction law to relate infrared and optical extinction \citep{Gordon2023}.

\begin{figure*}
\centering
\includegraphics[width=\textwidth]{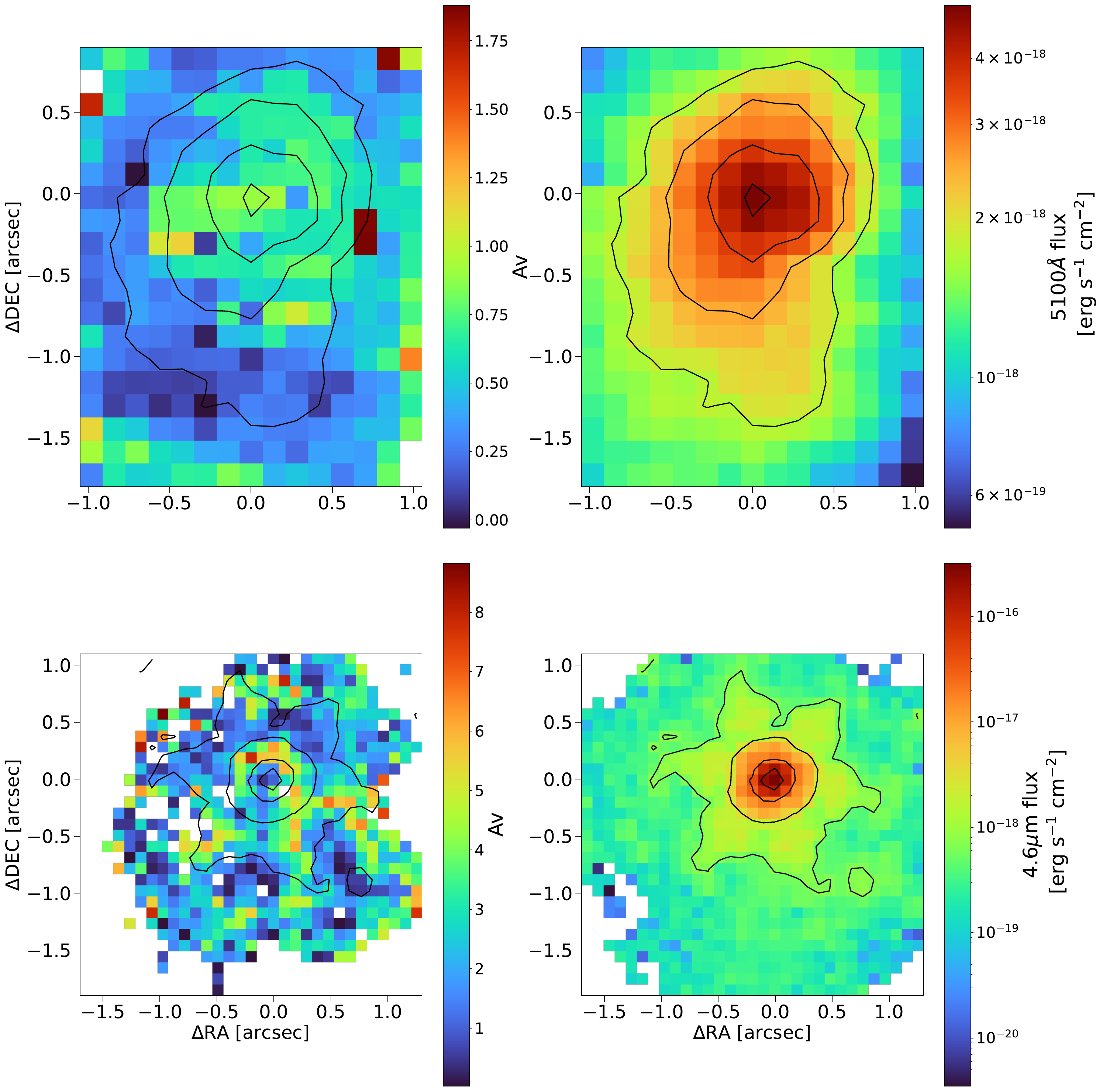}
\caption{Extinction and continuum maps of J1601. Top left: optical extinction, expressed as $A_V$, derived from the Balmer decrement measured with KCWI. Top right: line-free rest-frame 5100~\AA\ continuum from KCWI. Bottom left: near-infrared extinction, expressed as $A_V$, derived from the Pa$\alpha$/Br$\alpha$ ratio measured with JWST/NIRSpec. Bottom right: line-free 4.6~$\mu$m continuum from JWST/NIRSpec. Contours in the extinction panels show the corresponding line-free continuum emission. The extinction is spatially patchy, while the 4.6~$\mu$m continuum is compact and centrally concentrated. The recombination-line extinction maps trace only gas visible at optical or near-infrared wavelengths and therefore do not constrain material that is completely hidden behind much larger column densities.}
\label{fig:J1601_extinction}
\end{figure*}

Figure~\ref{fig:J1601_extinction} compares the optical and near-infrared extinction maps with the corresponding line-free continuum images. The KCWI Balmer-decrement map traces the extinction toward the optically visible ionized gas, while the NIRSpec Pa$\alpha$/Br$\alpha$ map traces the extinction toward the near-infrared recombination-line-emitting gas. The optical extinction is somewhat higher near the central region, but the absolute values remain modest. The near-infrared extinction is patchy and is systematically lower than that found in J1201. Neither recombination-line map shows the large, compact extinction peak coincident with the unresolved 4.6~$\mu$m continuum source that would be expected if the observed line emission were dominated by a heavily reddened nuclear source.

The extinction inferred from recombination-line ratios applies only to gas that remains visible at optical or near-infrared wavelengths. In a clumpy or deeply embedded geometry, a source hidden behind much larger column densities would contribute little or no observed recombination-line flux, and therefore would not necessarily appear as a high-extinction region in these maps. Thus, the measured extinction argues against strong obscuration of the observed line-emitting gas and against a visible or moderately obscured AGN dominating the nuclear spectrum, but it cannot by itself exclude a completely buried or quiescent black hole.

The extinction measurements therefore provide an important constraint on the geometry of the observed gas and dust in J1601. The line-emitting gas is not subject to the extreme attenuation required to hide a luminous AGN that otherwise produces normal optical or near-infrared emission-line signatures; such a scenario would require a more deeply buried component that is effectively invisible in the recombination-line maps.

\subsubsection{Stellar Signatures in the Nuclear Spectrum}
\label{subsec:stellar_population}

\begin{figure*}
\centering
\includegraphics[width=\textwidth]{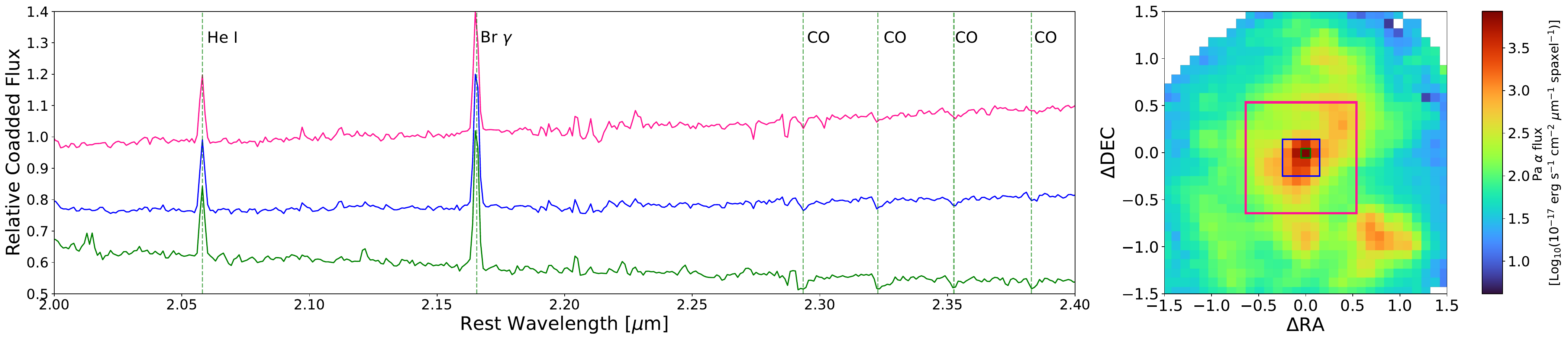}
\caption{Near-infrared spectra of J1601 highlighting the CO bandhead absorption features in the nuclear region. The CO bandheads are detected in the nuclear extractions, indicating that red supergiants/late-type stars contribute to the continuum emission in the central regions of the galaxy. In contrast, these features are weak or absent in the off-nuclear regions. The presence of CO absorption in J1601, but not in J1201, suggests a more developed recent starburst with red supergiants/ cool evolved stars in the nucleus of J1601. The spectrum shown in green is associated with the smallest single-spaxel aperture, the spectrum shown in blue corresponds to the next largest aperture, and the spectrum shown in red corresponds to the largest nuclear aperture.}
\label{fig:J1601_CO}
\end{figure*}

In addition to the emission-line diagnostics, the nuclear spectrum of J1601 shows clear stellar signatures. We detect CO absorption bandheads in the nuclear region (Figure~\ref{fig:J1601_CO}), indicating that red supergiants/late-type stars contribute to the near-infrared continuum. These features are not observed in J1201, highlighting an important difference between the stellar populations of the two systems.

\begin{figure*}
\centering
\includegraphics[width=\textwidth]{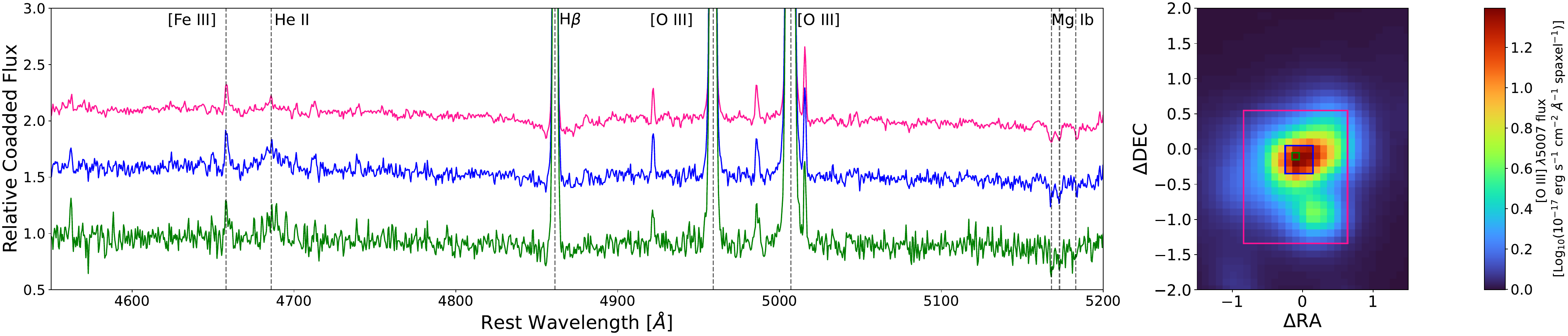}
\caption{Optical spectral region of J1601 showing the Wolf--Rayet features in the nuclear aperture. A faint broad \ion{He}{2} component associated with WR stars is detected in the smaller nuclear extractions and becomes diluted in the larger aperture, indicating recent massive star formation concentrated toward the central region. The WR features are weak and are not accompanied by the coronal-line emission expected from an AGN. The Mg~{\sc i}b triplet is detected in each aperture. The spectrum shown in green is associated with the smallest aperture, the spectrum shown in blue corresponds to the next largest aperture, and the spectrum shown in red corresponds to the largest nuclear aperture.}
\label{fig:J1601_WR}
\end{figure*}

We also identify weak Wolf--Rayet (WR) features in the optical nuclear spectrum (Figure~\ref{fig:J1601_WR}), including a faint broad \ion{He}{2} component associated with the WR population. The WR-feature emission has a compact, approximately circular distribution with a diameter of $\sim2.5$~kpc, and its peak is spatially coincident with the peak of the [\ion{O}{3}] $\lambda5007$ emission. The feature has a consistent FWHM of $\sim1200~{\rm km~s^{-1}}$ across the spaxels where it is detected.

Although \ion{He}{2} $2.189~\mu$m can be detected in some WR populations, a careful examination of the JWST/NIRSpec spaxels reveals no such emission line in J1601. This non-detection is not unexpected, given the strong K-band continuum in the nuclear spectrum and the intrinsically weaker near-infrared \ion{He}{2} transition compared to the optical \ion{He}{2} feature. The WR detection therefore indicates recent massive star formation and provides a plausible stellar source of moderately hard ionizing photons, while the absence of strong nebular \ion{He}{2} and coronal-line emission indicates that the nuclear radiation field is not dominated by a harder ionizing continuum expected from an AGN.

The optical spectrum also shows Mg~{\sc i}b absorption. The combination of CO bandhead absorption, Mg~{\sc i}b absorption, and WR emission points to a composite stellar population in the nuclear region, including both cool evolved stars and very young massive stars. These stellar signatures provide an important context for interpreting the red nuclear continuum and emission-line properties of J1601.

\subsection{Constraints on AGN Activity}
\label{subsec:agn_constraints}

The absence of coronal-line detections in the JWST/NIRSpec and KCWI data places strong constraints on the presence of an AGN in J1601. To quantify these constraints, we compare the upper limits on coronal-line luminosities with those expected for AGNs of similar mid-infrared luminosity. If the observed \textit{WISE} W2 luminosity were powered primarily by an AGN, then the corresponding high-ionization lines should follow the empirical relations observed in AGN samples.

\begin{figure*}
\centering
\includegraphics[width=0.49\textwidth]{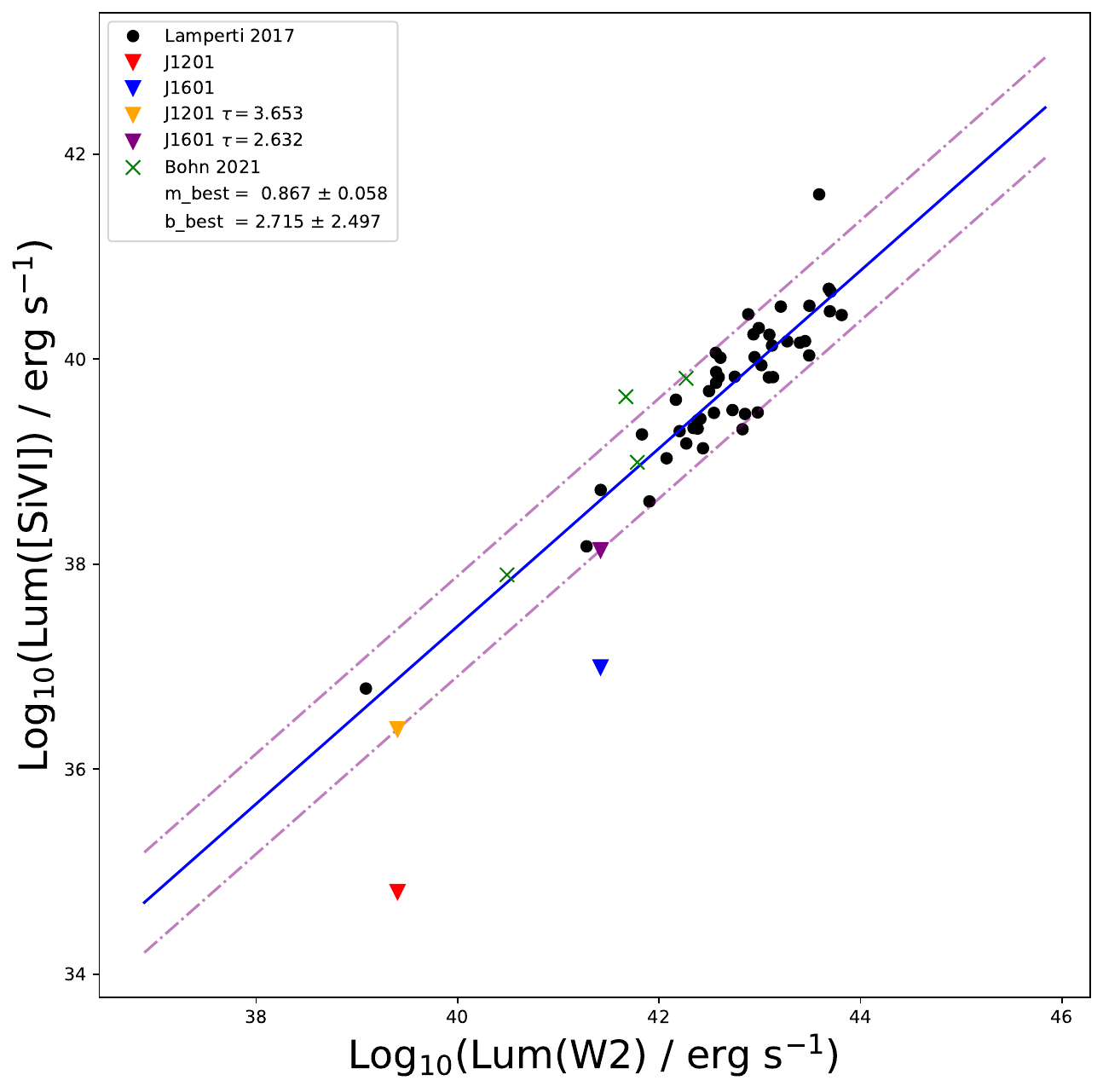}
\includegraphics[width=0.49\textwidth]{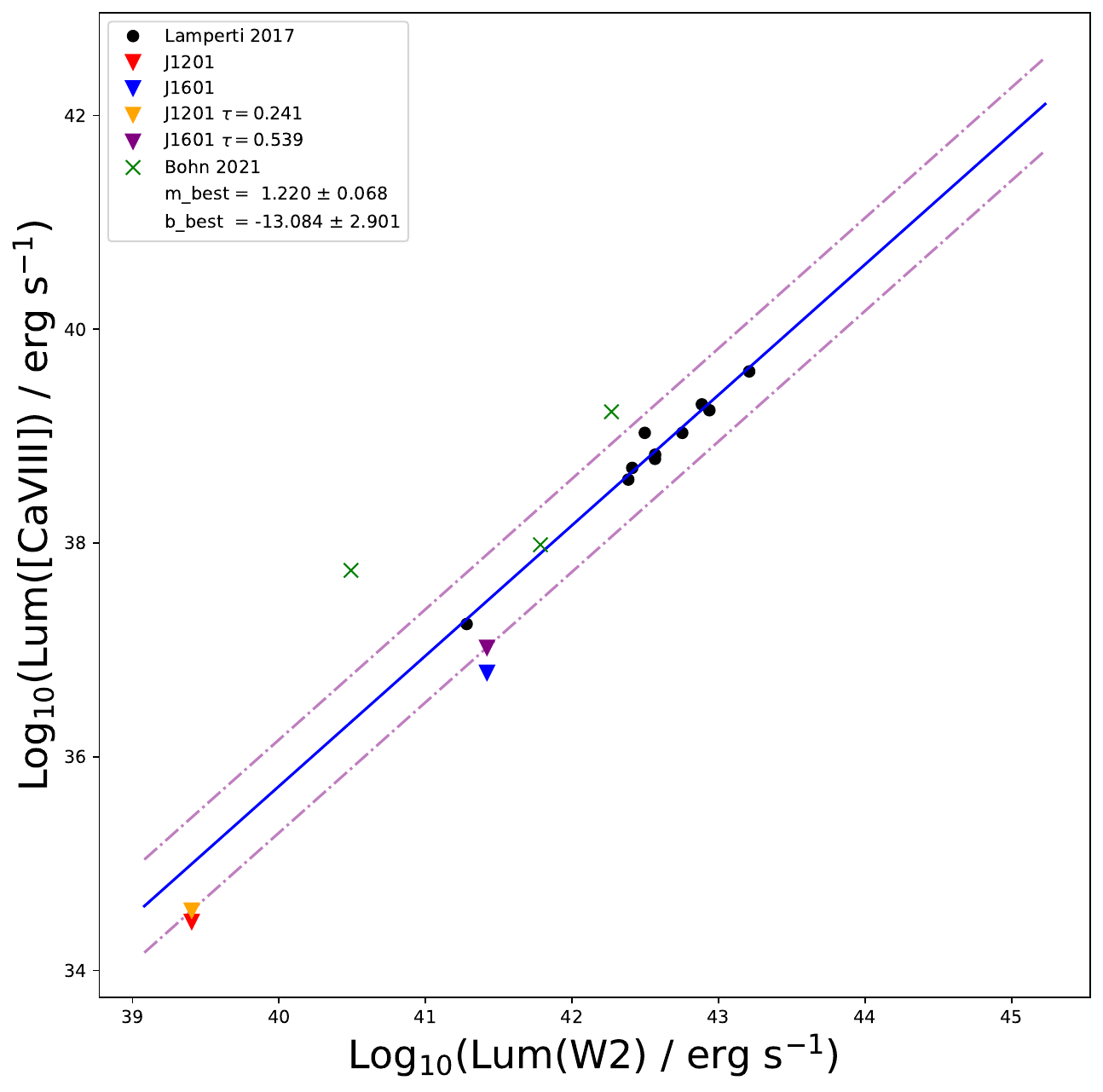}
\caption{Coronal-line luminosity as a function of \textit{WISE} W2 luminosity for J1201 and J1601 compared to AGN samples from \citet{2017MNRAS.467..540L} and \citet{Bohn2021}. The solid lines show the best-fit relations for AGNs, with dashed lines indicating the intrinsic scatter. Downward arrows denote the observed 3$\sigma$ upper limits for J1201 and J1601. In each plot, the value of the extinction at the wavelength of each line required to move the upper limit up to the lower bound in the local relations (yellow and purple triangles) is listed in the legend. The required optical depths are much larger than those inferred for the ionized gas from the near-infrared recombination-line ratios, indicating that the coronal-line deficit cannot be explained by the modest extinction inferred. If the W2 luminosities of J1201 and J1601 were powered primarily by AGN-heated dust, the corresponding [\ion{Si}{6}] and [\ion{Ca}{8}] luminosities would be expected to lie near the AGN relations; instead, the observed limits fall well below these correlations.}

\label{fig:cl_w2}
\end{figure*}

Figure~\ref{fig:cl_w2} shows the upper limits on [\ion{Si}{6}] and [\ion{Ca}{8}] luminosities as a function of W2 luminosity, compared to established AGN correlations from \citet{2017MNRAS.467..540L} and \citet{Bohn2021}. In AGNs, W2 emission traces hot dust heated by the central engine and correlates with coronal-line luminosity. For J1601 and J1201, however, the observed coronal-line upper limits lie well below the AGN relations and outside the intrinsic scatter. Thus, if the compact mid-infrared emission in these galaxies were powered by an AGN with luminosity comparable to that implied by W2, the corresponding infrared coronal lines should have been detected.

To test whether extinction could reconcile the observed upper limits with the AGN relations, we calculate the optical depth required to move each upper limit onto the corresponding local AGN relation. These extinction-corrected values are shown as yellow and purple triangles in Figure~\ref{fig:cl_w2}. The required optical depths (shown in the figure legend) are substantially larger than those inferred from the near-infrared recombination-line ratios in the observed ionized gas as shown in Section~\ref{subsubsec:extinction}.

Thus, the near-infrared coronal-line deficit is not easily explained by the modest extinction measured toward the observed line-emitting gas. As shown in Section~\ref{subsubsec:extinction}, the optical extinction in J1601 is somewhat enhanced near the central region but remains low in absolute value, and the near-infrared extinction is patchy rather than strongly peaked on the compact 4.6~$\mu$m continuum source. These recombination-line measurements do not rule out a completely buried or quiescent black hole, because they only probe gas visible at optical or near-infrared wavelengths. They do, however, argue against a visible or moderately obscured AGN dominating the observed nuclear spectrum.

\begin{figure}
\centering
\includegraphics[width=\columnwidth]{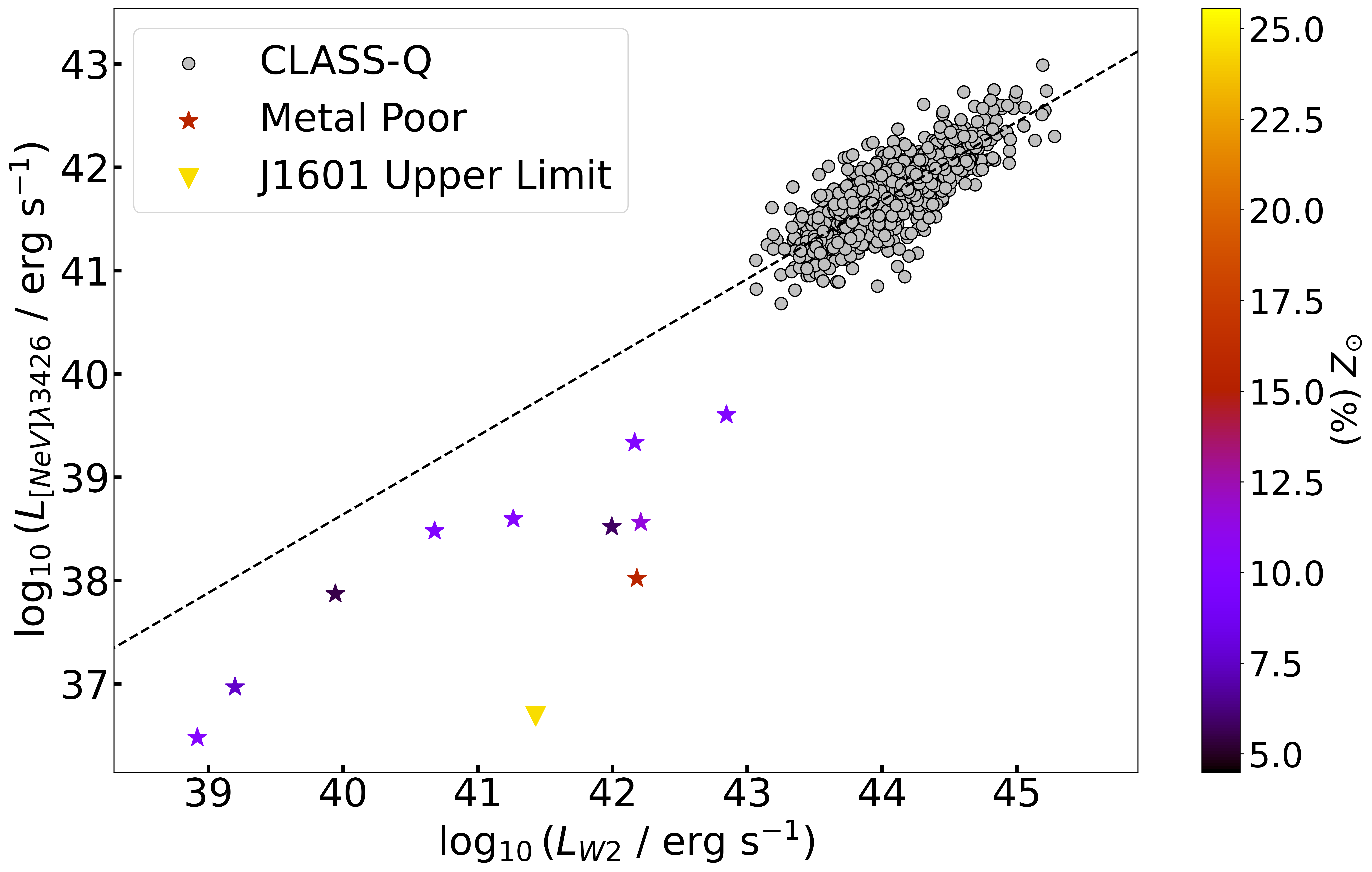}
\caption{Upper limit on the optical [\ion{Ne}{5}] $\lambda3426$ luminosity of J1601 compared with the quasar sample from \citet{Doan2025b} and metal-poor [\ion{Ne}{5}] emitters from \citet{2021MNRAS.508.2556I} and \citet{2024ApJ...966..170H}. The dashed line indicates the best-fit relation for the quasar sample. The color bar shows gas-phase metallicity in units of percent solar for the metal-poor comparison sample. J1601 lies significantly below the quasar relation and below the luminosities of detected metal-poor [\ion{Ne}{5}] emitters at comparable W2 luminosity. The non-detection of [\ion{Ne}{5}] in J1601 therefore cannot be explained solely by its low metallicity.}
\label{fig:opt_nev_w2}
\end{figure}

An independent constraint comes from the upper limit to the optical [\ion{Ne}{5}] $\lambda3426$ line. In Figure~\ref{fig:opt_nev_w2}, we compare the upper limit on the [\ion{Ne}{5}] $\lambda3426$ luminosity of J1601 as a function of the W2 luminosity with the quasar sample of \citet{Doan2025b} and with metal-poor galaxies in which [\ion{Ne}{5}] has been detected \citep{2021MNRAS.508.2556I,2024ApJ...966..170H}. J1601 lies well below the quasar relation and below the detected metal-poor [\ion{Ne}{5}] emitters at comparable W2 luminosity. Because the comparison sample includes galaxies with similarly low metallicities, the lack of [\ion{Ne}{5}] in J1601 cannot be attributed solely to its low metallicity.

The metal-poor [\ion{Ne}{5}] emitters themselves appear offset from the quasar relation, suggesting that low-metallicity systems may not follow the same [\ion{Ne}{5}]--mid-infrared scaling as luminous quasars. Even relative to this population, however, J1601 remains underluminous in [\ion{Ne}{5}] emission, indicating a deficit of observed high-ionization emission from the line-emitting gas.

These coronal-line constraints are consistent with the optical diagnostics presented above. The KCWI spaxels fall in the star-forming region of the BPT diagram, the narrow \ion{He}{2} $\lambda4686$/H$\beta$ ratios remain weak across the field, and the additional high-ionization optical diagnostics place J1601 outside the AGN and composite regions. Thus, the absence of coronal lines is not an isolated result, but part of a broader pattern in which the optical and near-infrared emission-line spectrum is dominated by stellar photoionization.

Additional multiwavelength constraints provide no positive evidence for accretion. J1601 was not detected in the Chandra observations presented by \citet{Latimer2021}, which give a hard X-ray upper limit of $\log L_{2-10,{\rm keV}} < 39.7$~erg~s$^{-1}$. This limit lies well below the X-ray luminosity expected from the observed W2 luminosity if the mid-infrared emission were powered by a standard AGN. The X-ray non-detection does not exclude a Compton-thick, intrinsically weak, or deeply buried accreting black hole, but it provides no evidence for an unobscured X-ray source associated with the compact nuclear continuum.

J1601 is also not detected in the Very Large Array Sky Survey (VLASS; \citealt{2020PASP..132c5001L}) at 3~GHz. Inspection of the VLASS image gives an rms noise of $113.1~\mu$Jy~beam$^{-1}$, corresponding to a $3\sigma$ upper limit of $0.34$~mJy. This upper limit is consistent with the radio emission expected from star formation based on the radio--SFR calibration of \citet{2011ApJ...737...67M}, and provides no evidence for a radio excess associated with an accreting black hole. Finally, we find no significant mid-infrared variability in the WISE light curve (Appendix~\ref{subsec:variability}). The lack of WISE variability does not rule out an AGN, particularly if the variable component is weak, obscured, or diluted by host-galaxy emission, but it is consistent with a stable dust-heated continuum dominated by star formation rather than by a variable accretion source.

The combined infrared, optical, X-ray, radio, and variability constraints show that J1601 lacks the high-ionization or non-stellar signatures expected if its compact mid-infrared source were powered by a visible or moderately obscured accreting black hole. A deeply buried, Compton-thick, radio-quiet, or quiescent black hole cannot be excluded, but the current data provide no independent evidence for such a component.

In the following section, we examine whether an accreting intermediate-mass black hole could nevertheless remain undetected by comparing the observed emission-line limits with predictions from photoionization models.

\subsection{The Nature of the Nuclear Source}
\subsubsection{A Compact Nuclear Starburst}
\label{subsubsec}
The preceding sections show that the compact nuclear source in J1601 is associated with hot dust, strong hydrogen recombination emission, CO bandhead absorption, Mg~{\sc i}b absorption, weak Wolf--Rayet features, and circumnuclear PAH emission. At the same time, the nuclear spectrum lacks the high-ionization coronal lines, broad AGN-like emission-line components, and optical line ratios expected from observable accretion. These diagnostics point to a starburst-dominated nuclear source rather than a visible or moderately obscured AGN.

The stellar signatures indicate a composite nuclear population. The CO bandheads and Mg~{\sc i}b absorption require a contribution from red supergiants or cool evolved stars, while the WR features indicate recent massive star formation. The compact 4.6~$\mu$m continuum therefore appears to arise from a centrally concentrated, dust-enshrouded star-forming component embedded within the larger clumpy star-forming system.

\subsubsection{Could a Deeply Buried AGN Contribute?}
\label{subsubsec:buried_agn}

The data do not exclude a completely buried or currently quiescent black hole. A deeply obscured AGN could evade the optical and near-infrared recombination-line diagnostics if it were hidden behind a column much larger than that sampled by the observed line-emitting gas. We therefore use CIGALE \citep{2022ApJ...927..192Y} to test whether a heavily obscured AGN component could reproduce the WISE photometry while remaining consistent with the coronal-line non-detections.

\begin{figure*}
\centering
\includegraphics[width=0.9\textwidth]{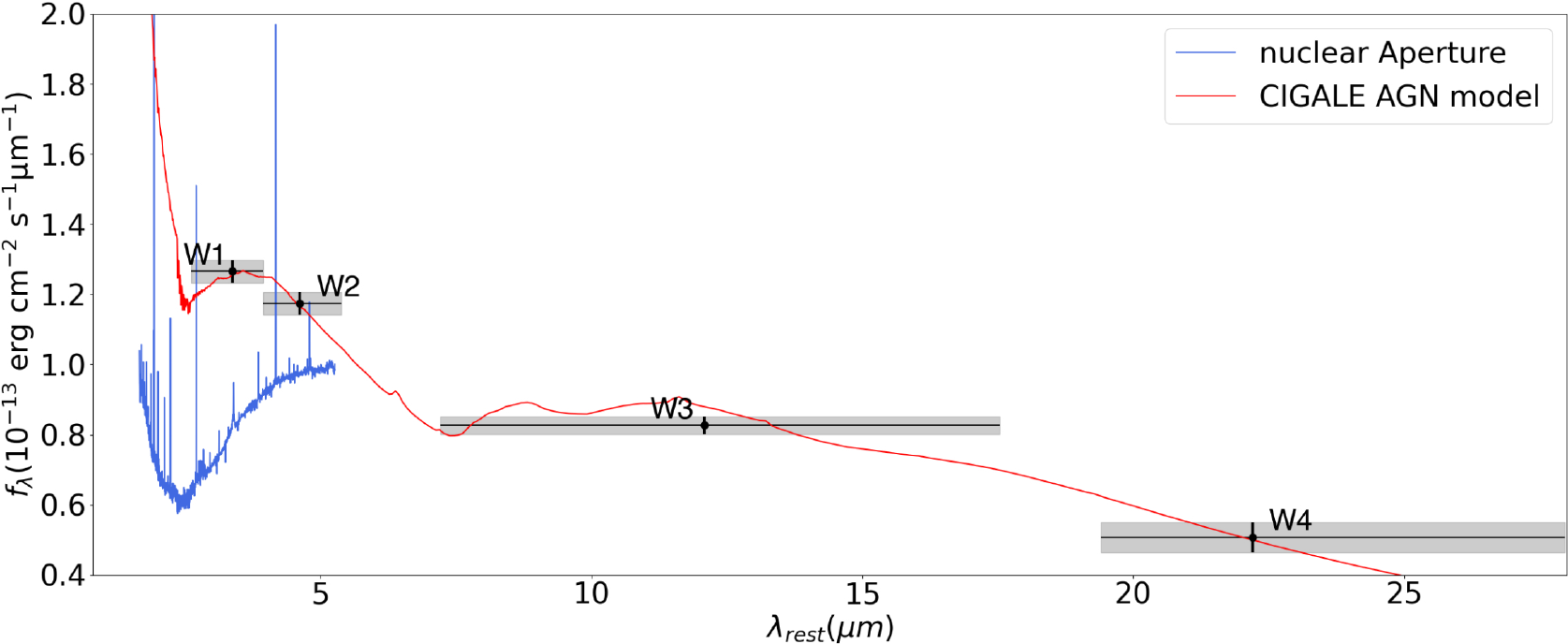}
\caption{JWST/NIRSpec nuclear spectrum of J1601 with WISE photometry overlaid. The red curve shows a CIGALE model including a heavily obscured AGN component using the SKIRTOR templates. The model reproduces the mid-infrared photometry, but requires substantial obscuration to remain consistent with the absence of near-infrared coronal lines. This fit should be interpreted as a test of whether a buried AGN could reproduce the infrared emission, not as unique evidence for accretion.}
\label{fig:J1601_cigale}
\end{figure*}

In Figure~\ref{fig:J1601_cigale}, we show the JWST/NIRSpec nuclear aperture spectrum with the WISE photometry overlaid. We model the infrared spectral energy distribution using the SKIRTOR AGN models \citep{2016MNRAS.458.2288S}, which include emission from the accretion disk, dusty torus, and polar dust component. For this test, we apply an additional obscuring screen with $\tau_{2~\mu{\rm m}}=2.715$, corresponding to the optical depth required to raise the observed near-infrared coronal-line upper limit to the local AGN relation in Figure~\ref{fig:cl_w2}. This optical depth is much larger than that inferred for the observed ionized gas from the near-infrared recombination-line ratios. We adopt the SMC extinction-law option in CIGALE for the obscuring dust.

The resulting model can reproduce the WISE photometry and yields a bolometric luminosity of $\sim5\times10^{42}$~erg~s$^{-1}$. If attributed entirely to Eddington-limited accretion, this luminosity would require a black hole mass of at least $\sim4\times10^{4}~M_\odot$. However, an AGN with this luminosity would be expected to produce strong high-ionization emission unless the coronal-line region is hidden behind a separate column that is not traced by the observed recombination-line emission. In particular, an AGN luminous enough to power the W2 emission would be expected to lie near the coronal-line--infrared relations defined by local AGN samples, producing detectable [\ion{Si}{6}] and [\ion{Ca}{8}] emission unless the line-emitting gas is much more heavily obscured than indicated by Pa$\alpha$/Br$\alpha$.

The CIGALE solution is therefore not unique, and the obscured AGN component should not be interpreted as evidence that accretion is required. Recent JWST studies of compact, dust-enshrouded stellar systems show that young star-forming regions can produce near- and mid-infrared excesses that are not fully captured by standard stellar-population templates. For example, \citet{Pedrini2025} find that CIGALE fits to young embedded clusters selected through strong Pa$\alpha$ and 3.3~$\mu$m PAH emission show a systematic flux excess at $1.5$--$2.5~\mu$m and favor dust parameters that are inconsistent with independent analyses of star-forming regions. This points to shortcomings in current template libraries in the $1$--$5~\mu$m regime. More broadly, JWST observations are revealing compact, highly embedded star-forming regions and emerging young massive clusters with steep infrared continua, hot dust emission, and substantial dust obscuration \citep[e.g.,][]{Knutas2025}. Such systems can resemble sources traditionally interpreted as AGNs in broadband infrared photometry while being powered by stellar processes.

Thus, the apparent best fit model for an obscured AGN component in the CIGALE fit may reflect limitations in the available stellar and dust templates rather than a physical requirement for an accreting black hole. Reconciling the CIGALE-inferred luminosity with the absence of coronal lines would require a geometry in which the AGN and its coronal-line region are hidden behind an obscuring column much larger than that affecting the observed ionized gas. While such a deeply buried AGN cannot be excluded, the recombination-line extinction maps show only modest extinction toward the observed gas, even where the optical extinction increases toward the central region. The available data therefore do not require an accreting black hole, and the compact infrared source is more naturally explained as a dust-enshrouded nuclear starburst.

\subsubsection{Photoionization Constraints on an Accreting Black Hole}

While the infrared continuum of J1601 can be reproduced with a heavily obscured AGN component (Section~5.6.2), the emission-line spectrum provides an independent and more stringent test of the presence of an accreting black hole. In particular, high-ionization and recombination line ratios are sensitive to the shape and hardness of the ionizing radiation field, and can be used to distinguish between stellar and accretion-powered sources.

To assess whether an intermediate-mass black hole (IMBH) could reproduce the observed emission-line properties, we construct a grid of photoionization calculations using \textsc{Cloudy} v25.00 \citep{Cloudyv25} with the \textsc{agnsed} spectral energy distribution from \citet{KubotaDone2018}, which self-consistently models emission from a standard optically thick and geometrically thin accretion disk, a warm Comptonizing region accounting for the soft X-ray excess, and a hot corona extending out to 100\,keV. These SEDs were generated in the X-ray spectral fitting package \textsc{Xspec} \citep{xspec}. We consider three black hole masses, $\log(M_{\mathrm{BH}}/\mathrm{M}_\odot)=3, 5,$ and $8$, and two accretion rates, $\log(\dot{m})=-3$ and $-1$. The $10^8,M_\odot$ model is included as a high-mass comparison case rather than as an IMBH model. For each model, the SED shape is set by the adopted black hole mass and accretion rate, and the luminosity is scaled to $L_{\mathrm{bol}}=5\times10^{42}$~erg~s$^{-1}$, consistent with the luminosity inferred from the obscured-AGN CIGALE model.

The gas-phase abundances and scaling are based on \citet{Nicholls2017}, whose reference abundances are taken from \cite{Nieva2012} based on observed metallicities of main-sequence B-stars in local galactic region, with oxygen scaled to match the metallicity measured from the KCWI data. Dust is included using Orion-type grains, with the dust abundance scaled to match the dust-to-gas ratio observed in \citet{RemyRuyer2014}. Element depletions are implemented using the prescription of \citet{Jenkins2009} and \citet{Gunasekera2022}, with $F_* = 0.427$, corresponding to an iron depletion of $-1.5$ (e.g., \citealp{Thomas2018}). We include dust and depletion explicitly because recent work has shown that coronal-line emission can be strongly suppressed in dusty gas, particularly for refractory species such as silicon and calcium \citep{McKaig2024}. In this sense, our models are conservative: they already account for one of the principal mechanisms that can reduce coronal-line strengths in AGN-ionized gas.

We adopt a constant-pressure equation of state and terminate the calculation at the hydrogen ionization front ($n(\mathrm{H}^{+}) / n(\mathrm{H}^{0})<0.01$). We model clouds of ionized gas spanning radii from the dust sublimation radius \citep{LaorDraine1993} to $\sim0.34$~kpc, corresponding to the radius of the nuclear aperture, and sum their emission assuming a total covering factor of 30\% which is consistent with the covering fractions of many nearby AGN (\citealp{Almeida2017}). The ionization parameter is varied between $\log U=-3$ and $0$, and is held constant across the cloud ensemble when computing integrated emission. This choice of ionization parameter leads to the ionized clouds spanning a wide range of densities from $1\,$cm$^{-3}$ to $10^{10}\,$cm$^{-3}$.


\begin{figure*}
\centering
\includegraphics[width=0.49\textwidth]{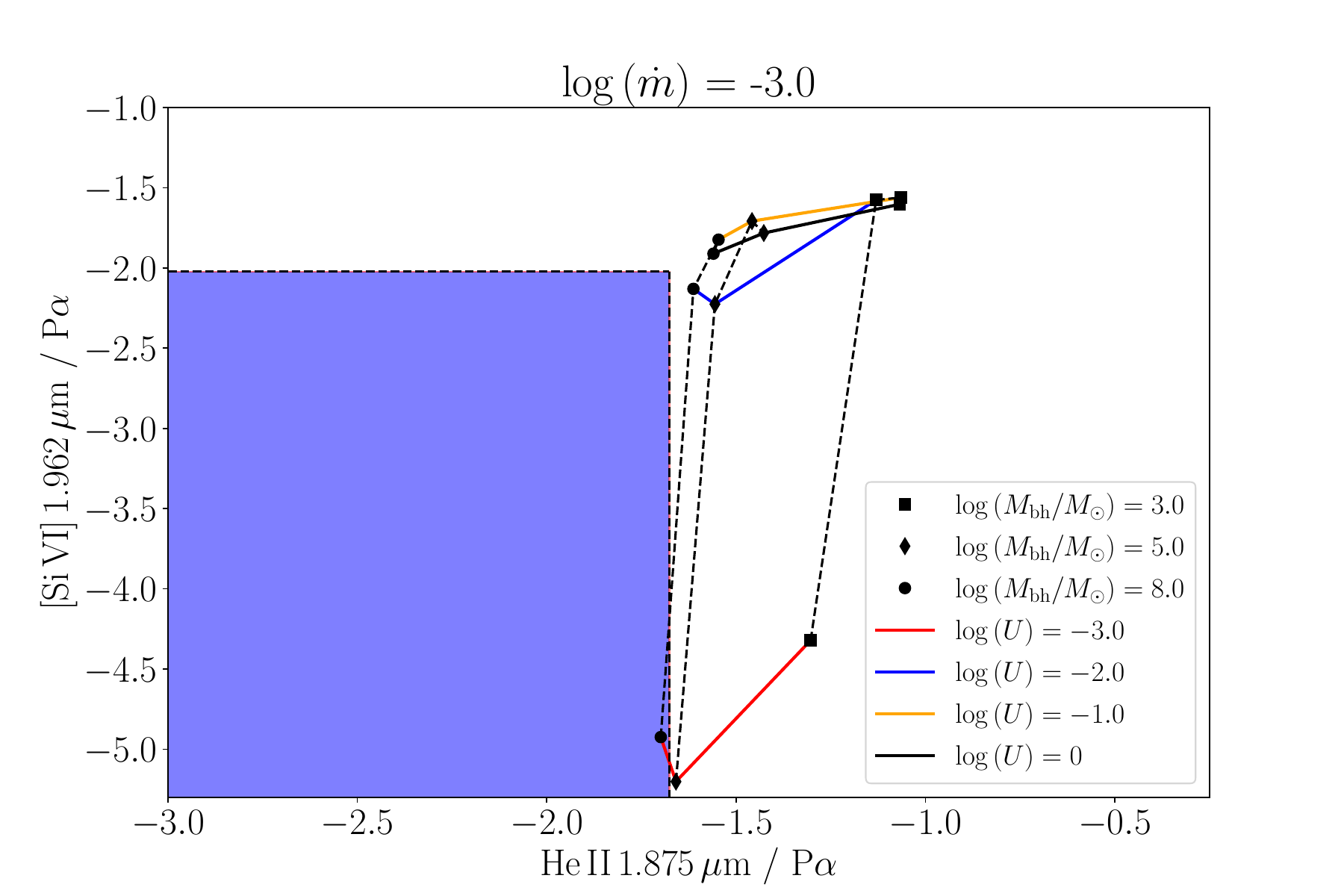}
\includegraphics[width=0.49\textwidth]{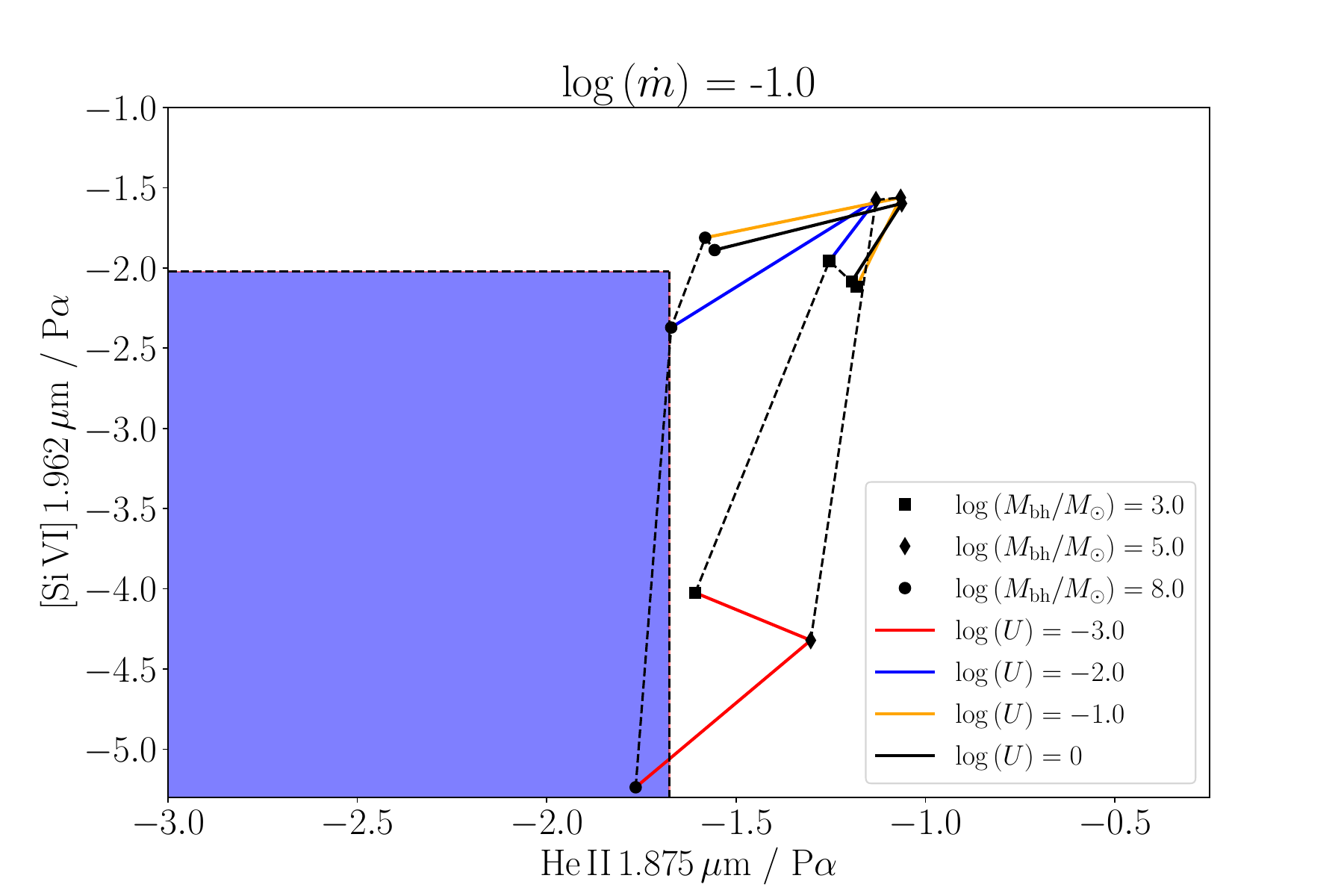}
\caption{Predicted emission-line ratios for AGN photoionization models as a function of ionization parameter. The left panel shows models with $\log(\dot{m})=-3$, and the right panel shows $\log(\dot{m})=-1$. Curves correspond to different black hole masses. The predicted \ion{He}{2}/Pa$\alpha$ and [\ion{Si}{6}]/Pa$\alpha$ ratios from the IMBH models are systematically higher than the observational upper limits, indicating that even weakly accreting IMBHs should produce detectable high-ionization emission. Moreover, models from the higher mass SMBH can be ruled out from the observed P$\alpha$ luminosity (see text). The blue shaded region represents the upper limits on the corresponding line ratios.}
\label{fig:agn_models}
\end{figure*}

Figure~\ref{fig:agn_models} compares the predicted near-infrared \ion{He}{2} $2.189~\mu$m/Pa$\alpha$ and [\ion{Si}{6}]/Pa$\alpha$ flux ratios with the JWST/NIRSpec upper limits. Most models lie outside the allowed region, predicting stronger high-ionization emission than observed. This behavior is especially pronounced for the lower-mass black hole models, whose accretion-disk SEDs are harder and shift to higher characteristic temperatures \citep{Cann2018}. Thus, for the adopted luminosity and cloud assumptions, accreting IMBH models predict high-ionization emission that should have been detected.

A small subset of the $10^8~M_\odot$ comparison models at low ionization parameter falls within the line-ratio limits. These cases, however, do not provide a compelling explanation for the central source in J1601. First, a $10^8,M_\odot$ black hole is not an IMBH and would be difficult to reconcile with the stellar mass of this dwarf galaxy. Second, these models must also reproduce the absolute recombination-line luminosity, not only the high-ionization line ratios. For the two high-mass, low-ionization cases that fall within the allowed line-ratio region, with ${\log(M_{\mathrm{BH}}/M_\odot),\log(\dot{m}),\log(U)}={8,-3,-3}$ and ${8,-1,-3}$, the predicted Pa$\alpha$ luminosities are $5.78\times10^{39}$~erg~s$^{-1}$ and $1.58\times10^{42}$~erg~s$^{-1}$, respectively. These correspond to predicted Pa$\alpha$ fluxes of $2.63\times10^{-15}$ and $7.19\times10^{-13}$~erg~cm$^{-2}$~s$^{-1}$, compared with the observed nuclear value of $1.50\times10^{-15}$~erg~cm$^{-2}$~s$^{-1}$ (Table~\ref{table:nuc_line_flux}). Thus, even the high-mass comparison cases that satisfy the high-ionization line-ratio limits do not naturally reproduce the observed nuclear spectrum for the adopted luminosity and covering factor.

The photoionization models therefore support the empirical coronal-line constraints. The absence of high-ionization emission is not simply a consequence of low metallicity, dust depletion, or an idealized dust-free cloud geometry. Rather, the observed line-emitting gas in J1601 does not appear to be exposed to the hard ionizing continuum expected from an accreting black hole capable of powering the compact infrared source. A quiescent black hole, or an accreting source hidden behind a separate column that is not sampled by the observed gas, cannot be excluded by these models alone.
\section{Discussion}

\subsection{Compact Nuclear Starbursts in Metal-Poor Dwarf Galaxies}

\begin{figure*}
\centering
\includegraphics[width=0.95\textwidth]{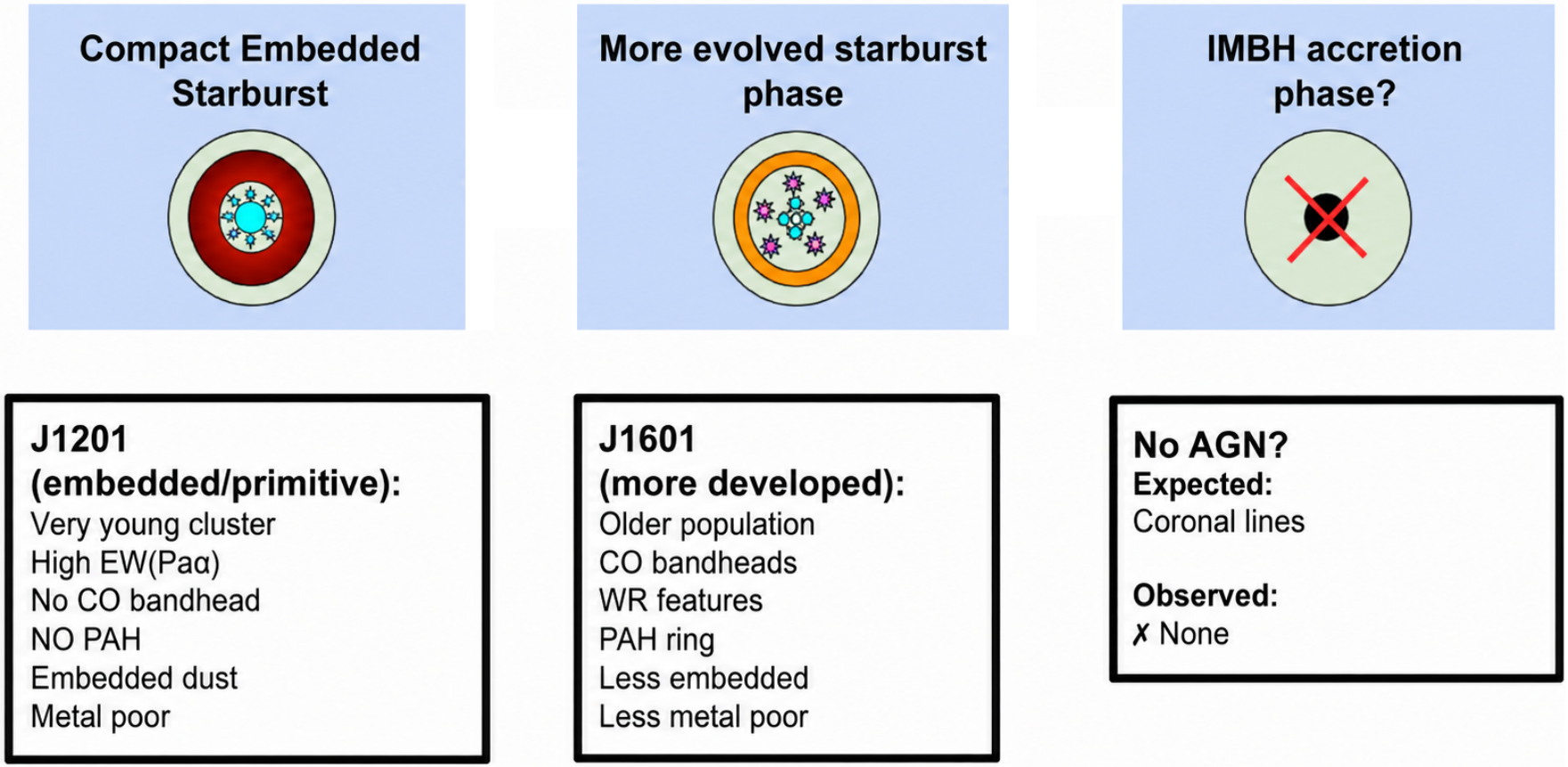}
\caption{Qualitative schematic comparing two possible states of compact nuclear starbursts in low-mass, metal-poor galaxies, based on the contrasting nuclear properties of J1201 and J1601. J1201 is shown as a more dust-enshrouded, chemically primitive nuclear starburst with high Pa$\alpha$ equivalent width, redder mid-infrared colors, and no detected CO bandheads, Wolf--Rayet features, or PAH emission. J1601 is shown as a less obscured state with signatures of a more developed recent starburst, including CO bandhead absorption, weak nuclear Wolf--Rayet features, and circumnuclear PAH emission. Despite these differences, neither system shows evidence for observable accretion, indicating that compact nuclear starbursts can produce AGN-like mid-infrared colors without clear AGN signatures. This figure is intended as a qualitative interpretation rather than a definitive evolutionary sequence.}
\label{fig}
\label{fig:schematic}
\end{figure*}

In Table~\ref{table:nucdiag} and Figure~\ref{fig:schematic}, we compare the nuclear properties of J1201 and J1601, two low-mass, metal-poor galaxies selected for their extreme mid-infrared colors but lacking clear AGN signatures. Although both systems show no evidence for accreting black holes, they differ systematically in their stellar population tracers, dust content, and chemical properties, suggesting that they may represent distinct states of compact nuclear starbursts.

\begin{table*}[t]
\centering
\caption{Comparison of Nuclear Diagnostics and Physical Properties for J1201 and J1601}
\label{tab:nucdiag}
\begin{tabular}{lcc}
\hline\hline
Property & J1201 & J1601 \\
\hline
$12+\log(\mathrm{O/H})$ & 7.67 & 8.08 \\
$\log L_{\mathrm{[Si\,VI]}}$ (erg s$^{-1}$) & $<34.8$ & $<37.5$ \\
EW(Pa$\alpha$) (\AA) & 980 & 176 \\
$\log Q(\mathrm{H})$ (photons s$^{-1}$) & 51.6 & 50.7 \\
PAH morphology & Absent & Circumnuclear \\
CO bandheads & Absent & Present \\
WR features & Absent & Weak / Nuclear \\
$\log L_{\mathrm{X}}$ (erg s$^{-1}$) & $<38.7$ & $<39.7$ \\
WISE $W2-W3$ color & Redder & Less red \\
Extinction & Higher & Lower \\
WISE variability & No & No \\
\hline
\end{tabular}

\begin{flushleft}
\textit{Notes:} Quantities are measured within nuclear apertures corresponding to the same physical scale ($\sim$0.12~kpc). EW(Pa$\alpha$) is the rest-frame equivalent width. $Q(\mathrm{H})$ is derived from extinction-corrected recombination-line luminosities assuming Case~B recombination. The [\ion{Si}{6}] luminosities are 3$\sigma$ upper limits. PAH morphology indicates the spatial distribution of PAH emission. CO bandheads and WR features trace evolved and young stellar populations, respectively. The WISE $W2-W3$ color and extinction are listed qualitatively to emphasize the relative differences in dust content and embeddedness between the two systems. Values for J1201 are taken from \citet{Doan2025}. The X-ray luminosity is from 2-10~keV and is taken from \citet{Doan2025} for J1201, and \citet{Latimer2021} for J1601.
\end{flushleft}
\label{table:nucdiag}
\end{table*}

J1201 is characterized by a lower metallicity, higher extinction, redder $W2-W3$ color, and a substantially larger Pa$\alpha$ equivalent width compared to J1601. In addition, J1201 shows no evidence for CO bandhead absorption, Wolf--Rayet (WR) features, or PAH emission in its nuclear spectrum. These properties are consistent with a compact, deeply embedded nuclear starburst with a high ionizing photon flux and little evidence for an evolved stellar population. By contrast, J1601 exhibits CO bandhead absorption, weak nuclear WR features, and circumnuclear PAH emission, along with a lower Pa$\alpha$ equivalent width and reduced extinction. These features point to a more complex stellar population that includes red supergiants and a less embedded nuclear environment, with dust redistributed into larger-scale structures.

The contrast between the two systems is qualitatively consistent with differences in stellar-population development and dust geometry, as illustrated schematically in Figure~\ref{fig:schematic}. In standard instantaneous-burst models, WR features typically emerge at ages of order $\sim3$--$6$~Myr, while near-infrared CO bandhead absorption becomes prominent once cool evolved stars, such as red supergiants, begin to contribute significantly to the continuum \citep[e.g.,][]{Leitherer2005}. However, the interpretation of these diagnostics depends strongly on metallicity, stellar evolution physics, and star formation history. In particular, BPASS models that include binary evolution predict that the production and visibility of WR features can be extended or modified relative to single-star models, especially at subsolar metallicity \citep[e.g.,][]{Eldridge2017,Stanway2018}. At the low metallicity of J1201, WR features may be intrinsically weak or absent even at ages where they would be expected at higher metallicity, and their detectability also depends on the total stellar mass of the cluster.

Similarly, the absence of CO bandheads in J1201 indicates a lack of a significant population of red supergiants, but does not uniquely constrain the age, since the development of red supergiants depends on both star formation history and metallicity. The combination of high Pa$\alpha$ equivalent width, lack of CO and WR features, and strong embedded dust emission in J1201 is therefore consistent with a very young and/or chemically primitive nuclear burst, but does not uniquely imply a specific age. In contrast, the presence of CO bandheads and weak WR features in J1601 suggests the presence of red supergiants, although again the precise age depends on the assumed star formation history and metallicity.

The important point is that both systems lack evidence for observable accretion despite occupying different parts of this compact-starburst parameter space. The absence of coronal lines, X-ray detections, and mid-infrared variability in both galaxies suggests that the lack of AGN signatures is not limited to a single set of starburst properties, dust geometries, or stellar-population indicators. Compact, dense, and highly ionized nuclear environments in low-mass, metal-poor galaxies can therefore produce AGN-like mid-infrared colors without necessarily showing evidence for an accreting black hole.

We emphasize that this interpretation is necessarily tentative, as it is based on a small number of systems. J1201 and J1601 may not form a single evolutionary sequence, but rather may represent different regions of parameter space in metallicity, total mass, star formation history, dust geometry, and gas conditions. Nevertheless, their systematic differences, combined with the persistent absence of AGN signatures, provide a useful empirical framework for understanding the range of compact nuclear starbursts in low-mass galaxies and their connection to searches for intermediate-mass black holes.

\subsection{Why Is Observable Accretion Absent?}
\label{subsec:why_no_accretion}

J1201 and J1601 were selected as promising accreting IMBH candidates because of their extreme mid-infrared properties and compact nuclear emission. However, as shown in Sections~5 and~6.1, neither system exhibits the optical, infrared, X-ray, radio, or variability signatures expected from observable accretion. In particular, both galaxies lack near-infrared and optical coronal lines, detectable X-ray emission, and any significant mid-infrared variability. These non-detections are notable because both systems contain compact, dense, highly ionized nuclear gas that should respond to a hard ionizing continuum if an AGN were present and visible to the observed line-emitting gas.

\begin{figure}
\centering
\includegraphics[width=\columnwidth]{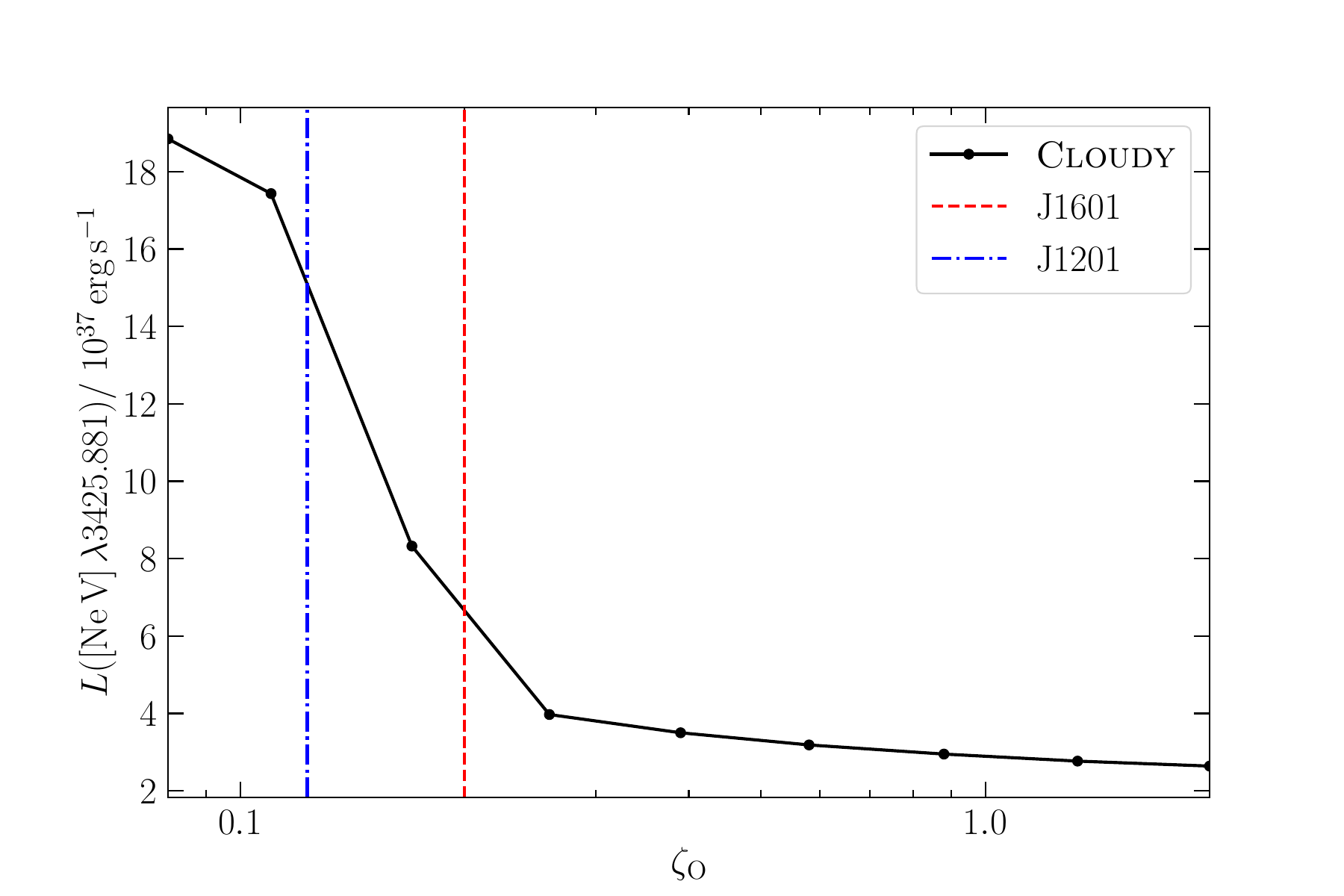}
\caption{Predicted [\ion{Ne}{5}] $\lambda3426$ luminosity as a function of oxygen abundance for an AGN ionizing spectrum. The models assume a fixed bolometric luminosity, ionizing spectral shape, and pressure ratio in order to isolate the effect of metallicity. The [\ion{Ne}{5}] luminosity does not decrease monotonically toward low metallicity, but instead increases over much of the metal-poor regime because of the declining dust-to-gas ratio and reduced absorption of ionizing photons by dust grains. The non-detection of [\ion{Ne}{5}] emission in J1601 and J1201 therefore cannot be explained by low metallicity alone.}
\label{fig:nev_metallicity}
\end{figure}

One possible explanation for the absence of high-ionization emission is that these galaxies are too metal poor to efficiently produce coronal lines. However, our photoionization models suggest that this explanation is insufficient. In Figure~\ref{fig:nev_metallicity}, we show the predicted [\ion{Ne}{5}] $\lambda3426$ luminosity as a function of oxygen abundance for an AGN ionizing spectrum. The models adopt the \textsc{agnsed} spectral energy distribution of \citet{KubotaDone2018} with $\log(M_{\mathrm{BH}}/M_\odot)=4.0$ and $\log(\dot{m})=-1.0$, corresponding to a bolometric luminosity of $L_{\mathrm{bol}}=1.445\times10^{41}$~erg~s$^{-1}$. The oxygen abundance is varied from 0.08 to 2 times the reference abundance set, while the pressure ratio is fixed at $P_{\mathrm{rad}}/P_{\mathrm{gas},0}=50$ to remain consistent with the approach used in \citet{McKaig2024}. We model a single cloud placed at $100$ times the dust sublimation radius \citep{LaorDraine1993} in order to isolate the effect of metallicity on the predicted [\ion{Ne}{5}] emission.

The resulting [\ion{Ne}{5}] luminosity does not decline monotonically toward low metallicity; instead, it increases over much of the metal-poor regime. This behavior arises because low-metallicity gas contains less dust, particularly below $\sim20\%$ solar metallicity, where the dust-to-gas ratio declines steeply with metallicity \citep{RemyRuyer2014}. Reduced dust absorption allows a larger fraction of ionizing photons to reach the gas, enhancing the predicted coronal-line emission \citep{McKaig2024}. In these calculations the ionizing spectral shape is held fixed to isolate the effect of metallicity. In practice, lower-mass accreting black holes produce intrinsically harder radiation fields, which would further enhance the production of high-ionization lines \citep{Cann2018}, including the [\ion{Ne}{5}] line. Thus, the weakness of [\ion{Ne}{5}] and the infrared coronal lines in J1201 and J1601 cannot be explained as a metallicity effect alone; it indicates that the observed line-emitting gas is not exposed to the hard ionizing continuum expected from an accreting black hole.

The absence of observable accretion may therefore reflect a low black hole occupation fraction, a low active fraction, inefficient fueling, or some combination of these effects. Active nuclei are known to exist in dwarf galaxies, but their observed incidence remains low and strongly dependent on selection method. For example, recent results from the DESI early-data census find that only $\sim2\%$ of dwarf galaxies host AGN candidates, compared to $\sim25\%$ of higher-mass galaxies \citep{Pucha2025}. Reviews of intermediate-mass black holes likewise emphasize that robust detections become increasingly rare toward lower stellar masses and that the black hole occupation fraction in the lowest-mass galaxies remains poorly constrained \citep{Greene2020}. The lack of observable accretion in J1201 and J1601 is therefore consistent with the possibility that some compact, metal-poor dwarf galaxies either do not host central black holes or host black holes that are quiescent, low-mass, or accreting below current detection limits.

A second, not mutually exclusive, possibility is that black holes are present but cannot sustain observable accretion. Compact nuclear starbursts may provide dense gas, but they also produce strong and rapidly evolving stellar feedback. Stellar winds, radiation pressure, and supernova explosions can heat, disrupt, or expel gas from the nuclear region on short timescales, preventing the formation of a stable, long-lived accretion flow \citep[e.g.,][]{Hopkins2012,Koudmani2019,Collins2022}. In low-mass galaxies, these processes are particularly effective because of the shallow gravitational potential, leading to bursty star formation histories, large-scale gas outflows, and highly time-variable inflow rates \citep[e.g.,][]{Habouzit2017,Koudmani2019,Hopkins2012,Sharma2020}. The high Pa$\alpha$ equivalent widths observed in J1201 and J1601 indicate recent or ongoing starburst activity on Myr timescales, whereas significant black hole growth requires gas to remain available and able to lose angular momentum over longer periods \citep[e.g.,][]{Volonteri2021,Hopkins2012}. If accretion episodes are repeatedly interrupted by feedback, the time-averaged accretion rate may remain too low to produce detectable coronal-line emission or X-ray emission.

The efficiency of black hole growth in such environments may also depend on the presence of a sufficiently massive and dynamically evolved nuclear star cluster (NSC). Recent simulations suggest that NSCs can enhance gas inflow and promote black hole growth in low-mass galaxies, but that this process is highly episodic and regulated by feedback \citep{Partmann2025}. If the nuclear clusters in systems like J1201 and J1601 are still forming, low in mass, or dynamically immature, they may not yet provide the long-lived central potential or gas inflow needed to sustain accretion. The lowest-mass, most metal-poor compact starbursts may lie below the regime in which black hole seeds can efficiently grow into observable IMBHs, even if black holes are present.

These considerations suggest that the absence of AGN signatures in J1201 and J1601 does not require a single explanation. The black holes may be absent, quiescent, accreting at very low rates, or unable to maintain a stable accretion flow in the presence of strong stellar feedback. What the data show more directly is that compact, dust-enshrouded nuclear starbursts in low-mass, metal-poor galaxies can produce AGN-like mid-infrared colors from compact regions without showing evidence for observable black hole growth. This result raises the possibility that the most compact and chemically primitive starbursting dwarfs are not necessarily the systems in which intermediate-mass black holes are most readily detectable, even though they satisfy widely-used mid-infrared AGN selection.

\subsection{Reassessing Mid-Infrared and High-Ionization AGN Diagnostics}
\label{subsec:diagnostics}

The results for J1201 and J1601 reveal an important degeneracy in mid-infrared AGN selection for metal-poor dwarf galaxies. Both systems were selected because their \textit{WISE} colors placed them in a region of color space commonly associated with hot dust heated by accretion. The JWST observations show that the red mid-infrared colors do not arise from diffuse, galaxy-wide star formation, but instead from compact nuclear emission on parsec scales. However, in both galaxies the compact infrared source lacks the coronal-line, X-ray, radio, variability, and both optical and infrared line-ratio signatures expected from observable accretion. Compactness alone is therefore not sufficient to identify an accreting black hole.

A likely missing ingredient in simple mid-infrared color selection is the structure of the star-forming region. Compact, dense, dust-enshrouded nuclear starbursts can heat dust on small spatial scales and produce steeply rising infrared continua that resemble AGN-heated dust in broadband photometry. This is especially important in metal-poor dwarfs, where the dust geometry, grain properties, star-formation intensity, and radiation field can differ substantially from those in the more massive galaxies used to calibrate many empirical AGN diagnostics. The results presented here therefore do not imply that mid-infrared selection is ineffective, but rather that mid-infrared colors alone cannot distinguish between accreting black holes and compact nuclear starbursts in this regime.

High-ionization lines provide a complementary but also complex diagnostic. Lines such as [\ion{Ne}{5}] $\lambda3426$ are widely regarded as robust tracers of active galactic nuclei, as they require ionizing photons with energies $\gtrsim97$~eV that are difficult to produce in normal stellar populations. As a result, detections of [\ion{Ne}{5}] emission in low-mass, metal-poor galaxies have often been interpreted as evidence for accreting black holes \citep[e.g.,][]{2024arXiv240218643C,2025ApJ...985..253M}. However, the origin of the hard ionizing radiation in metal-poor star-forming systems remains uncertain. Even when [\ion{Ne}{5}] is detected, there is not always definitive multiwavelength evidence for an accreting black hole, and the relevant physical mechanisms may differ across luminosity, metallicity, and star-formation regimes.

This ambiguity is evident in the comparison between metal-poor [\ion{Ne}{5}] emitters and classical AGNs. As shown in Figure~\ref{fig:opt_nev_w2}, classical AGNs follow a well-defined relation between the [\ion{Ne}{5}] luminosity and mid-infrared luminosity. By contrast, the metal-poor [\ion{Ne}{5}] emitters are offset from this relation, exhibiting lower [\ion{Ne}{5}] luminosities at fixed W2 luminosity. The upper limits for J1601 lie even further below the classical AGN relation, reinforcing the conclusion that its compact nuclear emission is not powered by a standard AGN.

The interpretation of detected [\ion{Ne}{5}] emission in metal-poor galaxies therefore requires caution. In well-studied nearby systems such as I~Zw~18 and SBS~0335$-$052E, the origin of the hard ionizing radiation remains debated, with no definitive evidence for an accreting black hole despite the presence of high-ionization lines \citep[e.g.,][]{2012MNRAS.427.1229I,Kehrig2018,2022ApJ...930..105S}. The more luminous systems discussed by \citet{2024arXiv240218643C} demonstrate that [\ion{Ne}{5}] can identify extreme hard-radiation environments, but they may not be direct analogs of the lower-luminosity, nearby metal-poor dwarfs considered here. In the local metal-poor systems, the physical origin of the [\ion{Ne}{5}] emission may be different and may include contributions from extreme stellar populations, shocks, X-ray binaries, or other sources of hard ionizing photons.

Recent JWST/MIRI observations further strengthen this point. High-ionization mid-infrared fine-structure emission, including [\ion{Ne}{5}] and [\ion{Ne}{6}], has now been reported from massive-star winds \citep{2026arXiv260623456H}, following the JWST/MIRI detection of [\ion{Ne}{5}], [\ion{Ne}{6}], and [\ion{O}{4}] in the O9~V star 10~Lacertae \citep{2024ApJ...976L..25L}. These detections show that massive stars can produce highly ionized neon emission under some conditions, opening the possibility that [\ion{Ne}{5}] detected in some metal-poor dwarf galaxies may have a stellar-wind contribution. This does not imply that all [\ion{Ne}{5}] emission in metal-poor galaxies is stellar in origin, and the connection between stellar wind lines and integrated nebular [\ion{Ne}{5}] emission remains uncertain. However, it reinforces the need to interpret high-ionization lines in metal-poor starbursts using line widths, spatial morphology, stellar-population diagnostics, and multiwavelength constraints rather than treating [\ion{Ne}{5}] as a stand-alone black-hole indicator.

Recent JWST observations of metal-poor galaxies have also highlighted the difficulty of interpreting high-ionization emission in such environments. The observed ionization conditions and line ratios can be difficult to reconcile with classical AGN models and may instead point to alternative mechanisms such as extreme stellar populations, shocks, or other non-standard sources of hard radiation \citep[e.g.,][]{Hunt2025,2025ApJ...985..253M}. Thus, [\ion{Ne}{5}] remains a powerful tracer of hard photons, but in metal-poor dwarf galaxies it should not be treated as an unambiguous AGN diagnostic without supporting evidence.

Our observations, together with recent JWST observations of other metal poor galaxies and massive stars, suggest that both mid-infrared colors and high-ionization lines must be interpreted with caution. A robust IMBH search in metal-poor dwarf galaxies requires information on the compactness of the infrared source, the spatial distribution of the ionized gas, coronal-line strengths, optical line ratios, stellar-population indicators, X-ray and radio constraints, and variability. JWST IFU spectroscopy is particularly powerful because it can not only provide access to a plethora of high ionization lines, but it can separate compact nuclear starbursts from extended star formation and directly test whether the compact source is accompanied by the high-ionization emission expected from accretion. Detailed modeling of the emission line spectrum is then required to confirm an accreting IMBH scenario.

These results refine, rather than invalidate, mid-infrared and high-ionization selection. These diagnostics remain valuable tools for identifying unusual compact nuclear sources and hard radiation fields in dwarf galaxies. However, J1201 and J1601 show that compact, metal-poor nuclear starbursts can occupy AGN-like mid-infrared color space without showing observable accretion. In this regime, spatially resolved spectroscopy and multiwavelength follow-up are essential for determining whether the compact source is powered by an accreting black hole or by an extreme nuclear starburst.

\section{Conclusions}

We have presented a detailed analysis of the low-mass, metal-poor galaxy J1601, selected for its extreme mid-infrared colors and compact nuclear emission, which made it a promising candidate for hosting an accreting intermediate-mass black hole (IMBH). Using JWST/NIRSpec IFU data, complemented by Keck/KCWI optical IFU data, \textit{WISE} photometry and variability constraints, published X-ray limits, and radio data, we find no evidence for observable accretion. We compare our results to those of J1201, a similar mid-infrared-selected, metal-poor dwarf galaxy analyzed previously \citep{Doan2025}.

Our main conclusions are as follows:

\begin{enumerate}

\item J1601 hosts a compact nuclear starburst with a steeply rising near-infrared continuum, strong recombination-line emission, molecular hydrogen emission, and stellar-population signatures, but shows no clear evidence for observable accretion. We detect no near-infrared coronal lines, no optical [\ion{Ne}{5}], no broad near-infrared recombination-line components, and no optical line-ratio evidence for AGN photoionization. The KCWI line ratios remain consistent with stellar photoionization across the field of view. Existing multiwavelength constraints are consistent with this picture: the published Chandra upper limit of $\log L_{2-10,{\rm keV}}<39.7$~erg~s$^{-1}$ lies below that expected from the W2 luminosity for a standard AGN, the VLASS non-detection shows no evidence for a radio excess, and the WISE light curve shows no significant mid-infrared variability. These constraints do not exclude a deeply buried, Compton-thick, radio-quiet, or quiescent black hole, but they provide no positive evidence for observable accretion.

\item The coronal-line upper limits are stringent relative to the mid-infrared luminosity. If the W2 luminosity of J1601 were powered by AGN-heated dust, the expected [\ion{Si}{6}], [\ion{Ca}{8}], and [\ion{Ne}{5}] luminosities would be much larger than observed based on standard AGN scaling relations. Reconciling the coronal-line non-detections with these relations would require obscuring columns much larger than those inferred toward the observed optical and near-infrared recombination-line-emitting gas.

\item Photoionization models support the empirical coronal-line constraints. For an accreting black hole luminous enough to power the compact infrared continuum, the models predict high-ionization emission that should be detectable in the observed line-emitting gas. These models include dust, depletion, and distributed cloud geometry, indicating that the weakness of the high-ionization lines cannot be explained by low metallicity or dust depletion alone. Additional metallicity-dependent models show that [\ion{Ne}{5}] emission is not expected to disappear at low metallicity; in fact, reduced dust absorption can enhance coronal-line emission in the metal-poor regime.

\item J1601 and J1201 were selected in a similar way, but they show markedly different nuclear starburst properties. J1601 has a stellar mass approximately an order of magnitude lower than that of the Large Magellanic Cloud and a metallicity of $\sim25\%$ solar, while J1201 is $\sim10^3$ times less massive than the Large Magellanic Cloud and has a metallicity of $\sim10\%$ solar. J1601 exhibits CO bandhead absorption, weak nuclear Wolf--Rayet features, and circumnuclear PAH emission, indicating both recent massive star formation and a contribution from red supergiants. In contrast, J1201 lacks CO bandheads, WR features, and PAH emission, and appears more dust-enshrouded and chemically primitive. The absence of observable accretion in both systems indicates that compact mid-infrared-bright nuclear starbursts can lack AGN signatures across a range of stellar-population, dust, and metallicity conditions.

\item The red mid-infrared colors used to identify these systems originate from compact nuclear emission rather than diffuse, galaxy-wide star formation. JWST shows that the hot dust continuum in J1601 is highly centrally concentrated, yet this compact nuclear source lacks the high-ionization, X-ray, radio, variability, and optical line-ratio signatures expected from an accreting black hole. This demonstrates that compact nuclear starbursts can mimic AGN-like mid-infrared colors in metal-poor dwarf galaxies.

\item More broadly, our results show that commonly used AGN diagnostics, including mid-infrared colors and high-ionization lines, are not by themselves sufficient to identify accreting black holes in metal-poor dwarf galaxies. Known metal-poor [\ion{Ne}{5}] emitters do not necessarily follow standard AGN scaling relations, and the origin of their hard ionizing radiation remains uncertain. Recent JWST results showing high-ionization neon emission from massive stars further reinforce the need to interpret such lines in spatial and multiwavelength context.

\end{enumerate}

These findings refine, rather than invalidate, mid-infrared and high-ionization selection for IMBH searches. Such diagnostics remain valuable for identifying unusual compact nuclear sources and hard radiation fields, but J1201 and J1601 show that compact, metal-poor nuclear starbursts can occupy AGN-like mid-infrared color space without showing observable accretion. Robust identification of accreting black holes in this regime will require spatially resolved spectroscopy together with X-ray, radio, variability, and stellar-population constraints.

\appendix

\section{Correction of Sinusoidal Modulations in NIRSpec IFU Data}
\label{sec:wiggles}

Spectra extracted from JWST/NIRSpec IFU data cubes can exhibit low-level sinusoidal modulations, particularly in individual spaxels or small apertures near bright sources. These features arise from undersampling of the point-spread function (PSF) and are commonly referred to as ``wiggles.'' While these modulations are negligible for spectra extracted over apertures larger than $\sim0\farcs2$, they can affect analyses at the spaxel level and therefore require correction.

To mitigate these effects, we applied a custom spaxel-level correction procedure to the data cube. The goal of this procedure is to model and subtract the sinusoidal component of the continuum while preserving genuine spectral features.

First, prominent emission lines and noise spikes were identified using a custom peak-finding algorithm. To prevent these features from biasing the continuum modeling, regions around each detected peak were masked by excluding three spectral channels on either side. The masked regions were then replaced with a local mean flux computed from neighboring channels. 

Because the spectra often exhibit a rising continuum toward longer wavelengths, we detrended each spectrum by fitting and subtracting a fifth-order Legendre polynomial. This step produces a flattened spectrum in which the sinusoidal modulations can be more robustly identified.

The detrended spectrum was then divided into wavelength segments of width $\sim0.2~\mu$m, and each segment was fit independently with a sinusoidal function of the form
\[
f(\lambda) = A \sin(\omega \lambda - \phi) + c,
\]
where $A$ is the amplitude, $\omega$ is the angular frequency, $\phi$ is the phase, and $c$ is a constant offset. Initial estimates of the dominant frequency were obtained using a discrete Fourier transform, and the fits were performed using the \texttt{curve\_fit} routine from \texttt{scipy.optimize}.

To ensure that the correction does not introduce artificial structure, we applied several quality-control criteria to the fitted models. Fits were rejected if the derived frequency was inconsistent with the dominant Fourier component, if the sinusoidal signal was not clearly present in the data, or if the correction did not significantly reduce the variance of the spectrum. These criteria were designed to avoid overfitting and to ensure that only genuine sinusoidal components were removed.

The resulting sinusoidal models were subtracted from the original spectra on a spaxel-by-spaxel basis. While this procedure does not perfectly remove all artifacts, it provides a consistent and robust method for mitigating spurious oscillations while preserving genuine spectral features. Importantly, all key results presented in this work are insensitive to the application of this correction, as the primary analysis relies on spectra extracted over larger apertures where the effect is minimal.

\subsection{Gemini Results}
\label{subsec:gemini_results}

J1601 was previously observed with Gemini/GNIRS by \citet{Cann2021}, who reported a marginal ($\sim$3$\sigma$) detection of the coronal line [Si\,VI] $\lambda1.96\,\mu$m. This is not confirmed by our JWST/NIRSpec observations, which show no evidence for coronal line emission. To investigate this discrepancy, we re-examined the original Gemini spectrum and compared it to a matched spectral extraction from the JWST data.

\begin{figure*}
    \centering
    \includegraphics[width = 1\textwidth]{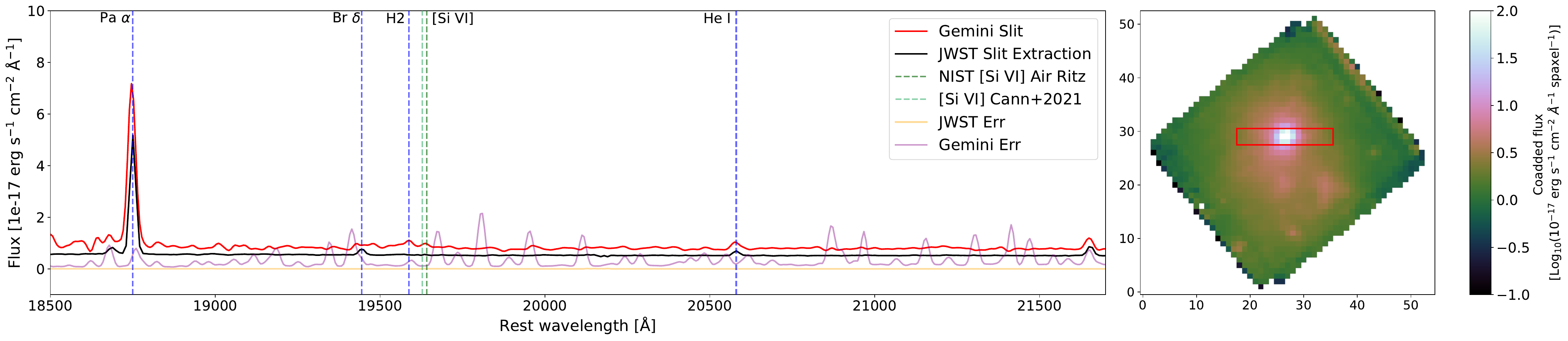}
    \caption{Comparison of the Gemini/GNIRS spectrum (red) with a matched extraction from JWST/NIRSpec (black). The JWST extraction aperture (right panel, red box) is chosen to approximate the $0.3'' \times 1.8''$ Gemini slit. Prominent emission lines are marked with dashed vertical lines. The Gemini spectrum has been convolved to the JWST spectral resolution.}
    \label{fig:jwst_gemini_comparison}
\end{figure*}

The JWST extraction reproduces the overall spectral features seen in the Gemini data, with recombination and He\,I line fluxes consistent within $\sim$2$\sigma$. The Pa$\alpha$ flux shows modest sensitivity to the exact extraction aperture but remains broadly consistent within systematic uncertainties. Crucially, however, no [Si\,VI] emission is detected in the JWST spectrum.

At $1.96\,\mu$m, the feature identified as [Si\,VI] in the Gemini spectrum is not present in the higher signal-to-noise JWST data. While the original detection was based on a careful analysis, the Gemini spectrum exhibits elevated noise in this wavelength region, and no additional coronal lines were detected to support the presence of a hard ionizing source. The absence of [Si\,VI] in JWST, combined with the lack of any other coronal line detections, indicates that the Gemini feature is not robust and is likely associated with the limited signal-to-noise of the data.

Variability is also unlikely to explain the discrepancy. The mid-infrared light curve from {\it ALLWISE} and {\it NEOWISE} shows no significant variability, indicating that there is no evidence for a transient or fading coronal-line source.

Taken together, the JWST data indicate that the previously reported [Si\,VI] detection was not robust, and that J1601 shows no evidence for coronal line emission.

\subsection{Mid-Infrared Variability from NEOWISE}
\label{subsec:variability}
We examined the mid-infrared light curve of J1601 using data from the {\it NEOWISE} mission in the W1 and W2 bands. The light curve spans multiple epochs over several years and shows no significant variability beyond the measurement uncertainties.

The absence of mid-infrared variability argues against the presence of a variable or transient AGN and supports the conclusion that the source is not powered by accretion.

\begin{figure}
    \centering
    \includegraphics[width=\columnwidth]{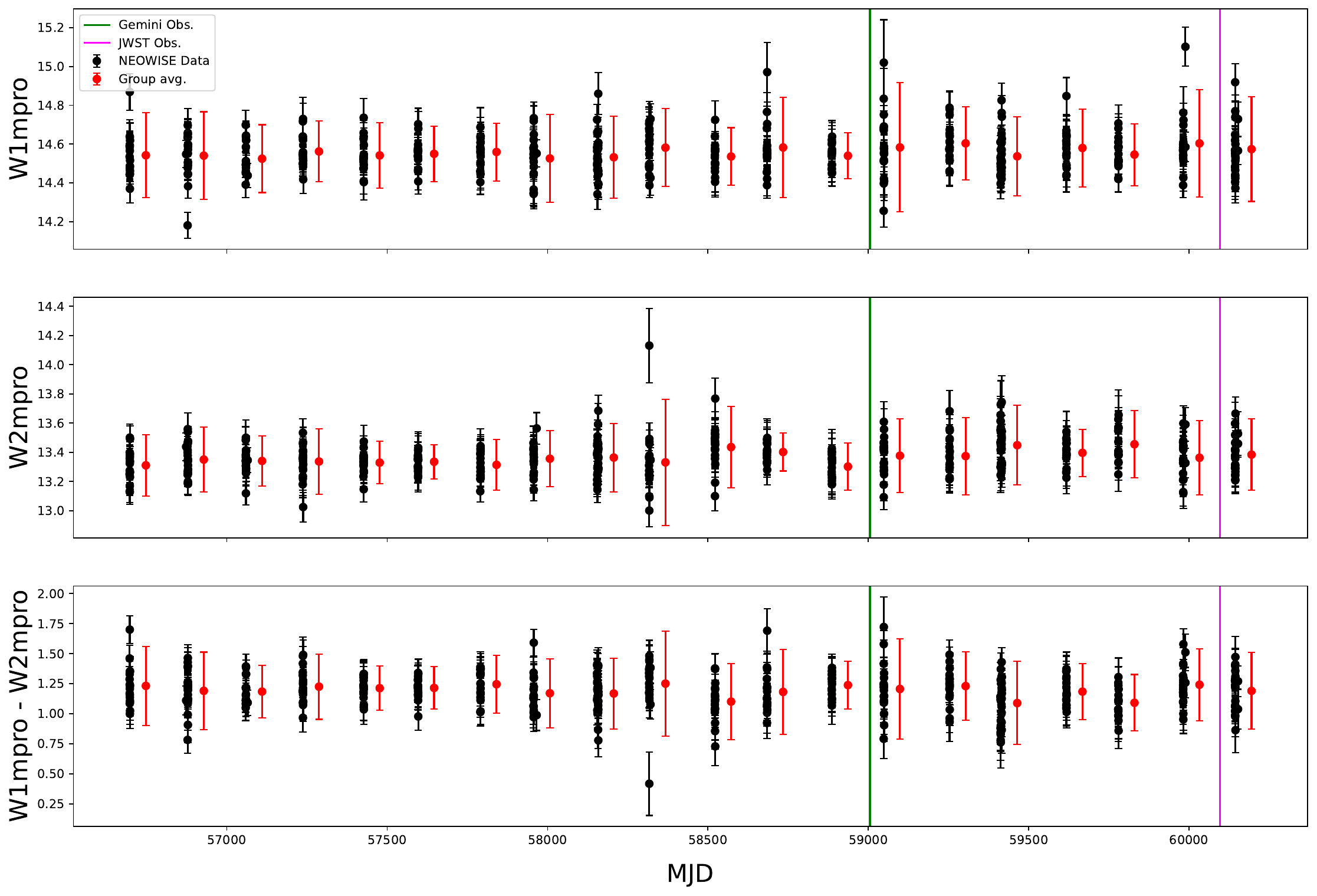}
    \caption{NEOWISE W1 (3.4~$\mu$m) and W2 (4.6~$\mu$m) light curves for J1601. The data span multiple epochs over several years. No significant variability is observed in either band beyond the measurement uncertainties. The absence of mid-infrared variability argues against the presence of a variable or transient AGN and supports the conclusion that the nuclear emission is not powered by accretion.}
    \label{fig:neowise.png}
\end{figure}

\section{Acknowledgments}

The authors are very grateful for the dedicated community who made the observatory a reality, and for the complex pipeline development work over the course of the year carried out by STScI, the ERS teams, and the community at large. 

This work is based on observations made with the NASA/ESA/CSA JWST. The data were obtained from the Mikulski Archive for Space Telescopes at the Space Telescope Science Institute, which is operated by the Association of Universities for Research in Astronomy, Inc., under NASA contract NAS 5-03127 for JWST. STScI is operated by the Association of Universities for Research in Astronomy, Inc., under NASA contract NAS5-26555. Support to MAST for these data is provided by the NASA Office of Space Science via grant NAG5-7584 and by other grants and contracts. These observations are associated with program \#1983. The JWST data used in this paper can be found in doi:10.17909/ct0b-sh28. S. D. and S. S. gratefully acknowledge funding from JWST Cycle 1 grant GO-01983.001-A. This work was supported by the National Science Foundation under Cooperative Agreement 2421782 and the Simons Foundation grant MPS-AI-00010515 awarded to the NSF-Simons AI Institute for Cosmic Origins — CosmicAI (\url{https://www.cosmicai.org/}). J.D.M's research was supported by an appointment to the NASA Postdoctoral Program at the NASA Goddard Space Flight Center, administered by Oak Ridge Associated Universities under contract with NASA.

This publication makes use of data products from the Wide-field Infrared Survey Explorer, which is a joint project of the University of California, Los Angeles, and the Jet Propulsion Laboratory/ California Institute of Technology, funded by the National Aeronautics and Space Administration. This research has made use of the NASA/IPAC Infrared Science Archive, which is funded by the National Aeronautics and Space Administration and operated by the California Institute of Technology. 

This work made use of v2.2.1 of the Binary Population and Spectral Synthesis (BPASS) models as described in Eldridge, Stanway et al. (2017) and Stanway \& Eldridge et al. (2018).

This research made use of Astropy,\footnote{\url{http://www.astropy.org}} a community-developed core Python package for Astronomy \citep{2013A&A...558A..33A}, as well as \textsc{topcat} \citep{2005ASPC..347...29T}. The pipeline processing, spectral fitting, and Cloudy simulations carried out in this work were run on ARGO and HOPPER, research computing clusters provided by the Office of Research Computing at George Mason University, VA. (\url{ http://orc.gmu.edu})

\bibliographystyle{yahapj}
\bibliography{bib}

@ARTICLE{2024ApJ...969L..18A,
       author = {{Ananna}, Tonima Tasnim and {Bogd{\'a}n}, {\'A}kos and {Kov{\'a}cs}, Orsolya E. and {Natarajan}, Priyamvada and {Hickox}, Ryan C.},
        title = "{X-Ray View of Little Red Dots: Do They Host Supermassive Black Holes?}",
      journal = {\apjl},
         year = 2024,
        month = jul,
       volume = {969},
       number = {1},
          eid = {L18},
        pages = {L18},
          doi = {10.3847/2041-8213/ad5669},
archivePrefix = {arXiv},
       eprint = {2404.19010},
 primaryClass = {astro-ph.GA},
       adsurl = {https://ui.adsabs.harvard.edu/abs/2024ApJ...969L..18A}
}

@ARTICLE{2013A&A...558A..33A,
       author = {{Astropy Collaboration} and {Robitaille}, Thomas P. and {Tollerud}, Erik J. and {Greenfield}, Perry and {Droettboom}, Michael and {Bray}, Erik and {Aldcroft}, Tom and {Davis}, Matt and {Ginsburg}, Adam and {Price-Whelan}, Adrian M. and {Kerzendorf}, Wolfgang E. and {Conley}, Alexander and {Crighton}, Neil and {Barbary}, Kyle and {Muna}, Demitri and {Ferguson}, Henry and {Grollier}, Fr{\'e}d{\'e}ric and {Parikh}, Madhura M. and {Nair}, Prasanth H. and {Unther}, Hans M. and {Deil}, Christoph and {Woillez}, Julien and {Conseil}, Simon and {Kramer}, Roban and {Turner}, James E.~H. and {Singer}, Leo and {Fox}, Ryan and {Weaver}, Benjamin A. and {Zabalza}, Victor and {Edwards}, Zachary I. and {Azalee Bostroem}, K. and {Burke}, D.~J. and {Casey}, Andrew R. and {Crawford}, Steven M. and {Dencheva}, Nadia and {Ely}, Justin and {Jenness}, Tim and {Labrie}, Kathleen and {Lim}, Pey Lian and {Pierfederici}, Francesco and {Pontzen}, Andrew and {Ptak}, Andy and {Refsdal}, Brian and {Servillat}, Mathieu and {Streicher}, Ole},
        title = "{Astropy: A community Python package for astronomy}",
      journal = {\aap},
         year = 2013,
        month = oct,
       volume = {558},
          eid = {A33},
        pages = {A33},
          doi = {10.1051/0004-6361/201322068},
archivePrefix = {arXiv},
       eprint = {1307.6212},
 primaryClass = {astro-ph.IM},
       adsurl = {https://ui.adsabs.harvard.edu/abs/2013A&A...558A..33A}
}

@ARTICLE{Bohn2021,
       author = {{Bohn}, Thomas and {Canalizo}, Gabriela and {Veilleux}, Sylvain and {Liu}, Weizhe},
        title = "{Near-infrared Coronal Line Observations of Dwarf Galaxies Hosting AGN-driven Outflows}",
      journal = {\apj},
         year = 2021,
        month = apr,
       volume = {911},
       number = {1},
          eid = {70},
        pages = {70},
          doi = {10.3847/1538-4357/abe70c},
archivePrefix = {arXiv},
       eprint = {2102.08397},
 primaryClass = {astro-ph.GA},
       adsurl = {https://ui.adsabs.harvard.edu/abs/2021ApJ...911...70B}
}

@ARTICLE{2026ApJ...996...48B,
       author = {{Begelman}, Mitchell C. and {Dexter}, Jason},
        title = "{Little Red Dots as Late-stage Quasi-stars}",
      journal = {\apj},
         year = 2026,
        month = jan,
       volume = {996},
       number = {1},
          eid = {48},
        pages = {48},
          doi = {10.3847/1538-4357/ae274a},
archivePrefix = {arXiv},
       eprint = {2507.09085},
 primaryClass = {astro-ph.GA},
       adsurl = {https://ui.adsabs.harvard.edu/abs/2026ApJ...996...48B}
}

@ARTICLE{2018MNRAS.478.3056B,
       author = {{Blecha}, Laura and {Snyder}, Gregory F. and {Satyapal}, Shobita and {Ellison}, Sara L.},
        title = "{The power of infrared AGN selection in mergers: a theoretical study}",
      journal = {\mnras},
         year = 2018,
        month = aug,
       volume = {478},
       number = {3},
        pages = {3056-3071},
          doi = {10.1093/mnras/sty1274},
archivePrefix = {arXiv},
       eprint = {1711.02094},
 primaryClass = {astro-ph.GA},
       adsurl = {https://ui.adsabs.harvard.edu/abs/2018MNRAS.478.3056B}
}

@ARTICLE{2022A&A...661A..82B,
       author = {{B{\"o}ker}, T. and {Arribas}, S. and {L{\"u}tzgendorf}, N. and {Alves de Oliveira}, C. and {Beck}, T.~L. and {Birkmann}, S. and {Bunker}, A.~J. and {Charlot}, S. and {de Marchi}, G. and {Ferruit}, P. and {Giardino}, G. and {Jakobsen}, P. and {Kumari}, N. and {L{\'o}pez-Caniego}, M. and {Maiolino}, R. and {Manjavacas}, E. and {Marston}, A. and {Moseley}, S.~H. and {Muzerolle}, J. and {Ogle}, P. and {Pirzkal}, N. and {Rauscher}, B. and {Rawle}, T. and {Rix}, H. -W. and {Sabbi}, E. and {Sargent}, B. and {Sirianni}, M. and {te Plate}, M. and {Valenti}, J. and {Willott}, C.~J. and {Zeidler}, P.},
        title = "{The Near-Infrared Spectrograph (NIRSpec) on the James Webb Space Telescope. III. Integral-field spectroscopy}",
      journal = {\aap},
         year = 2022,
        month = may,
       volume = {661},
          eid = {A82},
        pages = {A82},
          doi = {10.1051/0004-6361/202142589},
archivePrefix = {arXiv},
       eprint = {2202.03308},
 primaryClass = {astro-ph.IM},
       adsurl = {https://ui.adsabs.harvard.edu/abs/2022A&A...661A..82B}
}

@ARTICLE{2024NatAs...8..126B,
       author = {{Bogd{\'a}n}, {\'A}kos and {Goulding}, Andy D. and {Natarajan}, Priyamvada and {Kov{\'a}cs}, Orsolya E. and {Tremblay}, Grant R. and {Chadayammuri}, Urmila and {Volonteri}, Marta and {Kraft}, Ralph P. and {Forman}, William R. and {Jones}, Christine and {Churazov}, Eugene and {Zhuravleva}, Irina},
        title = "{Evidence for heavy-seed origin of early supermassive black holes from a z {\ensuremath{\approx}} 10 X-ray quasar}",
      journal = {Nature Astronomy},
         year = 2024,
        month = jan,
       volume = {8},
       number = {1},
        pages = {126-133},
          doi = {10.1038/s41550-023-02111-9},
archivePrefix = {arXiv},
       eprint = {2305.15458},
 primaryClass = {astro-ph.GA},
       adsurl = {https://ui.adsabs.harvard.edu/abs/2024NatAs...8..126B}
}

@ARTICLE{2021MNRAS.504..543B,
       author = {{Burke}, Colin J. and {Liu}, Xin and {Chen}, Yu-Ching and {Shen}, Yue and {Guo}, Hengxiao},
        title = "{On the AGN nature of broad balmer emission in four low-redshift metal-poor galaxies}",
      journal = {\mnras},
         year = 2021,
        month = jun,
       volume = {504},
       number = {1},
        pages = {543-550},
          doi = {10.1093/mnras/stab912},
archivePrefix = {arXiv},
       eprint = {2011.10053},
 primaryClass = {astro-ph.GA},
       adsurl = {https://ui.adsabs.harvard.edu/abs/2021MNRAS.504..543B}
}

@ARTICLE{2020ApJ...895..147C,
       author = {{Cann}, Jenna M. and {Satyapal}, Shobita and {Bohn}, Thomas and {Sexton}, Remington O. and {Pfeifle}, Ryan W. and {Manzano-King}, Christina and {Canalizo}, Gabriela and {Rothberg}, Barry and {Gliozzi}, Mario and {Secrest}, Nathan J. and {Blecha}, Laura},
        title = "{Multiwavelength Observations of SDSS J105621.45+313822.1, a Broad-line, Low-metallicity AGN}",
      journal = {\apj},
         year = 2020,
        month = jun,
       volume = {895},
       number = {2},
          eid = {147},
        pages = {147},
          doi = {10.3847/1538-4357/ab8b64},
archivePrefix = {arXiv},
       eprint = {2004.11295},
 primaryClass = {astro-ph.GA},
       adsurl = {https://ui.adsabs.harvard.edu/abs/2020ApJ...895..147C}
}

@ARTICLE{2025ApJ...994..146C,
       author = {{Cleri}, Nikko J. and {Olivier}, Grace M. and {Backhaus}, Bren E. and {Leja}, Joel and {Papovich}, Casey and {Trump}, Jonathan R. and {Arrabal Haro}, Pablo and {Buat}, V{\'e}ronique and {Burgarella}, Denis and {Burnham}, Emilie and {Calabr{\`o}}, Antonello and {Cohn}, Jonathan H. and {Cole}, Justin W. and {Davis}, Kelcey and {Dickinson}, Mark and {Finkelstein}, Steven L. and {Garner}, III, Ray and {Hirschmann}, Michaela and {Hu}, Weida and {Hutchison}, Taylor A. and {Kocevski}, Dale D. and {Koekemoer}, Anton M. and {Larson}, Rebecca L. and {Lewis}, Zach J. and {Maseda}, Michael V. and {Seill{\'e}}, Lise-Marie and {Simons}, Raymond C.},
        title = "{Optical Strong Line Ratios Cannot Distinguish between Stellar Populations and Accreting Black Holes at High Ionization Parameters and Low Metallicities}",
      journal = {\apj},
         year = 2025,
        month = dec,
       volume = {994},
       number = {2},
          eid = {146},
        pages = {146},
          doi = {10.3847/1538-4357/ae0f17},
archivePrefix = {arXiv},
       eprint = {2506.21660},
 primaryClass = {astro-ph.GA},
       adsurl = {https://ui.adsabs.harvard.edu/abs/2025ApJ...994..146C}
}

@ARTICLE{Cann2019,
       author = {{Cann}, Jenna M. and {Satyapal}, Shobita and {Abel}, Nicholas P. and
         {Blecha}, Laura and {Mushotzky}, Richard F. and
         {Reynolds}, Christopher S. and {Secrest}, Nathan J.},
        title = "{The Limitations of Optical Spectroscopic Diagnostics in Identifying Active Galactic Nuclei in the Low-mass Regime}",
      journal = {ApJL},
         year = 2019,
        month = jan,
       volume = {870},
       number = {1},
          eid = {L2},
        pages = {L2},
          doi = {10.3847/2041-8213/aaf88d},
archivePrefix = {arXiv},
       eprint = {1812.06170},
 primaryClass = {astro-ph.GA},
       adsurl = {https://ui.adsabs.harvard.edu/abs/2019ApJ...870L...2C}
}

@ARTICLE{Cann2018,
       author = {{Cann}, Jenna M. and {Satyapal}, Shobita and {Abel}, Nicholas P. and
         {Ricci}, Claudio and {Secrest}, Nathan J. and {Blecha}, Laura and
         {Gliozzi}, Mario},
        title = "{The Hunt for Intermediate-mass Black Holes in the JWST Era}",
      journal = {ApJ},
         year = 2018,
        month = jul,
       volume = {861},
       number = {2},
          eid = {142},
        pages = {142},
          doi = {10.3847/1538-4357/aac64a},
archivePrefix = {arXiv},
       eprint = {1805.09351},
 primaryClass = {astro-ph.GA},
       adsurl = {https://ui.adsabs.harvard.edu/abs/2018ApJ...861..142C}
}

@ARTICLE{Cann2021,
       author = {{Cann}, Jenna M. and {Satyapal}, Shobita and {Rothberg}, Barry and {Canalizo}, Gabriela and {Bohn}, Thomas and {LaMassa}, Stephanie and {Matzko}, William and {Blecha}, Laura and {Secrest}, Nathan J. and {Seth}, Anil and {B{\"o}ker}, Torsten and {Sexton}, Remington O. and {Kamal}, Lara and {Schmitt}, Henrique},
        title = "{Relics of Supermassive Black Hole Seeds: The Discovery of an Accreting Black Hole in an Optically Normal, Low Metallicity Dwarf Galaxy}",
      journal = {\apjl},
         year = 2021,
        month = may,
       volume = {912},
       number = {1},
          eid = {L2},
        pages = {L2},
          doi = {10.3847/2041-8213/abf56d},
archivePrefix = {arXiv},
       eprint = {2104.05689},
 primaryClass = {astro-ph.GA},
       adsurl = {https://ui.adsabs.harvard.edu/abs/2021ApJ...912L...2C}
}

@ARTICLE{2024arXiv240218643C,
       author = {{Chisholm}, J. and {Berg}, D.~A. and {Endsley}, R. and {Gazagnes}, S. and {Richardson}, C.~T. and {Lambrides}, E. and {Greene}, J. and {Finkelstein}, S. and {Flury}, S. and {Guseva}, N.~G. and {Henry}, A. and {Hutchison}, T.~A. and {Izotov}, Y.~I. and {Marques-Chaves}, R. and {Oesch}, P. and {Papovich}, C. and {Saldana-Lopez}, A. and {Schaerer}, D. and {Stephenson}, M.~G.},
        title = "{[Ne v] emission from a faint epoch of reionization-era galaxy: evidence for a narrow-line intermediate mass black hole}",
      journal = {arXiv e-prints},
         year = 2024,
        month = feb,
          eid = {arXiv:2402.18643},
        pages = {arXiv:2402.18643},
          doi = {10.48550/arXiv.2402.18643},
archivePrefix = {arXiv},
       eprint = {2402.18643},
 primaryClass = {astro-ph.GA},
       adsurl = {https://ui.adsabs.harvard.edu/abs/2024arXiv240218643C}
}

@ARTICLE{1991ApJ...378...65C,
       author = {{Condon}, J.~J. and {Huang}, Z.-P. and {Yin}, Q.~F. and {Thuan}, T.~X.},
        title = "{Compact Starbursts in Ultraluminous Infrared Galaxies}",
      journal = {\apj},
         year = 1991,
        month = sep,
       volume = {378},
        pages = {65},
          doi = {10.1086/170407},
       adsurl = {https://ui.adsabs.harvard.edu/abs/1991ApJ...378...65C}
}

@ARTICLE{2007ApJ...660..167D,
       author = {{Donley}, J.~L. and {Rieke}, G.~H. and {P{\'e}rez-Gonz{\'a}lez}, P.~G. and {Rigby}, J.~R. and {Alonso-Herrero}, A.},
        title = "{Spitzer Power-Law Active Galactic Nucleus Candidates in the Chandra Deep Field-North}",
      journal = {\apj},
         year = 2007,
        month = may,
       volume = {660},
       number = {1},
        pages = {167-190},
          doi = {10.1086/512798},
archivePrefix = {arXiv},
       eprint = {astro-ph/0701698},
 primaryClass = {astro-ph},
       adsurl = {https://ui.adsabs.harvard.edu/abs/2007ApJ...660..167D}
}

@ARTICLE{Doan2025,
       author = {{Doan}, Sara and {Satyapal}, Shobita and {Matzko}, William and {Abel}, Nicholas P. and {B{\"o}ker}, Torsten and {Bohn}, Thomas and {Canalizo}, Gabriela and {Cann}, Jenna M. and {Fischer}, Jacqueline and {LaMassa}, Stephanie and {Madden}, Suzanne C. and {McKaig}, Jeffrey D. and {Schaerer}, D. and {Secrest}, Nathan J. and {Seth}, Anil and {Blecha}, Laura and {Molina}, Mallory and {Rothberg}, Barry},
        title = "{Local Analogs of Primordial Galaxies: In Search of Intermediate-mass Black Holes with JWST NIRSpec}",
      journal = {\apj},
         year = 2025,
        month = jul,
       volume = {987},
       number = {1},
          eid = {99},
        pages = {99},
          doi = {10.3847/1538-4357/adcd59},
archivePrefix = {arXiv},
       eprint = {2408.04774},
 primaryClass = {astro-ph.GA},
       adsurl = {https://ui.adsabs.harvard.edu/abs/2025ApJ...987...99D}
}

@ARTICLE{2025A&A...701A.168D,
       author = {{de Graaff}, Anna and {Rix}, Hans-Walter and {Naidu}, Rohan P. and {Labb{\'e}}, Ivo and {Wang}, Bingjie and {Leja}, Joel and {Matthee}, Jorryt and {Katz}, Harley and {Greene}, Jenny E. and {Hviding}, Raphael E. and {Baggen}, Josephine and {Bezanson}, Rachel and {Boogaard}, Leindert A. and {Brammer}, Gabriel and {Dayal}, Pratika and {van Dokkum}, Pieter and {Goulding}, Andy D. and {Hirschmann}, Michaela and {Maseda}, Michael V. and {McConachie}, Ian and {Miller}, Tim B. and {Nelson}, Erica and {Oesch}, Pascal A. and {Setton}, David J. and {Shivaei}, Irene and {Weibel}, Andrea and {Whitaker}, Katherine E. and {Williams}, Christina C.},
        title = "{A remarkable ruby: Absorption in dense gas, rather than evolved stars, drives the extreme Balmer break of a little red dot at z = 3.5}",
      journal = {\aap},
         year = 2025,
        month = sep,
       volume = {701},
          eid = {A168},
        pages = {A168},
          doi = {10.1051/0004-6361/202554681},
archivePrefix = {arXiv},
       eprint = {2503.16600},
 primaryClass = {astro-ph.GA},
       adsurl = {https://ui.adsabs.harvard.edu/abs/2025A&A...701A.168D}
}

@ARTICLE{2013PASP..125..306F,
       author = {{Foreman-Mackey}, Daniel and {Hogg}, David W. and {Lang}, Dustin and {Goodman}, Jonathan},
        title = "{emcee: The MCMC Hammer}",
      journal = {\pasp},
         year = 2013,
        month = mar,
       volume = {125},
       number = {925},
        pages = {306},
          doi = {10.1086/670067},
archivePrefix = {arXiv},
       eprint = {1202.3665},
 primaryClass = {astro-ph.IM},
       adsurl = {https://ui.adsabs.harvard.edu/abs/2013PASP..125..306F}
}

@ARTICLE{2026arXiv260606575G,
       author = {{Gentile}, Fabrizio and {Giavalisco}, Mauro and {Daddi}, Emanuele and {Elbaz}, David and {Billand}, Jean-Baptiste and {Franco}, Maximilen and {Magnelli}, Benjamin and {Barro}, Guillermo and {Cheng}, Yingjie and {Cleri}, Nikko J. and {Davis}, Kelcey and {Delvecchio}, Ivan and {Dickinson}, Mark and {Finkelstein}, Steven L. and {Gandolfi}, Giovanni and {Hirschmann}, Michaela and {Hu}, Weida and {Kocevski}, Dale and {Koekemoer}, Anton M. and {Lucas}, Ray and {Mascia}, Sara and {Napolitano}, Lorenzo and {Papovich}, Casey and {P{\'e}rez-D{\'\i}az}, Borja and {Perez-Gonzalez}, Pablo and {Trump}, Jonathan R. and {Wang}, Xin and {Yung}, L.~Y. Aaron},
        title = "{The quasi-star model for Little Red Dots: potential and challenges}",
      journal = {arXiv e-prints},
         year = 2026,
        month = jun,
          eid = {arXiv:2606.06575},
        pages = {arXiv:2606.06575},
          doi = {10.48550/arXiv.2606.06575},
archivePrefix = {arXiv},
       eprint = {2606.06575},
 primaryClass = {astro-ph.GA},
       adsurl = {https://ui.adsabs.harvard.edu/abs/2026arXiv260606575G}
}

@ARTICLE{2019MNRAS.488.5438G,
       author = {{Graziani}, R. and {Courtois}, H.~M. and {Lavaux}, G. and {Hoffman}, Y. and {Tully}, R.~B. and {Copin}, Y. and {Pomar{\`e}de}, D.},
        title = "{The peculiar velocity field up to z {\ensuremath{\sim}} 0.05 by forward-modelling Cosmicflows-3 data}",
      journal = {\mnras},
         year = 2019,
        month = oct,
       volume = {488},
       number = {4},
        pages = {5438-5451},
          doi = {10.1093/mnras/stz078},
archivePrefix = {arXiv},
       eprint = {1901.01818},
 primaryClass = {astro-ph.CO},
       adsurl = {https://ui.adsabs.harvard.edu/abs/2019MNRAS.488.5438G}
}

@ARTICLE{2020ARA&A..58..257G,
       author = {{Greene}, Jenny E. and {Strader}, Jay and {Ho}, Luis C.},
        title = "{Intermediate-Mass Black Holes}",
      journal = {\araa},
         year = 2020,
        month = aug,
       volume = {58},
        pages = {257-312},
          doi = {10.1146/annurev-astro-032620-021835},
archivePrefix = {arXiv},
       eprint = {1911.09678},
 primaryClass = {astro-ph.GA},
       adsurl = {https://ui.adsabs.harvard.edu/abs/2020ARA&A..58..257G}
}

@ARTICLE{Groves2006,
       author = {{Groves}, Brent A. and {Heckman}, Timothy M. and {Kauffmann}, Guinevere},
        title = "{Emission-line diagnostics of low-metallicity active galactic nuclei}",
      journal = {MNRAS},
         year = 2006,
        month = oct,
       volume = {371},
       number = {4},
        pages = {1559-1569},
          doi = {10.1111/j.1365-2966.2006.10812.x},
archivePrefix = {arXiv},
       eprint = {astro-ph/0607311},
 primaryClass = {astro-ph},
       adsurl = {https://ui.adsabs.harvard.edu/abs/2006MNRAS.371.1559G}
}

@ARTICLE{Greene2020,
       author = {{Greene}, Jenny E. and {Strader}, Jay and {Ho}, Luis C.},
        title = "{Intermediate-Mass Black Holes}",
      journal = {\araa},
         year = 2020,
        month = aug,
       volume = {58},
        pages = {257-312},
          doi = {10.1146/annurev-astro-032620-021835},
archivePrefix = {arXiv},
       eprint = {1911.09678},
 primaryClass = {astro-ph.GA},
       adsurl = {https://ui.adsabs.harvard.edu/abs/2020ARA&A..58..257G}
}

@ARTICLE{2023ApJ...959...39H,
       author = {{Harikane}, Yuichi and {Zhang}, Yechi and {Nakajima}, Kimihiko and {Ouchi}, Masami and {Isobe}, Yuki and {Ono}, Yoshiaki and {Hatano}, Shun and {Xu}, Yi and {Umeda}, Hiroya},
        title = "{A JWST/NIRSpec First Census of Broad-line AGNs at z = 4-7: Detection of 10 Faint AGNs with M $_{BH}$ {}10$^{6}$-{}10$^{8}$ M $_{{\ensuremath{\odot}}}$ and Their Host Galaxy Properties}",
      journal = {\apj},
         year = 2023,
        month = dec,
       volume = {959},
       number = {1},
          eid = {39},
        pages = {39},
          doi = {10.3847/1538-4357/ad029e},
archivePrefix = {arXiv},
       eprint = {2303.11946},
 primaryClass = {astro-ph.GA},
       adsurl = {https://ui.adsabs.harvard.edu/abs/2023ApJ...959...39H}
}

@ARTICLE{2024ApJ...966..170H,
       author = {{Hatano}, Shun and {Ouchi}, Masami and {Umeda}, Hiroya and {Nakajima}, Kimihiko and {Kawaguchi}, Toshihiro and {Isobe}, Yuki and {Aoyama}, Shohei and {Watanabe}, Kuria and {Harikane}, Yuichi and {Kusakabe}, Haruka and {Matsumoto}, Akinori and {Moriya}, Takashi J. and {Nishigaki}, Moka and {Ono}, Yoshiaki and {Onodera}, Masato and {Sugahara}, Yuma and {Suzuki}, Akihiro and {Xu}, Yi and {Zhang}, Yechi},
        title = "{EMPRESS. XIV. Strong High-ionization Lines of Young Galaxies at z = 0{\textendash}8: Ionizing Spectra Consistent with the Intermediate-mass Black Holes with M $_{BH}$ {\ensuremath{\sim}} {}10$^{3}${\textendash}{}10$^{6}$ M $_{{\ensuremath{\odot}}}$}",
      journal = {\apj},
         year = 2024,
        month = may,
       volume = {966},
       number = {2},
          eid = {170},
        pages = {170},
          doi = {10.3847/1538-4357/ad335c},
archivePrefix = {arXiv},
       eprint = {2305.02189},
 primaryClass = {astro-ph.GA},
       adsurl = {https://ui.adsabs.harvard.edu/abs/2024ApJ...966..170H}
}

@ARTICLE{2023arXiv230403726H,
       author = {{Hatano}, Shun and {Ouchi}, Masami and {Nakajima}, Kimihiko and {Kawaguchi}, Toshihiro and {Kokubo}, Mitsuru and {Kikuta}, Satoshi and {Tominaga}, Nozomu and {Xu}, Yi and {Watanabe}, Kuria and {Harikane}, Yuichi and {Isobe}, Yuki and {Matsumoto}, Akinori and {Nishigaki}, Moka and {Ono}, Yoshiaki and {Onodera}, Masato and {Sugahara}, Yuma and {Umeda}, Hiroya and {Zhang}, Yechi},
        title = "{Active Massive Black Hole Found in the Young Star-Forming Dwarf Galaxy SBS 0335-052E}",
      journal = {arXiv e-prints},
         year = 2023,
        month = apr,
          eid = {arXiv:2304.03726},
        pages = {arXiv:2304.03726},
          doi = {10.48550/arXiv.2304.03726},
archivePrefix = {arXiv},
       eprint = {2304.03726},
 primaryClass = {astro-ph.GA},
       adsurl = {https://ui.adsabs.harvard.edu/abs/2023arXiv230403726H}
}

@ARTICLE{2026arXiv260623456H,
       author = {{Hawcroft}, Calum and {Law}, David R. and {Smith}, Linda J. and {Fullerton}, Alexander W. and {Gordon}, Karl D. and {Crowther}, Paul A. and {Decleir}, Marjorie and {Zeegers}, Sascha T. and {Erba}, Christiana and {Ignace}, Richard and {Hillier}, D. John},
        title = "{Stellar winds of O-type stars traced by high ionization fine-structure emission lines with JWST/MIRI}",
      journal = {arXiv e-prints},
         year = 2026,
        month = jun,
          eid = {arXiv:2606.23456},
        pages = {arXiv:2606.23456},
archivePrefix = {arXiv},
       eprint = {2606.23456},
 primaryClass = {astro-ph.SR},
       adsurl = {https://ui.adsabs.harvard.edu/abs/2026arXiv260623456H}
}

@ARTICLE{Hunt2025,
       author = {{Hunt}, L.~K. and {Draine}, B.~T. and {Navarro}, M.~G. and {Aloisi}, A. and {Vaught}, R.~J. Rickards and {Adamo}, A. and {Annibali}, F. and {Calzetti}, D. and {Hernandez}, S. and {James}, B.~L. and {Mingozzi}, M. and {Schneider}, R. and {Tosi}, M. and {Brandl}, B. and {del Valle-Espinosa}, M.~G. and {Donnan}, F. and {Hirschauer}, A.~S. and {Meixner}, M. and {Rigopoulou}, D.},
        title = "{The Interstellar Medium in I Zw 18 Seen with JWST/MIRI. II. Warm Molecular Hydrogen and Warm Dust}",
      journal = {\apj},
         year = 2025,
        month = nov,
       volume = {993},
       number = {1},
          eid = {84},
        pages = {84},
          doi = {10.3847/1538-4357/ae0191},
archivePrefix = {arXiv},
       eprint = {2509.02690},
 primaryClass = {astro-ph.GA},
       adsurl = {https://ui.adsabs.harvard.edu/abs/2025ApJ...993...84H}
}

@ARTICLE{2023RNAAS...7...99H,
       author = {{Herenz}, E.~C. and {Micheva}, G. and {Weilbacher}, P.~M. and {Monreal-Ibero}, A. and {Hayes}, M. and {Anders}, F. and {Rivinius}, T.},
        title = "{On the Recent Discovery of Coronal [Fe X]{\ensuremath{\lambda}}6374 Emission in the Low-metallicity Dwarf Galaxy SDSS J0944-0038}",
      journal = {Research Notes of the American Astronomical Society},
         year = 2023,
        month = may,
       volume = {7},
       number = {5},
          eid = {99},
        pages = {99},
          doi = {10.3847/2515-5172/acd69e},
       adsurl = {https://ui.adsabs.harvard.edu/abs/2023RNAAS...7...99H}
}

@ARTICLE{2007ApJ...671.1297I,
       author = {{Izotov}, Yuri I. and {Thuan}, Trinh X. and {Guseva}, Natalia G.},
        title = "{Broad-Line Emission in Low-Metallicity Blue Compact Dwarf Galaxies: Evidence for Stellar Wind, Supernova, and Possible AGN Activity}",
      journal = {\apj},
         year = 2007,
        month = dec,
       volume = {671},
       number = {2},
        pages = {1297-1320},
          doi = {10.1086/522923},
archivePrefix = {arXiv},
       eprint = {0709.3643},
 primaryClass = {astro-ph},
       adsurl = {https://ui.adsabs.harvard.edu/abs/2007ApJ...671.1297I}
}

@ARTICLE{2008ApJ...687..133I,
       author = {{Izotov}, Yuri I. and {Thuan}, Trinh X.},
        title = "{Active Galactic Nuclei in Four Metal-poor Dwarf Emission-Line Galaxies}",
      journal = {\apj},
         year = 2008,
        month = nov,
       volume = {687},
       number = {1},
        pages = {133-140},
          doi = {10.1086/591660},
archivePrefix = {arXiv},
       eprint = {0807.2029},
 primaryClass = {astro-ph},
       adsurl = {https://ui.adsabs.harvard.edu/abs/2008ApJ...687..133I}
}

@ARTICLE{2012MNRAS.427.1229I,
       author = {{Izotov}, Y.~I. and {Thuan}, T.~X. and {Privon}, G.},
        title = "{The detection of [Ne V] emission in five blue compact dwarf galaxies}",
      journal = {\mnras},
         year = 2012,
        month = dec,
       volume = {427},
       number = {2},
        pages = {1229-1237},
          doi = {10.1111/j.1365-2966.2012.22051.x},
archivePrefix = {arXiv},
       eprint = {1209.5265},
 primaryClass = {astro-ph.CO},
       adsurl = {https://ui.adsabs.harvard.edu/abs/2012MNRAS.427.1229I}
}

@ARTICLE{2021MNRAS.508.2556I,
       author = {{Izotov}, Y.~I. and {Thuan}, T.~X. and {Guseva}, N.~G.},
        title = "{Large binocular telescope observations of new six compact star-forming galaxies with [Ne V] {\ensuremath{\lambda}}3426 {\r{A}} emission}",
      journal = {\mnras},
         year = 2021,
        month = dec,
       volume = {508},
       number = {2},
        pages = {2556-2574},
          doi = {10.1093/mnras/stab2798},
archivePrefix = {arXiv},
       eprint = {2109.12971},
 primaryClass = {astro-ph.GA},
       adsurl = {https://ui.adsabs.harvard.edu/abs/2021MNRAS.508.2556I}
}

@ARTICLE{2022A&A...661A..80J,
       author = {{Jakobsen}, P. and {Ferruit}, P. and {Alves de Oliveira}, C. and {Arribas}, S. and {Bagnasco}, G. and {Barho}, R. and {Beck}, T.~L. and {Birkmann}, S. and {B{\"o}ker}, T. and {Bunker}, A.~J. and {Charlot}, S. and {de Jong}, P. and {de Marchi}, G. and {Ehrenwinkler}, R. and {Falcolini}, M. and {Fels}, R. and {Franx}, M. and {Franz}, D. and {Funke}, M. and {Giardino}, G. and {Gnata}, X. and {Holota}, W. and {Honnen}, K. and {Jensen}, P.~L. and {Jentsch}, M. and {Johnson}, T. and {Jollet}, D. and {Karl}, H. and {Kling}, G. and {K{\"o}hler}, J. and {Kolm}, M. -G. and {Kumari}, N. and {Lander}, M.~E. and {Lemke}, R. and {L{\'o}pez-Caniego}, M. and {L{\"u}tzgendorf}, N. and {Maiolino}, R. and {Manjavacas}, E. and {Marston}, A. and {Maschmann}, M. and {Maurer}, R. and {Messerschmidt}, B. and {Moseley}, S.~H. and {Mosner}, P. and {Mott}, D.~B. and {Muzerolle}, J. and {Pirzkal}, N. and {Pittet}, J. -F. and {Plitzke}, A. and {Posselt}, W. and {Rapp}, B. and {Rauscher}, B.~J. and {Rawle}, T. and {Rix}, H. -W. and {R{\"o}del}, A. and {Rumler}, P. and {Sabbi}, E. and {Salvignol}, J. -C. and {Schmid}, T. and {Sirianni}, M. and {Smith}, C. and {Strada}, P. and {te Plate}, M. and {Valenti}, J. and {Wettemann}, T. and {Wiehe}, T. and {Wiesmayer}, M. and {Willott}, C.~J. and {Wright}, R. and {Zeidler}, P. and {Zincke}, C.},
        title = "{The Near-Infrared Spectrograph (NIRSpec) on the James Webb Space Telescope. I. Overview of the instrument and its capabilities}",
      journal = {\aap},
         year = 2022,
        month = may,
       volume = {661},
          eid = {A80},
        pages = {A80},
          doi = {10.1051/0004-6361/202142663},
archivePrefix = {arXiv},
       eprint = {2202.03305},
 primaryClass = {astro-ph.IM},
       adsurl = {https://ui.adsabs.harvard.edu/abs/2022A&A...661A..80J}
}

@ARTICLE{jarrett2011,
       author = {{Jarrett}, T.~H. and {Cohen}, M. and {Masci}, F. and {Wright}, E. and
         {Stern}, D. and {Benford}, D. and {Blain}, A. and {Carey}, S. and
         {Cutri}, R.~M. and {Eisenhardt}, P. and {Lonsdale}, C. and
         {Mainzer}, A. and {Marsh}, K. and {Padgett}, D. and {Petty}, S. and
         {Ressler}, M. and {Skrutskie}, M. and {Stanford}, S. and {Surace}, J. and
         {Tsai}, C.~W. and {Wheelock}, S. and {Yan}, D.~L.},
        title = "{The Spitzer-WISE Survey of the Ecliptic Poles}",
      journal = {ApJ},
         year = 2011,
        month = jul,
       volume = {735},
       number = {2},
          eid = {112},
        pages = {112},
          doi = {10.1088/0004-637X/735/2/112},
       adsurl = {https://ui.adsabs.harvard.edu/abs/2011ApJ...735..112J}
}

@ARTICLE{2026MNRAS.545f2235J,
       author = {{Ji}, Xihan and {D'Eugenio}, Francesco and {Juod{\v{z}}balis}, Ignas and {Walton}, Dominic J. and {Fabian}, Andrew C. and {Maiolino}, Roberto and {Ramos Almeida}, Cristina and {Acosta Pulido}, Jose A. and {Belokurov}, Vasily A. and {Isobe}, Yuki and {Jones}, Gareth and {Maraston}, Claudia and {Scholtz}, Jan and {Simmonds}, Charlotte and {Tacchella}, Sandro and {Terlevich}, Elena and {Terlevich}, Roberto},
        title = "{Lord of LRDs: insights into a 'Little Red Dot' with a low-ionization spectrum at z = 0.1}",
      journal = {\mnras},
         year = 2026,
        month = jan,
       volume = {545},
       number = {3},
          eid = {staf2235},
        pages = {staf2235},
          doi = {10.1093/mnras/staf2235},
archivePrefix = {arXiv},
       eprint = {2507.23774},
 primaryClass = {astro-ph.GA},
       adsurl = {https://ui.adsabs.harvard.edu/abs/2026MNRAS.545f2235J}
}

@ARTICLE{2019ApJ...885...96J,
       author = {{Jaskot}, Anne E. and {Dowd}, Tara and {Oey}, M.~S. and {Scarlata}, Claudia and {McKinney}, Jed},
        title = "{New Insights on Ly{\ensuremath{\alpha}} and Lyman Continuum Radiative Transfer in the Greenest Peas}",
      journal = {\apj},
         year = 2019,
        month = nov,
       volume = {885},
       number = {1},
          eid = {96},
        pages = {96},
          doi = {10.3847/1538-4357/ab3d3b},
archivePrefix = {arXiv},
       eprint = {1908.09763},
 primaryClass = {astro-ph.GA},
       adsurl = {https://ui.adsabs.harvard.edu/abs/2019ApJ...885...96J}
}

@ARTICLE{2003MNRAS.346.1055K,
       author = {{Kauffmann}, Guinevere and {Heckman}, Timothy M. and {Tremonti}, Christy and {Brinchmann}, Jarle and {Charlot}, St{\'e}phane and {White}, Simon D.~M. and {Ridgway}, Susan E. and {Brinkmann}, Jon and {Fukugita}, Masataka and {Hall}, Patrick B. and {Ivezi{\'c}}, {\v{Z}}eljko and {Richards}, Gordon T. and {Schneider}, Donald P.},
        title = "{The host galaxies of active galactic nuclei}",
      journal = {\mnras},
         year = 2003,
        month = dec,
       volume = {346},
       number = {4},
        pages = {1055-1077},
          doi = {10.1111/j.1365-2966.2003.07154.x},
archivePrefix = {arXiv},
       eprint = {astro-ph/0304239},
 primaryClass = {astro-ph},
       adsurl = {https://ui.adsabs.harvard.edu/abs/2003MNRAS.346.1055K}
}

@ARTICLE{2001ApJ...556..121K,
       author = {{Kewley}, L.~J. and {Dopita}, M.~A. and {Sutherland}, R.~S. and {Heisler}, C.~A. and {Trevena}, J.},
        title = "{Theoretical Modeling of Starburst Galaxies}",
      journal = {\apj},
         year = 2001,
        month = jul,
       volume = {556},
       number = {1},
        pages = {121-140},
          doi = {10.1086/321545},
archivePrefix = {arXiv},
       eprint = {astro-ph/0106324},
 primaryClass = {astro-ph},
       adsurl = {https://ui.adsabs.harvard.edu/abs/2001ApJ...556..121K}
}

@ARTICLE{2020AJ....159...67K,
       author = {{Kourkchi}, Ehsan and {Courtois}, H{\'e}l{\`e}ne M. and {Graziani}, Romain and {Hoffman}, Yehuda and {Pomar{\`e}de}, Daniel and {Shaya}, Edward J. and {Tully}, R. Brent},
        title = "{Cosmicflows-3: Two Distance-Velocity Calculators}",
      journal = {\aj},
         year = 2020,
        month = feb,
       volume = {159},
       number = {2},
          eid = {67},
        pages = {67},
          doi = {10.3847/1538-3881/ab620e},
archivePrefix = {arXiv},
       eprint = {1912.07214},
 primaryClass = {astro-ph.CO},
       adsurl = {https://ui.adsabs.harvard.edu/abs/2020AJ....159...67K}
}

@ARTICLE{2020PASP..132c5001L,
       author = {{Lacy}, M. and {Baum}, S.~A. and {Chandler}, C.~J. and {Chatterjee}, S. and {Clarke}, T.~E. and {Deustua}, S. and {English}, J. and {Farnes}, J. and {Gaensler}, B.~M. and {Gugliucci}, N. and {Hallinan}, G. and {Kent}, B.~R. and {Kimball}, A. and {Law}, C.~J. and {Lazio}, T.~J.~W. and {Marvil}, J. and {Mao}, S.~A. and {Medlin}, D. and {Mooley}, K. and {Murphy}, E.~J. and {Myers}, S. and {Osten}, R. and {Richards}, G.~T. and {Rosolowsky}, E. and {Rudnick}, L. and {Schinzel}, F. and {Sivakoff}, G.~R. and {Sjouwerman}, L.~O. and {Taylor}, R. and {White}, R.~L. and {Wrobel}, J. and {Andernach}, H. and {Beasley}, A.~J. and {Berger}, E. and {Bhatnager}, S. and {Birkinshaw}, M. and {Bower}, G.~C. and {Brandt}, W.~N. and {Brown}, S. and {Burke-Spolaor}, S. and {Butler}, B.~J. and {Comerford}, J. and {Demorest}, P.~B. and {Fu}, H. and {Giacintucci}, S. and {Golap}, K. and {G{\"u}th}, T. and {Hales}, C.~A. and {Hiriart}, R. and {Hodge}, J. and {Horesh}, A. and {Ivezi{\'c}}, {\v{Z}}. and {Jarvis}, M.~J. and {Kamble}, A. and {Kassim}, N. and {Liu}, X. and {Loinard}, L. and {Lyons}, D.~K. and {Masters}, J. and {Mezcua}, M. and {Moellenbrock}, G.~A. and {Mroczkowski}, T. and {Nyland}, K. and {O'Dea}, C.~P. and {O'Sullivan}, S.~P. and {Peters}, W.~M. and {Radford}, K. and {Rao}, U. and {Robnett}, J. and {Salcido}, J. and {Shen}, Y. and {Sobotka}, A. and {Witz}, S. and {Vaccari}, M. and {van Weeren}, R.~J. and {Vargas}, A. and {Williams}, P.~K.~G. and {Yoon}, I.},
        title = "{The Karl G. Jansky Very Large Array Sky Survey (VLASS). Science Case and Survey Design}",
      journal = {\pasp},
         year = 2020,
        month = mar,
       volume = {132},
       number = {1009},
          eid = {035001},
        pages = {035001},
          doi = {10.1088/1538-3873/ab63eb},
archivePrefix = {arXiv},
       eprint = {1907.01981},
 primaryClass = {astro-ph.IM},
       adsurl = {https://ui.adsabs.harvard.edu/abs/2020PASP..132c5001L}
}

@ARTICLE{2017MNRAS.467..540L,
       author = {{Lamperti}, Isabella and {Koss}, Michael and {Trakhtenbrot}, Benny and {Schawinski}, Kevin and {Ricci}, Claudio and {Oh}, Kyuseok and {Landt}, Hermine and {Riffel}, Rog{\'e}rio and {Rodr{\'\i}guez-Ardila}, Alberto and {Gehrels}, Neil and {Harrison}, Fiona and {Masetti}, Nicola and {Mushotzky}, Richard and {Treister}, Ezequiel and {Ueda}, Yoshihiro and {Veilleux}, Sylvain},
        title = "{BAT AGN Spectroscopic Survey - IV: Near-Infrared Coronal Lines, Hidden Broad Lines, and Correlation with Hard X-ray Emission}",
      journal = {\mnras},
         year = 2017,
        month = may,
       volume = {467},
       number = {1},
        pages = {540-572},
          doi = {10.1093/mnras/stx055},
archivePrefix = {arXiv},
       eprint = {1701.02755},
 primaryClass = {astro-ph.GA},
       adsurl = {https://ui.adsabs.harvard.edu/abs/2017MNRAS.467..540L}
}

@ARTICLE{Latimer2021,
       author = {{Latimer}, Lilikoi J. and {Reines}, Amy E. and {Hainline}, Kevin N. and {Greene}, Jenny E. and {Stern}, Daniel},
        title = "{A Chandra and HST View of WISE-selected AGN Candidates in Dwarf Galaxies}",
      journal = {\apj},
         year = 2021,
        month = jun,
       volume = {914},
       number = {2},
          eid = {133},
        pages = {133},
          doi = {10.3847/1538-4357/abfe0c},
archivePrefix = {arXiv},
       eprint = {2105.05876},
 primaryClass = {astro-ph.GA},
       adsurl = {https://ui.adsabs.harvard.edu/abs/2021ApJ...914..133L}
}

@ARTICLE{2024ApJ...976L..25L,
       author = {{Law}, David R. and {Hawcroft}, Calum and {Smith}, Linda J. and {Fullerton}, Alexander W. and {Evans}, Christopher J. and {Gordon}, Karl D. and {Kumari}, Nimisha and {Leitherer}, Claus},
        title = "{JWST/MIRI Detection of [Ne V], [Ne VI], and [O IV] Wind Emission in the O9 V Star 10 Lacertae}",
      journal = {\apjl},
         year = 2024,
        month = dec,
       volume = {976},
       number = {2},
          eid = {L25},
        pages = {L25},
          doi = {10.3847/2041-8213/ad91a6},
archivePrefix = {arXiv},
       eprint = {2410.05469},
 primaryClass = {astro-ph.SR},
       adsurl = {https://ui.adsabs.harvard.edu/abs/2024ApJ...976L..25L}
}

@ARTICLE{2001ApJ...560..630L,
       author = {{Legrand}, Francois and {Tenorio-Tagle}, Guillermo and {Silich}, Sergey and {Kunth}, Daniel and {Cervi{\~n}o}, Miguel},
        title = "{On the Metallicity of Star-forming Dwarf Galaxies}",
      journal = {\apj},
         year = 2001,
        month = oct,
       volume = {560},
       number = {2},
        pages = {630-635},
          doi = {10.1086/322960},
archivePrefix = {arXiv},
       eprint = {astro-ph/0106431},
 primaryClass = {astro-ph},
       adsurl = {https://ui.adsabs.harvard.edu/abs/2001ApJ...560..630L}
}

@ARTICLE{2026arXiv260521574L,
       author = {{Lin}, Xiaojing and {Fan}, Xiaohui and {Cai}, Zheng and {Liu}, Yichen and {Sun}, Fengwu and {Bian}, Fuyan and {Li}, Mingyu and {Mao}, Junjie and {Greene}, Jenny E. and {Liu}, Hanpu and {Li}, Jiaxuan and {Liu}, Weizhe and {Ma}, Yilun and {Sun}, Zechang and {Zhang}, Zijian},
        title = "{(LRDs)$^2$: The Low-ReDshift Little Red Dots Survey. II. DESI DR1 Sample}",
      journal = {arXiv e-prints},
         year = 2026,
        month = may,
          eid = {arXiv:2605.21574},
        pages = {arXiv:2605.21574},
          doi = {10.48550/arXiv.2605.21574},
archivePrefix = {arXiv},
       eprint = {2605.21574},
 primaryClass = {astro-ph.GA},
       adsurl = {https://ui.adsabs.harvard.edu/abs/2026arXiv260521574L}
}

@ARTICLE{2019NatAs...3....6M,
       author = {{Mezcua}, Mar},
        title = "{Dwarf galaxies might not be the birth sites of supermassive black holes}",
      journal = {Nature Astronomy},
         year = 2019,
        month = jan,
       volume = {3},
        pages = {6-7},
          doi = {10.1038/s41550-018-0662-2},
archivePrefix = {arXiv},
       eprint = {1901.02895},
 primaryClass = {astro-ph.GA},
       adsurl = {https://ui.adsabs.harvard.edu/abs/2019NatAs...3....6M}
}

@ARTICLE{2018ApJ...864...93M,
       author = {{Morrissey}, Patrick and {Matuszewski}, Matuesz and {Martin}, D. Christopher and {Neill}, James D. and {Epps}, Harland and {Fucik}, Jason and {Weber}, Bob and {Darvish}, Behnam and {Adkins}, Sean and {Allen}, Steve and {Bartos}, Randy and {Belicki}, Justin and {Cabak}, Jerry and {Callahan}, Shawn and {Cowley}, Dave and {Crabill}, Marty and {Deich}, Willian and {Delecroix}, Alex and {Doppman}, Greg and {Hilyard}, David and {James}, Ean and {Kaye}, Steve and {Kokorowski}, Michael and {Kwok}, Shui and {Lanclos}, Kyle and {Milner}, Steve and {Moore}, Anna and {O'Sullivan}, Donal and {Parihar}, Prachi and {Park}, Sam and {Phillips}, Andrew and {Rizzi}, Luca and {Rockosi}, Constance and {Rodriguez}, Hector and {Salaun}, Yves and {Seaman}, Kirk and {Sheikh}, David and {Weiss}, Jason and {Zarzaca}, Ray},
        title = "{The Keck Cosmic Web Imager Integral Field Spectrograph}",
      journal = {\apj},
         year = 2018,
        month = sep,
       volume = {864},
       number = {1},
          eid = {93},
        pages = {93},
          doi = {10.3847/1538-4357/aad597},
archivePrefix = {arXiv},
       eprint = {1807.10356},
 primaryClass = {astro-ph.IM},
       adsurl = {https://ui.adsabs.harvard.edu/abs/2018ApJ...864...93M}
}

@ARTICLE{2024A&A...691A.145M,
       author = {{Maiolino}, Roberto and {Scholtz}, Jan and {Curtis-Lake}, Emma and {Carniani}, Stefano and {Baker}, William and {de Graaff}, Anna and {Tacchella}, Sandro and {{\"U}bler}, Hannah and {D'Eugenio}, Francesco and {Witstok}, Joris and {Curti}, Mirko and {Arribas}, Santiago and {Bunker}, Andrew J. and {Charlot}, St{\'e}phane and {Chevallard}, Jacopo and {Eisenstein}, Daniel J. and {Egami}, Eiichi and {Ji}, Zhiyuan and {Jones}, Gareth C. and {Lyu}, Jianwei and {Rawle}, Tim and {Robertson}, Brant and {Rujopakarn}, Wiphu and {Perna}, Michele and {Sun}, Fengwu and {Venturi}, Giacomo and {Williams}, Christina C. and {Willott}, Chris},
        title = "{JADES: The diverse population of infant black holes at 4 < z < 11: Merging, tiny, poor, but mighty}",
      journal = {\aap},
         year = 2024,
        month = nov,
       volume = {691},
          eid = {A145},
        pages = {A145},
          doi = {10.1051/0004-6361/202347640},
archivePrefix = {arXiv},
       eprint = {2308.01230},
 primaryClass = {astro-ph.GA},
       adsurl = {https://ui.adsabs.harvard.edu/abs/2024A&A...691A.145M}
}

@ARTICLE{2025MNRAS.538.1921M,
       author = {{Maiolino}, Roberto and {Risaliti}, Guido and {Signorini}, Matilde and {Trefoloni}, Bartolomeo and {Juod{\v{z}}balis}, Ignas and {Scholtz}, Jan and {{\"U}bler}, Hannah and {D'Eugenio}, Francesco and {Carniani}, Stefano and {Fabian}, Andy and {Ji}, Xihan and {Mazzolari}, Giovanni and {Bertola}, Elena and {Brusa}, Marcella and {Bunker}, Andrew J. and {Charlot}, Stephane and {Comastri}, Andrea and {Cresci}, Giovanni and {DeCoursey}, Christa Noel and {Egami}, Eiichi and {Fiore}, Fabrizio and {Gilli}, Roberto and {Perna}, Michele and {Tacchella}, Sandro and {Venturi}, Giacomo},
        title = "{JWST meets Chandra: a large population of Compton thick, feedback-free, and intrinsically X-ray weak AGN, with a sprinkle of SNe}",
      journal = {\mnras},
         year = 2025,
        month = apr,
       volume = {538},
       number = {3},
        pages = {1921-1943},
          doi = {10.1093/mnras/staf359},
archivePrefix = {arXiv},
       eprint = {2405.00504},
 primaryClass = {astro-ph.GA},
       adsurl = {https://ui.adsabs.harvard.edu/abs/2025MNRAS.538.1921M}
}

@ARTICLE{2024ApJ...963..129M,
       author = {{Matthee}, Jorryt and {Naidu}, Rohan P. and {Brammer}, Gabriel and {Chisholm}, John and {Eilers}, Anna-Christina and {Goulding}, Andy and {Greene}, Jenny and {Kashino}, Daichi and {Labbe}, Ivo and {Lilly}, Simon J. and {Mackenzie}, Ruari and {Oesch}, Pascal A. and {Weibel}, Andrea and {Wuyts}, Stijn and {Xiao}, Mengyuan and {Bordoloi}, Rongmon and {Bouwens}, Rychard and {van Dokkum}, Pieter and {Illingworth}, Garth and {Kramarenko}, Ivan and {Maseda}, Michael V. and {Mason}, Charlotte and {Meyer}, Romain A. and {Nelson}, Erica J. and {Reddy}, Naveen A. and {Shivaei}, Irene and {Simcoe}, Robert A. and {Yue}, Minghao},
        title = "{Little Red Dots: An Abundant Population of Faint Active Galactic Nuclei at z {\ensuremath{\sim}} 5 Revealed by the EIGER and FRESCO JWST Surveys}",
      journal = {\apj},
         year = 2024,
        month = mar,
       volume = {963},
       number = {2},
          eid = {129},
        pages = {129},
          doi = {10.3847/1538-4357/ad2345},
archivePrefix = {arXiv},
       eprint = {2306.05448},
 primaryClass = {astro-ph.GA},
       adsurl = {https://ui.adsabs.harvard.edu/abs/2024ApJ...963..129M}
}

@ARTICLE{2013MNRAS.434..941M,
       author = {{Mateos}, S. and {Alonso-Herrero}, A. and {Carrera}, F.~J. and {Blain}, A. and {Severgnini}, P. and {Caccianiga}, A. and {Ruiz}, A.},
        title = "{Uncovering obscured luminous AGN with WISE}",
      journal = {\mnras},
         year = 2013,
        month = sep,
       volume = {434},
       number = {2},
        pages = {941-955},
          doi = {10.1093/mnras/stt953},
archivePrefix = {arXiv},
       eprint = {1305.7237},
 primaryClass = {astro-ph.CO},
       adsurl = {https://ui.adsabs.harvard.edu/abs/2013MNRAS.434..941M}
}

@ARTICLE{Mazzolari2025,
       author = {{Mazzolari}, Giovanni and {Scholtz}, Jan and {Maiolino}, Roberto and {Gilli}, Roberto and {Traina}, Alberto and {L{\'o}pez}, Ivan E. and {{\"U}bler}, Hannah and {Trefoloni}, Bartolomeo and {D'Eugenio}, Francesco and {Ji}, Xihan and {Mignoli}, Marco and {Vito}, Fabio and {Vignali}, Cristian and {Brusa}, Marcella},
        title = "{Narrow-line AGN selection in CEERS: Spectroscopic selection, physical properties, and X-ray and radio analysis}",
      journal = {\aap},
         year = 2025,
        month = aug,
       volume = {700},
          eid = {A12},
        pages = {A12},
          doi = {10.1051/0004-6361/202451860},
archivePrefix = {arXiv},
       eprint = {2408.15615},
 primaryClass = {astro-ph.GA},
       adsurl = {https://ui.adsabs.harvard.edu/abs/2025A&A...700A..12M}
}

@ARTICLE{2021ApJ...922..155M,
       author = {{Molina}, Mallory and {Reines}, Amy E. and {Latimer}, Lilikoi J. and {Baldassare}, Vivienne and {Salehirad}, Sheyda},
        title = "{A Sample of Massive Black Holes in Dwarf Galaxies Detected via [Fe X] Coronal Line Emission: Active Galactic Nuclei and/or Tidal Disruption Events}",
      journal = {\apj},
         year = 2021,
        month = dec,
       volume = {922},
       number = {2},
          eid = {155},
        pages = {155},
          doi = {10.3847/1538-4357/ac1ffa},
archivePrefix = {arXiv},
       eprint = {2108.09307},
 primaryClass = {astro-ph.GA},
       adsurl = {https://ui.adsabs.harvard.edu/abs/2021ApJ...922..155M}
}

@ARTICLE{2025A&A...693A..50N,
       author = {{Napolitano}, L. and {Castellano}, M. and {Pentericci}, L. and {Arrabal Haro}, P. and {Fontana}, A. and {Treu}, T. and {Bergamini}, P. and {Calabr{\`o}}, A. and {Mascia}, S. and {Morishita}, T. and {Roberts-Borsani}, G. and {Santini}, P. and {Vanzella}, E. and {Vulcani}, B. and {Zakharova}, D. and {Bakx}, T. and {Dickinson}, M. and {Grillo}, C. and {Leethochawalit}, N. and {Llerena}, M. and {Merlin}, E. and {Paris}, D. and {Rojas-Ruiz}, S. and {Rosati}, P. and {Wang}, X. and {Yoon}, I. and {Zavala}, J.},
        title = "{Seven wonders of Cosmic Dawn: JWST confirms a high abundance of galaxies and AGN at z ≃ 9{\textendash}11 in the GLASS field}",
      journal = {\aap},
         year = 2025,
        month = jan,
       volume = {693},
          eid = {A50},
        pages = {A50},
          doi = {10.1051/0004-6361/202452090},
archivePrefix = {arXiv},
       eprint = {2410.10967},
 primaryClass = {astro-ph.GA},
       adsurl = {https://ui.adsabs.harvard.edu/abs/2025A&A...693A..50N}
}

@ARTICLE{2015ApJ...798...38S,
       author = {{Secrest}, N.~J. and {Satyapal}, S. and {Gliozzi}, M. and {Rothberg}, B. and {Ellison}, S.~L. and {Mowry}, W.~S. and {Rosenberg}, J.~L. and {Fischer}, J. and {Schmitt}, H.},
        title = "{An Optically Obscured AGN in a Low Mass, Irregular Dwarf Galaxy: A Multi-Wavelength Analysis of J1329+3234}",
      journal = {\apj},
         year = 2015,
        month = jan,
       volume = {798},
       number = {1},
          eid = {38},
        pages = {38},
          doi = {10.1088/0004-637X/798/1/38},
archivePrefix = {arXiv},
       eprint = {1412.6709},
 primaryClass = {astro-ph.HE},
       adsurl = {https://ui.adsabs.harvard.edu/abs/2015ApJ...798...38S}
}

@ARTICLE{2026arXiv260514233P,
       author = {{Park}, Kevin and {Torralba}, Alberto and {Matthee}, Jorryt and {Mascia}, Sara and {Haiman}, Zolt{\'a}n and {Naidu}, Rohan P. and {de Graaff}, Anna},
        title = "{A new sample of Little Red Dots at $z<0.45$ in DESI DR1: Broad Balmer lines, low ionization spectrum and no variability}",
      journal = {arXiv e-prints},
         year = 2026,
        month = may,
          eid = {arXiv:2605.14233},
        pages = {arXiv:2605.14233},
          doi = {10.48550/arXiv.2605.14233},
archivePrefix = {arXiv},
       eprint = {2605.14233},
 primaryClass = {astro-ph.GA},
       adsurl = {https://ui.adsabs.harvard.edu/abs/2026arXiv260514233P}
}

@ARTICLE{Pucha2025,
       author = {{Pucha}, Ragadeepika and {Juneau}, S. and {Dey}, Arjun and {Siudek}, M. and {Mezcua}, M. and {Moustakas}, J. and {BenZvi}, S. and {Hainline}, K. and {Hviding}, R. and {Mao}, Yao-Yuan and {Alexander}, D.~M. and {Alfarsy}, R. and {Circosta}, C. and {Guo}, Wei-Jian and {Manwadkar}, V. and {Martini}, P. and {Weaver}, B.~A. and {Aguilar}, J. and {Ahlen}, S. and {Bianchi}, D. and {Brooks}, D. and {Canning}, R. and {Claybaugh}, T. and {Dawson}, K. and {de la Macorra}, A. and {Dey}, Biprateep and {Doel}, P. and {Font-Ribera}, A. and {Forero-Romero}, J.~E. and {Gazta{\~n}aga}, E. and {Gontcho A Gontcho}, S. and {Gutierrez}, G. and {Honscheid}, K. and {Kehoe}, R. and {Koposov}, S.~E. and {Lambert}, A. and {Landriau}, M. and {Le Guillou}, L. and {Meisner}, A. and {Miquel}, R. and {Prada}, F. and {Rossi}, G. and {Sanchez}, E. and {Schlegel}, D. and {Schubnell}, M. and {Seo}, H. and {Sprayberry}, D. and {Tarl{\'e}}, G. and {Zou}, H.},
        title = "{Tripling the Census of Dwarf AGN Candidates Using DESI Early Data}",
      journal = {\apj},
         year = 2025,
        month = mar,
       volume = {982},
       number = {1},
          eid = {10},
        pages = {10},
          doi = {10.3847/1538-4357/adb1dd},
archivePrefix = {arXiv},
       eprint = {2411.00091},
 primaryClass = {astro-ph.GA},
       adsurl = {https://ui.adsabs.harvard.edu/abs/2025ApJ...982...10P}
}

@ARTICLE{2022ApJ...936..140R,
       author = {{Reefe}, Michael and {Satyapal}, Shobita and {Sexton}, Remington O. and {Doan}, Sara M. and {Secrest}, Nathan J. and {Cann}, Jenna M.},
        title = "{CLASS: Coronal Line Activity Spectroscopic Survey}",
      journal = {\apj},
         year = 2022,
        month = sep,
       volume = {936},
       number = {2},
          eid = {140},
        pages = {140},
          doi = {10.3847/1538-4357/ac8981},
archivePrefix = {arXiv},
       eprint = {2208.10532},
 primaryClass = {astro-ph.GA},
       adsurl = {https://ui.adsabs.harvard.edu/abs/2022ApJ...936..140R}
}

@ARTICLE{2013ApJ...775..116R,
       author = {{Reines}, Amy E. and {Greene}, Jenny E. and {Geha}, Marla},
        title = "{Dwarf Galaxies with Optical Signatures of Active Massive Black Holes}",
      journal = {\apj},
         year = 2013,
        month = oct,
       volume = {775},
       number = {2},
          eid = {116},
        pages = {116},
          doi = {10.1088/0004-637X/775/2/116},
archivePrefix = {arXiv},
       eprint = {1308.0328},
 primaryClass = {astro-ph.CO},
       adsurl = {https://ui.adsabs.harvard.edu/abs/2013ApJ...775..116R}
}

@ARTICLE{2022ApJ...933..160S,
       author = {{Sargent}, Andrew J. and {Johnson}, Megan C. and {Reines}, Amy E. and {Secrest}, Nathan J. and {van der Horst}, Alexander J. and {Cigan}, Phil J. and {Darling}, Jeremy and {Greene}, Jenny E.},
        title = "{Wandering Black Hole Candidates in Dwarf Galaxies at VLBI Resolution}",
      journal = {\apj},
         year = 2022,
        month = jul,
       volume = {933},
       number = {2},
          eid = {160},
        pages = {160},
          doi = {10.3847/1538-4357/ac74be},
archivePrefix = {arXiv},
       eprint = {2205.16006},
 primaryClass = {astro-ph.GA},
       adsurl = {https://ui.adsabs.harvard.edu/abs/2022ApJ...933..160S}
}

@ARTICLE{satyapal2018,
       author = {{Satyapal}, Shobita and {Abel}, Nicholas P. and {Secrest}, Nathan J.},
        title = "{Star-forming Galaxies as AGN Imposters? A Theoretical Investigation of the Mid-infrared Colors of AGNs and Extreme Starbursts}",
      journal = {ApJ},
         year = 2018,
        month = may,
       volume = {858},
       number = {1},
          eid = {38},
        pages = {38},
          doi = {10.3847/1538-4357/aab7f8},
archivePrefix = {arXiv},
       eprint = {1803.05980},
 primaryClass = {astro-ph.GA},
       adsurl = {https://ui.adsabs.harvard.edu/abs/2018ApJ...858...38S}
}

@article{sexton_2020,
   title={Bayesian AGN Decomposition Analysis for SDSS spectra: a correlation analysis of [O iii] λ5007 outflow kinematics with AGN and host galaxy properties},
   volume={500},
   ISSN={1365-2966},
   url={http://dx.doi.org/10.1093/mnras/staa3278},
   DOI={10.1093/mnras/staa3278},
   number={3},
   journal={Monthly Notices of the Royal Astronomical Society},
   publisher={Oxford University Press (OUP)},
   author={Sexton, Remington O and Matzko, William and Darden, Nicholas and Canalizo, Gabriela and Gorjian, Varoujan},
   year={2020},
   month={Oct},
   pages={2871–2895} }

@ARTICLE{2012MNRAS.421.1043S,
       author = {{Shirazi}, Maryam and {Brinchmann}, Jarle},
        title = "{Strongly star forming galaxies in the local Universe with nebular He II{\ensuremath{\lambda}}4686 emission}",
      journal = {\mnras},
         year = 2012,
        month = apr,
       volume = {421},
       number = {2},
        pages = {1043-1063},
          doi = {10.1111/j.1365-2966.2012.20439.x},
archivePrefix = {arXiv},
       eprint = {1201.1290},
 primaryClass = {astro-ph.CO},
       adsurl = {https://ui.adsabs.harvard.edu/abs/2012MNRAS.421.1043S}
}

@ARTICLE{2016A&A...596A..64S,
       author = {{Simmonds}, C. and {Bauer}, F.~E. and {Thuan}, T.~X. and {Izotov}, Y.~I. and {Stern}, D. and {Harrison}, F.~A.},
        title = "{Do some AGN lack X-ray emission?}",
      journal = {\aap},
         year = 2016,
        month = dec,
       volume = {596},
          eid = {A64},
        pages = {A64},
          doi = {10.1051/0004-6361/201629310},
archivePrefix = {arXiv},
       eprint = {1609.07619},
 primaryClass = {astro-ph.GA},
       adsurl = {https://ui.adsabs.harvard.edu/abs/2016A&A...596A..64S}
}

@ARTICLE{2016MNRAS.458.2288S,
       author = {{Stalevski}, Marko and {Ricci}, Claudio and {Ueda}, Yoshihiro and {Lira}, Paulina and {Fritz}, Jacopo and {Baes}, Maarten},
        title = "{The dust covering factor in active galactic nuclei}",
      journal = {\mnras},
         year = 2016,
        month = may,
       volume = {458},
       number = {3},
        pages = {2288-2302},
          doi = {10.1093/mnras/stw444},
archivePrefix = {arXiv},
       eprint = {1602.06954},
 primaryClass = {astro-ph.GA},
       adsurl = {https://ui.adsabs.harvard.edu/abs/2016MNRAS.458.2288S}
}

@ARTICLE{2012ApJ...753...30S,
       author = {{Stern}, Daniel and {Assef}, Roberto J. and {Benford}, Dominic J. and {Blain}, Andrew and {Cutri}, Roc and {Dey}, Arjun and {Eisenhardt}, Peter and {Griffith}, Roger L. and {Jarrett}, T.~H. and {Lake}, Sean and {Masci}, Frank and {Petty}, Sara and {Stanford}, S.~A. and {Tsai}, Chao-Wei and {Wright}, E.~L. and {Yan}, Lin and {Harrison}, Fiona and {Madsen}, Kristin},
        title = "{Mid-infrared Selection of Active Galactic Nuclei with the Wide-Field Infrared Survey Explorer. I. Characterizing WISE-selected Active Galactic Nuclei in COSMOS}",
      journal = {\apj},
         year = 2012,
        month = jul,
       volume = {753},
       number = {1},
          eid = {30},
        pages = {30},
          doi = {10.1088/0004-637X/753/1/30},
archivePrefix = {arXiv},
       eprint = {1205.0811},
 primaryClass = {astro-ph.CO},
       adsurl = {https://ui.adsabs.harvard.edu/abs/2012ApJ...753...30S}
}

@ARTICLE{2005ApJS..161..240T,
       author = {{Thuan}, Trinh X. and {Izotov}, Yuri I.},
        title = "{High-Ionization Emission in Metal-deficient Blue Compact Dwarf Galaxies}",
      journal = {\apjs},
         year = 2005,
        month = dec,
       volume = {161},
       number = {2},
        pages = {240-270},
          doi = {10.1086/491657},
archivePrefix = {arXiv},
       eprint = {astro-ph/0507209},
 primaryClass = {astro-ph},
       adsurl = {https://ui.adsabs.harvard.edu/abs/2005ApJS..161..240T}
}

@INPROCEEDINGS{2005ASPC..347...29T,
       author = {{Taylor}, M.~B.},
        title = "{TOPCAT \& STIL: Starlink Table/VOTable Processing Software}",
    booktitle = {Astronomical Data Analysis Software and Systems XIV},
         year = 2005,
       editor = {{Shopbell}, P. and {Britton}, M. and {Ebert}, R.},
       series = {Astronomical Society of the Pacific Conference Series},
       volume = {347},
        month = dec,
        pages = {29},
       adsurl = {https://ui.adsabs.harvard.edu/abs/2005ASPC..347...29T}
}

@ARTICLE{2010AJ....140.1868W,
       author = {{Wright}, Edward L. and {Eisenhardt}, Peter R.~M. and {Mainzer}, Amy K. and {Ressler}, Michael E. and {Cutri}, Roc M. and {Jarrett}, Thomas and {Kirkpatrick}, J. Davy and {Padgett}, Deborah and {McMillan}, Robert S. and {Skrutskie}, Michael and {Stanford}, S.~A. and {Cohen}, Martin and {Walker}, Russell G. and {Mather}, John C. and {Leisawitz}, David and {Gautier}, III, Thomas N. and {McLean}, Ian and {Benford}, Dominic and {Lonsdale}, Carol J. and {Blain}, Andrew and {Mendez}, Bryan and {Irace}, William R. and {Duval}, Valerie and {Liu}, Fengchuan and {Royer}, Don and {Heinrichsen}, Ingolf and {Howard}, Joan and {Shannon}, Mark and {Kendall}, Martha and {Walsh}, Amy L. and {Larsen}, Mark and {Cardon}, Joel G. and {Schick}, Scott and {Schwalm}, Mark and {Abid}, Mohamed and {Fabinsky}, Beth and {Naes}, Larry and {Tsai}, Chao-Wei},
        title = "{The Wide-field Infrared Survey Explorer (WISE): Mission Description and Initial On-orbit Performance}",
      journal = {\aj},
         year = 2010,
        month = dec,
       volume = {140},
       number = {6},
        pages = {1868-1881},
          doi = {10.1088/0004-6256/140/6/1868},
archivePrefix = {arXiv},
       eprint = {1008.0031},
 primaryClass = {astro-ph.IM},
       adsurl = {https://ui.adsabs.harvard.edu/abs/2010AJ....140.1868W}
}

@ARTICLE{2022ApJ...927..192Y,
       author = {{Yang}, Guang and {Boquien}, M{\'e}d{\'e}ric and {Brandt}, W.~N. and {Buat}, V{\'e}ronique and {Burgarella}, Denis and {Ciesla}, Laure and {Lehmer}, Bret D. and {Ma{\l}ek}, Katarzyna and {Mountrichas}, George and {Papovich}, Casey and {Pons}, Estelle and {Stalevski}, Marko and {Theul{\'e}}, Patrice and {Zhu}, Shifu},
        title = "{Fitting AGN/Galaxy X-Ray-to-radio SEDs with CIGALE and Improvement of the Code}",
      journal = {\apj},
         year = 2022,
        month = mar,
       volume = {927},
       number = {2},
          eid = {192},
        pages = {192},
          doi = {10.3847/1538-4357/ac4971},
archivePrefix = {arXiv},
       eprint = {2201.03718},
 primaryClass = {astro-ph.GA},
       adsurl = {https://ui.adsabs.harvard.edu/abs/2022ApJ...927..192Y}
}

@misc{jwst_pipeline,
  doi = {10.5281/ZENODO.10022973},
  url = {https://zenodo.org/doi/10.5281/zenodo.10022973},
  author = {Bushouse,  Howard and Eisenhamer,  Jonathan and Dencheva,  Nadia and Davies,  James and Greenfield,  Perry and Morrison,  Jane and Hodge,  Phil and Simon,  Bernie and Grumm,  David and Droettboom,  Michael and Slavich,  Edward and Sosey,  Megan and Pauly,  Tyler and Miller,  Todd and Jedrzejewski,  Robert and Hack,  Warren and Davis,  David and Crawford,  Steven and Law,  David and Gordon,  Karl and Regan,  Michael and Cara,  Mihai and MacDonald,  Ken and Bradley,  Larry and Shanahan,  Clare and Jamieson,  William and Teodoro,  Mairan and Williams,  Thomas and Pena-Guerrero,  Maria},
  title = {JWST Calibration Pipeline},
  publisher = {Zenodo},
  year = {2023},
  copyright = {Creative Commons Attribution 4.0 International}
}

@ARTICLE{Jiang2019,
       author = {{Jiang}, Tianxing and {Malhotra}, Sangeeta and {Rhoads}, James E. and {Yang}, Huan},
        title = "{Direct T $_{ e }$ Metallicity Calibration of R23 in Strong Line Emitters}",
      journal = {\apj},
         year = 2019,
        month = feb,
       volume = {872},
       number = {2},
          eid = {145},
        pages = {145},
          doi = {10.3847/1538-4357/aaee8a},
archivePrefix = {arXiv},
       eprint = {1811.05796},
 primaryClass = {astro-ph.GA},
       adsurl = {https://ui.adsabs.harvard.edu/abs/2019ApJ...872..145J}
}

@ARTICLE{Izotov2006,
       author = {{Izotov}, Y.~I. and {Stasi{\'n}ska}, G. and {Meynet}, G. and {Guseva}, N.~G. and {Thuan}, T.~X.},
        title = "{The chemical composition of metal-poor emission-line galaxies in the Data Release 3 of the Sloan Digital Sky Survey}",
      journal = {\aap},
         year = 2006,
        month = mar,
       volume = {448},
       number = {3},
        pages = {955-970},
          doi = {10.1051/0004-6361:20053763},
archivePrefix = {arXiv},
       eprint = {astro-ph/0511644},
 primaryClass = {astro-ph},
       adsurl = {https://ui.adsabs.harvard.edu/abs/2006A&A...448..955I}
}

@ARTICLE{Sanders2016,
       author = {{Sanders}, Ryan L. and {Shapley}, Alice E. and {Kriek}, Mariska and {Reddy}, Naveen A. and {Freeman}, William R. and {Coil}, Alison L. and {Siana}, Brian and {Mobasher}, Bahram and {Shivaei}, Irene and {Price}, Sedona H. and {de Groot}, Laura},
        title = "{The MOSDEF Survey: Electron Density and Ionization Parameter at z \raisebox{-0.5ex}\textasciitilde 2.3}",
      journal = {\apj},
         year = 2016,
        month = jan,
       volume = {816},
       number = {1},
          eid = {23},
        pages = {23},
          doi = {10.3847/0004-637X/816/1/23},
archivePrefix = {arXiv},
       eprint = {1509.03636},
 primaryClass = {astro-ph.GA},
       adsurl = {https://ui.adsabs.harvard.edu/abs/2016ApJ...816...23S}
}

@ARTICLE{2024ApJ...961..173Z,
       author = {{Zou}, Hu and {Sui}, Jipeng and {Saintonge}, Am{\'e}lie and {Scholte}, Dirk and {Moustakas}, John and {Siudek}, Malgorzata and {Dey}, Arjun and {Juneau}, Stephanie and {Guo}, Weijian and {Canning}, Rebecca and {Aguilar}, J. and {Ahlen}, S. and {Brooks}, D. and {Claybaugh}, T. and {Dawson}, K. and {de la Macorra}, A. and {Doel}, P. and {Forero-Romero}, J.~E. and {Gontcho A Gontcho}, S. and {Honscheid}, K. and {Landriau}, M. and {Le Guillou}, L. and {Manera}, M. and {Meisner}, A. and {Miquel}, R. and {Nie}, Jundan and {Poppett}, C. and {Rezaie}, M. and {Rossi}, G. and {Sanchez}, E. and {Schubnell}, M. and {Seo}, H. and {Tarl{\'e}}, G. and {Zhou}, Zhimin and {Zou}, Siwei},
        title = "{A Large Sample of Extremely Metal-poor Galaxies at z < 1 Identified from the DESI Early Data}",
      journal = {\apj},
         year = 2024,
        month = feb,
       volume = {961},
       number = {2},
          eid = {173},
        pages = {173},
          doi = {10.3847/1538-4357/ad1409},
archivePrefix = {arXiv},
       eprint = {2312.00300},
 primaryClass = {astro-ph.GA},
       adsurl = {https://ui.adsabs.harvard.edu/abs/2024ApJ...961..173Z}
}

@ARTICLE{2025arXiv250113082J,
       author = {{Ji}, Xihan and {Maiolino}, Roberto and {{\"U}bler}, Hannah and {Scholtz}, Jan and {D'Eugenio}, Francesco and {Sun}, Fengwu and {Perna}, Michele and {Turner}, Hannah and {Arribas}, Santiago and {Bennett}, Jake S. and {Bunker}, Andrew and {Carniani}, Stefano and {Charlot}, St{\'e}phane and {Cresci}, Giovanni and {Curti}, Mirko and {Egami}, Eiichi and {Fabian}, Andy and {Inayoshi}, Kohei and {Isobe}, Yuki and {Jones}, Gareth and {Juod{\v{z}}balis}, Ignas and {Kumari}, Nimisha and {Lyu}, Jianwei and {Mazzolari}, Giovanni and {Parlanti}, Eleonora and {Robertson}, Brant and {Rodr{\'\i}guez Del Pino}, Bruno and {Schneider}, Raffaella and {Sijacki}, Debora and {Tacchella}, Sandro and {Trinca}, Alessandro and {Valiante}, Rosa and {Venturi}, Giacomo and {Volonteri}, Marta and {Willott}, Chris and {Witten}, Callum and {Witstok}, Joris},
        title = "{BlackTHUNDER -- A non-stellar Balmer break in a black hole-dominated little red dot at $z=7.04$}",
      journal = {arXiv e-prints},
         year = 2025,
        month = jan,
          eid = {arXiv:2501.13082},
        pages = {arXiv:2501.13082},
          doi = {10.48550/arXiv.2501.13082},
archivePrefix = {arXiv},
       eprint = {2501.13082},
 primaryClass = {astro-ph.GA},
       adsurl = {https://ui.adsabs.harvard.edu/abs/2025arXiv250113082J}
}

@ARTICLE{2024arXiv240704777K,
       author = {{Kokubo}, Mitsuru and {Harikane}, Yuichi},
        title = "{Challenging the AGN scenario for JWST/NIRSpec LRD and non-LRD broad H$α$ emitters in light of non-detection of NIRCam photometric variability and X-ray}",
      journal = {arXiv e-prints},
         year = 2024,
        month = jul,
          eid = {arXiv:2407.04777},
        pages = {arXiv:2407.04777},
          doi = {10.48550/arXiv.2407.04777},
archivePrefix = {arXiv},
       eprint = {2407.04777},
 primaryClass = {astro-ph.GA},
       adsurl = {https://ui.adsabs.harvard.edu/abs/2024arXiv240704777K}
}

@ARTICLE{2024Natur.627...59M,
       author = {{Maiolino}, Roberto and {Scholtz}, Jan and {Witstok}, Joris and {Carniani}, Stefano and {D'Eugenio}, Francesco and {de Graaff}, Anna and {{\"U}bler}, Hannah and {Tacchella}, Sandro and {Curtis-Lake}, Emma and {Arribas}, Santiago and {Bunker}, Andrew and {Charlot}, St{\'e}phane and {Chevallard}, Jacopo and {Curti}, Mirko and {Looser}, Tobias J. and {Maseda}, Michael V. and {Rawle}, Timothy D. and {Rodr{\'\i}guez del Pino}, Bruno and {Willott}, Chris J. and {Egami}, Eiichi and {Eisenstein}, Daniel J. and {Hainline}, Kevin N. and {Robertson}, Brant and {Williams}, Christina C. and {Willmer}, Christopher N.~A. and {Baker}, William M. and {Boyett}, Kristan and {DeCoursey}, Christa and {Fabian}, Andrew C. and {Helton}, Jakob M. and {Ji}, Zhiyuan and {Jones}, Gareth C. and {Kumari}, Nimisha and {Laporte}, Nicolas and {Nelson}, Erica J. and {Perna}, Michele and {Sandles}, Lester and {Shivaei}, Irene and {Sun}, Fengwu},
        title = "{A small and vigorous black hole in the early Universe}",
      journal = {\nat},
         year = 2024,
        month = mar,
       volume = {627},
       number = {8002},
        pages = {59-63},
          doi = {10.1038/s41586-024-07052-5},
archivePrefix = {arXiv},
       eprint = {2305.12492},
 primaryClass = {astro-ph.GA},
       adsurl = {https://ui.adsabs.harvard.edu/abs/2024Natur.627...59M}
}

@ARTICLE{2024arXiv241204224M,
       author = {{Mazzolari}, G. and {Gilli}, R. and {Maiolino}, R. and {Prandoni}, I. and {Delvecchio}, I. and {Norman}, C. and {Jimenez-Andrade}, E.~F. and {Belladitta}, S. and {Vito}, F. and {Momjian}, E. and {Chiaberge}, M. and {Trefoloni}, B. and {Signorini}, M. and {Ji}, X. and {D'Amato}, Q. and {Risaliti}, G. and {Baldi}, R.~D. and {Fabian}, A. and {{\"U}bler}, H. and {D'Eugenio}, F. and {Scholtz}, J. and {Juod{\v{z}}balis}, I. and {Mignoli}, M. and {Brusa}, M. and {Murphy}, E. and {Muxlow}, T.~W.~B.},
        title = "{The radio properties of the JWST-discovered AGN}",
      journal = {arXiv e-prints},
         year = 2024,
        month = dec,
          eid = {arXiv:2412.04224},
        pages = {arXiv:2412.04224},
          doi = {10.48550/arXiv.2412.04224},
archivePrefix = {arXiv},
       eprint = {2412.04224},
 primaryClass = {astro-ph.GA},
       adsurl = {https://ui.adsabs.harvard.edu/abs/2024arXiv241204224M}
}

@ARTICLE{2025ApJ...985..253M,
       author = {{Mingozzi}, Matilde and {Garcia del Valle-Espinosa}, Macarena and {James}, Bethan L. and {Rickards Vaught}, Ryan J. and {Hayes}, Matthew and {Amor{\'\i}n}, Ricardo O. and {Leitherer}, Claus and {Aloisi}, Alessandra and {Hunt}, Leslie and {Law}, David and {Richardson}, Chris T. and {Pidgeon}, Aidan and {Arellano-C{\'o}rdova}, Karla Z. and {Berg}, Danielle A. and {Chisholm}, John and {Hernandez}, Svea and {Jones}, Logan and {Kumari}, Nimisha and {Martin}, Crystal L. and {Ravindranath}, Swara and {Vallini}, Livia and {Xu}, Xinfeng},
        title = "{Exploring the Mysterious High-ionization Source Powering [Ne V] in High-z Analog SBS0335-052 E with JWST/MIRI}",
      journal = {\apj},
         year = 2025,
        month = jun,
       volume = {985},
       number = {2},
          eid = {253},
        pages = {253},
          doi = {10.3847/1538-4357/adc996},
archivePrefix = {arXiv},
       eprint = {2502.07662},
 primaryClass = {astro-ph.GA},
       adsurl = {https://ui.adsabs.harvard.edu/abs/2025ApJ...985..253M}
}

@ARTICLE{2025arXiv250316596N,
       author = {{Naidu}, Rohan P. and {Matthee}, Jorryt and {Katz}, Harley and {de Graaff}, Anna and {Oesch}, Pascal and {Smith}, Aaron and {Greene}, Jenny E. and {Brammer}, Gabriel and {Weibel}, Andrea and {Hviding}, Raphael and {Chisholm}, John and {Labb\textbackslash'e}, Ivo and {Simcoe}, Robert A. and {Witten}, Callum and {Atek}, Hakim and {Baggen}, Josephine F.~W. and {Belli}, Sirio and {Bezanson}, Rachel and {Boogaard}, Leindert A. and {Bose}, Sownak and {Covelo-Paz}, Alba and {Dayal}, Pratika and {Fudamoto}, Yoshinobu and {Furtak}, Lukas J. and {Giovinazzo}, Emma and {Goulding}, Andy and {Gronke}, Max and {Heintz}, Kasper E. and {Hirschmann}, Michaela and {Illingworth}, Garth and {Inoue}, Akio K. and {Johnson}, Benjamin D. and {Leja}, Joel and {Leonova}, Ecaterina and {McConachie}, Ian and {Maseda}, Michael V. and {Natarajan}, Priyamvada and {Nelson}, Erica and {Setton}, David J. and {Shivaei}, Irene and {Sobral}, David and {Stefanon}, Mauro and {Tacchella}, Sandro and {Toft}, Sune and {Torralba}, Alberto and {van Dokkum}, Pieter and {van der Wel}, Arjen and {Volonteri}, Marta and {Walter}, Fabian and {Wang}, Bingjie and {Watson}, Darach},
        title = "{A ``Black Hole Star'' Reveals the Remarkable Gas-Enshrouded Hearts of the Little Red Dots}",
      journal = {arXiv e-prints},
         year = 2025,
        month = mar,
          eid = {arXiv:2503.16596},
        pages = {arXiv:2503.16596},
          doi = {10.48550/arXiv.2503.16596},
archivePrefix = {arXiv},
       eprint = {2503.16596},
 primaryClass = {astro-ph.GA},
       adsurl = {https://ui.adsabs.harvard.edu/abs/2025arXiv250316596N}
}

@ARTICLE{2022ApJ...930..105S,
       author = {{Senchyna}, Peter and {Stark}, Daniel P. and {Charlot}, St{\'e}phane and {Plat}, Adele and {Chevallard}, Jacopo and {Chen}, Zuyi and {Jones}, Tucker and {Sanders}, Ryan L. and {Rudie}, Gwen C. and {Cooper}, Thomas J. and {Bruzual}, Gustavo},
        title = "{Direct Constraints on the Extremely Metal-poor Massive Stars Underlying Nebular C IV Emission from Ultra-deep HST/COS Ultraviolet Spectroscopy}",
      journal = {\apj},
         year = 2022,
        month = may,
       volume = {930},
       number = {2},
          eid = {105},
        pages = {105},
          doi = {10.3847/1538-4357/ac5d38},
archivePrefix = {arXiv},
       eprint = {2111.11508},
 primaryClass = {astro-ph.GA},
       adsurl = {https://ui.adsabs.harvard.edu/abs/2022ApJ...930..105S}
}

@ARTICLE{2025ApJ...991L..10S,
       author = {{Setton}, David J. and {Greene}, Jenny E. and {Spilker}, Justin S. and {Williams}, Christina C. and {Labb{\'e}}, Ivo and {Ma}, Yilun and {Wang}, Bingjie and {Whitaker}, Katherine E. and {Leja}, Joel and {de Graaff}, Anna and {Alberts}, Stacey and {Bezanson}, Rachel and {Boogaard}, Leindert A. and {Brammer}, Gabriel and {Cutler}, Sam E. and {Cleri}, Nikko J. and {Cooper}, Olivia R. and {Dayal}, Pratika and {Fujimoto}, Seiji and {Furtak}, Lukas J. and {Goulding}, Andy D. and {Hirschmann}, Michaela and {Kokorev}, Vasily and {Maseda}, Michael V. and {McConachie}, Ian and {Matthee}, Jorryt and {Miller}, Tim B. and {Naidu}, Rohan P. and {Oesch}, Pascal A. and {Pan}, Richard and {Price}, Sedona H. and {Suess}, Katherine A. and {Weaver}, John R. and {Xiao}, Mengyuan and {Zhang}, Yunchong and {Zitrin}, Adi},
        title = "{A Confirmed Deficit of Hot and Cold Dust Emission in the Most Luminous Little Red Dots}",
      journal = {\apjl},
         year = 2025,
        month = sep,
       volume = {991},
       number = {1},
          eid = {L10},
        pages = {L10},
          doi = {10.3847/2041-8213/ade78b},
archivePrefix = {arXiv},
       eprint = {2503.02059},
 primaryClass = {astro-ph.GA},
       adsurl = {https://ui.adsabs.harvard.edu/abs/2025ApJ...991L..10S}
}

@ARTICLE{2025arXiv250818358W,
       author = {{Wang}, Bingjie and {Leja}, Joel and {Katz}, Harley and {Inayoshi}, Kohei and {Cleri}, Nikko J. and {de Graaff}, Anna and {Hviding}, Raphael E. and {van Dokkum}, Pieter and {Greene}, Jenny E. and {Labb{\'e}}, Ivo and {Matthee}, Jorryt and {McConachie}, Ian and {Naidu}, Rohan P. and {Nelson}, Erica J.},
        title = "{The Missing Hard Photons of Little Red Dots: Their Incident Ionizing Spectra Resemble Massive Stars}",
      journal = {arXiv e-prints},
         year = 2025,
        month = aug,
          eid = {arXiv:2508.18358},
        pages = {arXiv:2508.18358},
          doi = {10.48550/arXiv.2508.18358},
archivePrefix = {arXiv},
       eprint = {2508.18358},
 primaryClass = {astro-ph.GA},
       adsurl = {https://ui.adsabs.harvard.edu/abs/2025arXiv250818358W}
}

@misc{2014ascl.soft09004R,
       author = {{Rupke}, David S.~N.},
        title = "{IFSRED: Data Reduction for Integral Field Spectrographs}",
 howpublished = {Astrophysics Source Code Library, record ascl:1409.004},
         year = 2014,
        month = sep,
          eid = {ascl:1409.004},
archivePrefix = {ascl},
       eprint = {1409.004},
       adsurl = {https://ui.adsabs.harvard.edu/abs/2014ascl.soft09004R}
}

@ARTICLE{Hummer1987,
       author = {{Hummer}, D.~G. and {Storey}, P.~J.},
        title = "{Recombination-line intensities for hydrogenic ions - I. Case B calculations for H I and He II.}",
      journal = {\mnras},
         year = 1987,
        month = feb,
       volume = {224},
        pages = {801-820},
          doi = {10.1093/mnras/224.3.801},
       adsurl = {https://ui.adsabs.harvard.edu/abs/1987MNRAS.224..801H}
}

@ARTICLE{Gordon2023,
       author = {{Gordon}, Karl D. and {Clayton}, Geoffrey C. and {Decleir}, Marjorie and {Fitzpatrick}, E.~L. and {Massa}, Derck and {Misselt}, Karl A. and {Tollerud}, Erik J.},
        title = "{One Relation for All Wavelengths: The Far-ultraviolet to Mid-infrared Milky Way Spectroscopic R(V)-dependent Dust Extinction Relationship}",
      journal = {\apj},
         year = 2023,
        month = jun,
       volume = {950},
       number = {2},
          eid = {86},
        pages = {86},
          doi = {10.3847/1538-4357/accb59},
archivePrefix = {arXiv},
       eprint = {2304.01991},
 primaryClass = {astro-ph.GA},
       adsurl = {https://ui.adsabs.harvard.edu/abs/2023ApJ...950...86G}
}

@ARTICLE{Fitzpatrick2009,
       author = {{Fitzpatrick}, E.~L. and {Massa}, D.},
        title = "{An Analysis of the Shapes of Interstellar Extinction Curves. VI. The Near-IR Extinction Law}",
      journal = {\apj},
         year = 2009,
        month = jul,
       volume = {699},
       number = {2},
        pages = {1209-1222},
          doi = {10.1088/0004-637X/699/2/1209},
archivePrefix = {arXiv},
       eprint = {0905.0133},
 primaryClass = {astro-ph.GA},
       adsurl = {https://ui.adsabs.harvard.edu/abs/2009ApJ...699.1209F}
}

@ARTICLE{Doan2025b,
       author = {{Doan}, Sara and {Satyapal}, Shobita and {Reefe}, Michael and {Sexton}, Remington O. and {Matzko}, William and {McKaig}, Jeffrey D. and {Secrest}, Nathan J. and {Cann}, Jenna M. and {Laor}, Ari and {Canalizo}, Gabriela},
        title = "{The CLASS Quasar Catalog: Coronal Line Activity in Type 1 SDSS Quasars}",
      journal = {\apjs},
         year = 2025,
        month = oct,
       volume = {280},
       number = {2},
          eid = {57},
        pages = {57},
          doi = {10.3847/1538-4365/ade304},
archivePrefix = {arXiv},
       eprint = {2501.17067},
 primaryClass = {astro-ph.GA},
       adsurl = {https://ui.adsabs.harvard.edu/abs/2025ApJS..280...57D}
}

@ARTICLE{Pedrini2025,
       author = {{Pedrini}, Alex and {Adamo}, Angela and {Bik}, Arjan and {Calzetti}, Daniela and {Linden}, Sean T. and {Gregg}, Benjamin and {Bajaj}, Varun and {Ryon}, Jenna E. and {Buckner}, Anne S.~M. and {Bortolini}, Giacomo and {Cignoni}, Michele and {Correnti}, Matteo and {Duarte-Cabral}, Ana and {Elmegreen}, Bruce G. and {Faustino Vieira}, Helena and {Gallagher}, John S. and {Grasha}, Kathryn and {Johnson}, Kelsey E. and {Krumholz}, Mark R. and {Lapeer}, Drew and {Lai}, Thomas S.-Y. and {Messa}, Matteo and {{\"O}stlin}, G{\"o}ran and {Roos}, Linn and {Smith}, Linda J. and {Tosi}, Monica},
        title = "{The Near Infrared Spectral Energy Distribution of Young Star Clusters in the FEAST Galaxies: Missing Ingredients at 1─5 {\ensuremath{\mu}}m}",
      journal = {\apj},
         year = 2025,
        month = oct,
       volume = {992},
       number = {1},
          eid = {96},
        pages = {96},
          doi = {10.3847/1538-4357/ae0182},
archivePrefix = {arXiv},
       eprint = {2509.01670},
 primaryClass = {astro-ph.GA},
       adsurl = {https://ui.adsabs.harvard.edu/abs/2025ApJ...992...96P}
}

@ARTICLE{Knutas2025,
       author = {{Knutas}, Alice and {Adamo}, Angela and {Pedrini}, Alex and {Linden}, Sean T. and {Bajaj}, Varun and {Ryon}, Jenna E. and {Gregg}, Benjamin and {Ali}, Ahmad A. and {Andersson}, Eric P. and {Bik}, Arjan and {Bortolini}, Giacomo and {Buckner}, Anne S.~M. and {Calzetti}, Daniela and {Duarte-Cabral}, Ana and {Elmegreen}, Bruce G. and {Faustino Vieira}, Helena and {Gallagher}, John S. and {Grasha}, Kathryn and {Johnson}, Kelsey and {Lai}, Thomas S.-Y. and {Lapeer}, Drew and {Messa}, Matteo and {{\"O}stlin}, G{\"o}ran and {Sabbi}, Elena and {Smith}, Linda J. and {Tosi}, Monica},
        title = "{FEAST: JWST Uncovers the Emerging Timescales of Young Star Clusters in M83}",
      journal = {\apj},
         year = 2025,
        month = nov,
       volume = {993},
       number = {1},
          eid = {13},
        pages = {13},
          doi = {10.3847/1538-4357/ae018c},
archivePrefix = {arXiv},
       eprint = {2505.08874},
 primaryClass = {astro-ph.GA},
       adsurl = {https://ui.adsabs.harvard.edu/abs/2025ApJ...993...13K}
}

@ARTICLE{KubotaDone2018,
       author = {{Kubota}, Aya and {Done}, Chris},
        title = "{A physical model of the broad-band continuum of AGN and its implications for the UV/X relation and optical variability}",
      journal = {\mnras},
         year = 2018,
        month = oct,
       volume = {480},
       number = {1},
        pages = {1247-1262},
          doi = {10.1093/mnras/sty1890},
archivePrefix = {arXiv},
       eprint = {1804.00171},
 primaryClass = {astro-ph.HE},
       adsurl = {https://ui.adsabs.harvard.edu/abs/2018MNRAS.480.1247K}
}

@ARTICLE{Nicholls2017,
       author = {{Nicholls}, David C. and {Sutherland}, Ralph S. and {Dopita}, Michael A. and {Kewley}, Lisa J. and {Groves}, Brent A.},
        title = "{Abundance scaling in stars, nebulae and galaxies}",
      journal = {\mnras},
         year = 2017,
        month = apr,
       volume = {466},
       number = {4},
        pages = {4403-4422},
          doi = {10.1093/mnras/stw3235},
archivePrefix = {arXiv},
       eprint = {1612.03546},
 primaryClass = {astro-ph.GA},
       adsurl = {https://ui.adsabs.harvard.edu/abs/2017MNRAS.466.4403N}
}

@ARTICLE{RemyRuyer2014,
       author = {{R{\'e}my-Ruyer}, A. and {Madden}, S.~C. and {Galliano}, F. and {Galametz}, M. and {Takeuchi}, T.~T. and {Asano}, R.~S. and {Zhukovska}, S. and {Lebouteiller}, V. and {Cormier}, D. and {Jones}, A. and {Bocchio}, M. and {Baes}, M. and {Bendo}, G.~J. and {Boquien}, M. and {Boselli}, A. and {DeLooze}, I. and {Doublier-Pritchard}, V. and {Hughes}, T. and {Karczewski}, O. {\L}. and {Spinoglio}, L.},
        title = "{Gas-to-dust mass ratios in local galaxies over a 2 dex metallicity range}",
      journal = {\aap},
         year = 2014,
        month = mar,
       volume = {563},
          eid = {A31},
        pages = {A31},
          doi = {10.1051/0004-6361/201322803},
archivePrefix = {arXiv},
       eprint = {1312.3442},
 primaryClass = {astro-ph.GA},
       adsurl = {https://ui.adsabs.harvard.edu/abs/2014A&A...563A..31R}
}

@ARTICLE{Jenkins2009,
       author = {{Jenkins}, Edward B.},
        title = "{A Unified Representation of Gas-Phase Element Depletions in the Interstellar Medium}",
      journal = {\apj},
         year = 2009,
        month = aug,
       volume = {700},
       number = {2},
        pages = {1299-1348},
          doi = {10.1088/0004-637X/700/2/1299},
archivePrefix = {arXiv},
       eprint = {0905.3173},
 primaryClass = {astro-ph.GA},
       adsurl = {https://ui.adsabs.harvard.edu/abs/2009ApJ...700.1299J}
}

@ARTICLE{Thomas2018,
       author = {{Thomas}, Adam D. and {Kewley}, Lisa J. and {Dopita}, Michael A. and {Groves}, Brent A. and {Hopkins}, Andrew M. and {Sutherland}, Ralph S.},
        title = "{Mixing between Seyfert and H II Region Excitation in Local Active Galaxies}",
      journal = {\apjl},
         year = 2018,
        month = jul,
       volume = {861},
       number = {1},
          eid = {L2},
        pages = {L2},
          doi = {10.3847/2041-8213/aacce7},
archivePrefix = {arXiv},
       eprint = {1806.06364},
 primaryClass = {astro-ph.GA},
       adsurl = {https://ui.adsabs.harvard.edu/abs/2018ApJ...861L...2T}
}

@ARTICLE{McKaig2024,
       author = {{McKaig}, Jeffrey D. and {Satyapal}, Shobita and {Laor}, Ari and {Abel}, Nicholas P. and {Doan}, Sara M. and {Ricci}, Claudio and {Cann}, Jenna M.},
        title = "{Why Are Optical Coronal Lines Faint in Active Galactic Nuclei?}",
      journal = {\apj},
         year = 2024,
        month = nov,
       volume = {976},
       number = {1},
          eid = {130},
        pages = {130},
          doi = {10.3847/1538-4357/ad7a79},
archivePrefix = {arXiv},
       eprint = {2408.15229},
 primaryClass = {astro-ph.GA},
       adsurl = {https://ui.adsabs.harvard.edu/abs/2024ApJ...976..130M}
}

@ARTICLE{LaorDraine1993,
       author = {{Laor}, Ari and {Draine}, Bruce T.},
        title = "{Spectroscopic Constraints on the Properties of Dust in Active Galactic Nuclei}",
      journal = {\apj},
         year = 1993,
        month = jan,
       volume = {402},
        pages = {441},
          doi = {10.1086/172149},
       adsurl = {https://ui.adsabs.harvard.edu/abs/1993ApJ...402..441L}
}

@INPROCEEDINGS{Leitherer2005,
       author = {{Leitherer}, Claus},
        title = "{Age-Dating of Starburst Galaxies}",
    booktitle = {The Evolution of Starbursts},
         year = 2005,
       editor = {{H{\"u}ttmeister}, Susanne and {Manthey}, Eva and {Bomans}, Dominik and {Weis}, Kerstin},
       series = {American Institute of Physics Conference Series},
       volume = {783},
        month = aug,
    publisher = {AIP},
        pages = {280-295},
          doi = {10.1063/1.2034996},
archivePrefix = {arXiv},
       eprint = {astro-ph/0409407},
 primaryClass = {astro-ph},
       adsurl = {https://ui.adsabs.harvard.edu/abs/2005AIPC..783..280L}
}

@ARTICLE{Eldridge2017,
       author = {{Eldridge}, J.~J. and {Stanway}, E.~R. and {Xiao}, L. and {McClelland}, L.~A.~S. and {Taylor}, G. and {Ng}, M. and {Greis}, S.~M.~L. and {Bray}, J.~C.},
        title = "{Binary Population and Spectral Synthesis Version 2.1: Construction, Observational Verification, and New Results}",
      journal = {\pasa},
         year = 2017,
        month = nov,
       volume = {34},
          eid = {e058},
        pages = {e058},
          doi = {10.1017/pasa.2017.51},
archivePrefix = {arXiv},
       eprint = {1710.02154},
 primaryClass = {astro-ph.SR},
       adsurl = {https://ui.adsabs.harvard.edu/abs/2017PASA...34...58E}
}

@ARTICLE{Stanway2018,
       author = {{Stanway}, E.~R. and {Eldridge}, J.~J.},
        title = "{Re-evaluating old stellar populations}",
      journal = {\mnras},
         year = 2018,
        month = sep,
       volume = {479},
       number = {1},
        pages = {75-93},
          doi = {10.1093/mnras/sty1353},
archivePrefix = {arXiv},
       eprint = {1805.08784},
 primaryClass = {astro-ph.GA},
       adsurl = {https://ui.adsabs.harvard.edu/abs/2018MNRAS.479...75S}
}

@ARTICLE{Hopkins2012,
       author = {{Hopkins}, Philip F. and {Quataert}, Eliot and {Murray}, Norman},
        title = "{Stellar feedback in galaxies and the origin of galaxy-scale winds}",
      journal = {\mnras},
         year = 2012,
        month = apr,
       volume = {421},
       number = {4},
        pages = {3522-3537},
          doi = {10.1111/j.1365-2966.2012.20593.x},
archivePrefix = {arXiv},
       eprint = {1110.4638},
 primaryClass = {astro-ph.CO},
       adsurl = {https://ui.adsabs.harvard.edu/abs/2012MNRAS.421.3522H}
}

@ARTICLE{Koudmani2019,
       author = {{Koudmani}, Sophie and {Sijacki}, Debora and {Bourne}, Martin A. and {Smith}, Matthew C.},
        title = "{Fast and energetic AGN-driven outflows in simulated dwarf galaxies}",
      journal = {\mnras},
         year = 2019,
        month = apr,
       volume = {484},
       number = {2},
        pages = {2047-2066},
          doi = {10.1093/mnras/stz097},
archivePrefix = {arXiv},
       eprint = {1812.04629},
 primaryClass = {astro-ph.GA},
       adsurl = {https://ui.adsabs.harvard.edu/abs/2019MNRAS.484.2047K}
}

@ARTICLE{Collins2022,
       author = {{Collins}, Michelle L.~M. and {Read}, Justin I.},
        title = "{Observational constraints on stellar feedback in dwarf galaxies}",
      journal = {Nature Astronomy},
         year = 2022,
        month = may,
       volume = {6},
        pages = {647-658},
          doi = {10.1038/s41550-022-01657-4},
archivePrefix = {arXiv},
       eprint = {2205.06825},
 primaryClass = {astro-ph.GA},
       adsurl = {https://ui.adsabs.harvard.edu/abs/2022NatAs...6..647C}
}

@ARTICLE{Partmann2025,
       author = {{Partmann}, Christian and {Naab}, Thorsten and {Lah{\'e}n}, Natalia and {Rantala}, Antti and {Hirschmann}, Michaela and {Hislop}, Jessica M. and {Petersson}, Jonathan and {Johansson}, Peter H.},
        title = "{The importance of nuclear star clusters for massive black hole growth and nuclear star formation in simulated low-mass galaxies}",
      journal = {\mnras},
         year = 2025,
        month = feb,
       volume = {537},
       number = {2},
        pages = {956-977},
          doi = {10.1093/mnras/staf002},
archivePrefix = {arXiv},
       eprint = {2409.18096},
 primaryClass = {astro-ph.GA},
       adsurl = {https://ui.adsabs.harvard.edu/abs/2025MNRAS.537..956P}
}

@ARTICLE{Volonteri2021,
       author = {{Volonteri}, Marta and {Habouzit}, M{\'e}lanie and {Colpi}, Monica},
        title = "{The origins of massive black holes}",
      journal = {Nature Reviews Physics},
         year = 2021,
        month = sep,
       volume = {3},
       number = {11},
        pages = {732-743},
          doi = {10.1038/s42254-021-00364-9},
archivePrefix = {arXiv},
       eprint = {2110.10175},
 primaryClass = {astro-ph.GA},
       adsurl = {https://ui.adsabs.harvard.edu/abs/2021NatRP...3..732V}
}

@ARTICLE{Habouzit2017,
       author = {{Habouzit}, M{\'e}lanie and {Volonteri}, Marta and {Dubois}, Yohan},
        title = "{Blossoms from black hole seeds: properties and early growth regulated by supernova feedback}",
      journal = {\mnras},
         year = 2017,
        month = jul,
       volume = {468},
       number = {4},
        pages = {3935-3948},
          doi = {10.1093/mnras/stx666},
archivePrefix = {arXiv},
       eprint = {1605.09394},
 primaryClass = {astro-ph.GA},
       adsurl = {https://ui.adsabs.harvard.edu/abs/2017MNRAS.468.3935H}
}

@ARTICLE{Sharma2020,
       author = {{Sharma}, Mahavir and {Theuns}, Tom},
        title = "{The I{\ensuremath{\kappa}}{\ensuremath{\in}}{\ensuremath{\alpha}} model of feedback-regulated galaxy formation}",
      journal = {\mnras},
         year = 2020,
        month = feb,
       volume = {492},
       number = {2},
        pages = {2418-2436},
          doi = {10.1093/mnras/stz2909},
archivePrefix = {arXiv},
       eprint = {1906.10135},
 primaryClass = {astro-ph.GA},
       adsurl = {https://ui.adsabs.harvard.edu/abs/2020MNRAS.492.2418S}
}

@ARTICLE{Kehrig2018,
       author = {{Kehrig}, C. and {V{\'\i}lchez}, J.~M. and {Guerrero}, M.~A. and {Iglesias-P{\'a}ramo}, J. and {Hunt}, L.~K. and {Duarte-Puertas}, S. and {Ramos-Larios}, G.},
        title = "{The extended He II {\ensuremath{\lambda}}4686 emission in the extremely metal-poor galaxy SBS 0335 - 052E seen with MUSE}",
      journal = {\mnras},
         year = 2018,
        month = oct,
       volume = {480},
       number = {1},
        pages = {1081-1095},
          doi = {10.1093/mnras/sty1920},
archivePrefix = {arXiv},
       eprint = {1807.09307},
 primaryClass = {astro-ph.GA},
       adsurl = {https://ui.adsabs.harvard.edu/abs/2018MNRAS.480.1081K}
}

@ARTICLE{Nieva2012,
       author = {{Nieva}, M.-F. and {Przybilla}, N.},
        title = "{Present-day cosmic abundances. A comprehensive study of nearby early B-type stars and implications for stellar and Galactic evolution and interstellar dust models}",
      journal = {\aap},
         year = 2012,
        month = mar,
       volume = {539},
          eid = {A143},
        pages = {A143},
          doi = {10.1051/0004-6361/201118158},
archivePrefix = {arXiv},
       eprint = {1203.5787},
 primaryClass = {astro-ph.SR},
       adsurl = {https://ui.adsabs.harvard.edu/abs/2012A&A...539A.143N}
}

@ARTICLE{Gunasekera2022,
       author = {{Gunasekera}, Chamani M. and {Ji}, Xihan and {Chatzikos}, Marios and {Yan}, Renbin and {Ferland}, Gary},
        title = "{Self-consistent grain depletions and abundances I: the Orion Nebula as a test case}",
      journal = {\mnras},
         year = 2022,
        month = may,
       volume = {512},
       number = {2},
        pages = {2310-2317},
          doi = {10.1093/mnras/stac022},
archivePrefix = {arXiv},
       eprint = {2201.02882},
 primaryClass = {astro-ph.GA},
       adsurl = {https://ui.adsabs.harvard.edu/abs/2022MNRAS.512.2310G}
}

@ARTICLE{Cloudyv25,
       author = {{Gunasekera}, C.~M. and {van Hoof}, P.~A.~M. and {Dehghanian}, M. and {Chakraborty}, P. and {Shaw}, G. and {Bianchi}, S. and {Chatzikos}, M. and {Tsujimoto}, M. and {Ferland}, G.~J.},
        title = "{The 2025 release of Cloudy}",
      journal = {\rmxaa},
         year = 2025,
        month = nov,
       volume = {61},
        pages = {120-133},
          doi = {10.22201/ia.01851101p.2025.61.03.01},
archivePrefix = {arXiv},
       eprint = {2508.01102},
 primaryClass = {astro-ph.GA},
       adsurl = {https://ui.adsabs.harvard.edu/abs/2025RMxAA..61c.120G}
}

@INPROCEEDINGS{xspec,
       author = {{Arnaud}, K.~A.},
        title = "{XSPEC: The First Ten Years}",
    booktitle = {Astronomical Data Analysis Software and Systems V},
         year = 1996,
       editor = {{Jacoby}, George H. and {Barnes}, Jeannette},
       series = {Astronomical Society of the Pacific Conference Series},
       volume = {101},
        month = jan,
        pages = {17},
       adsurl = {https://ui.adsabs.harvard.edu/abs/1996ASPC..101...17A}
}

@ARTICLE{Almeida2017,
       author = {{Ramos Almeida}, Cristina and {Ricci}, Claudio},
        title = "{Nuclear obscuration in active galactic nuclei}",
      journal = {Nature Astronomy},
         year = 2017,
        month = oct,
       volume = {1},
        pages = {679-689},
          doi = {10.1038/s41550-017-0232-z},
archivePrefix = {arXiv},
       eprint = {1709.00019},
 primaryClass = {astro-ph.GA},
       adsurl = {https://ui.adsabs.harvard.edu/abs/2017NatAs...1..679R}
}

@ARTICLE{2025A&A...700A..12M,
       author = {{Mazzolari}, Giovanni and {Scholtz}, Jan and {Maiolino}, Roberto and {Gilli}, Roberto and {Traina}, Alberto and {L{\'o}pez}, Ivan E. and {{\"U}bler}, Hannah and {Trefoloni}, Bartolomeo and {D'Eugenio}, Francesco and {Ji}, Xihan and {Mignoli}, Marco and {Vito}, Fabio and {Vignali}, Cristian and {Brusa}, Marcella},
        title = "{Narrow-line AGN selection in CEERS: Spectroscopic selection, physical properties, and X-ray and radio analysis}",
      journal = {\aap},
         year = 2025,
        month = aug,
       volume = {700},
          eid = {A12},
        pages = {A12},
          doi = {10.1051/0004-6361/202451860},
archivePrefix = {arXiv},
       eprint = {2408.15615},
 primaryClass = {astro-ph.GA},
       adsurl = {https://ui.adsabs.harvard.edu/abs/2025A&A...700A..12M}
}

@ARTICLE{2011ApJ...737...67M,
       author = {{Murphy}, E.~J. and {Condon}, J.~J. and {Schinnerer}, E. and {Kennicutt}, R.~C. and {Calzetti}, D. and {Armus}, L. and {Helou}, G. and {Turner}, J.~L. and {Aniano}, G. and {Beir{\~a}o}, P. and {Bolatto}, A.~D. and {Brandl}, B.~R. and {Croxall}, K.~V. and {Dale}, D.~A. and {Donovan Meyer}, J.~L. and {Draine}, B.~T. and {Engelbracht}, C. and {Hunt}, L.~K. and {Hao}, C.-N. and {Koda}, J. and {Roussel}, H. and {Skibba}, R. and {Smith}, J.-D.~T.},
        title = "{Calibrating Extinction-free Star Formation Rate Diagnostics with 33 GHz Free-free Emission in NGC 6946}",
      journal = {\apj},
         year = 2011,
        month = aug,
       volume = {737},
       number = {2},
          eid = {67},
        pages = {67},
          doi = {10.1088/0004-637X/737/2/67},
archivePrefix = {arXiv},
       eprint = {1105.4877},
 primaryClass = {astro-ph.CO},
       adsurl = {https://ui.adsabs.harvard.edu/abs/2011ApJ...737...67M}
}

\end{document}